\documentclass{nature}
\usepackage{amsmath}
\usepackage[T1]{fontenc}
\usepackage[latin1]{inputenc}
\usepackage{graphicx}
\usepackage{amssymb}
\usepackage{dcolumn}
\usepackage{color}
\usepackage{bm}
\usepackage{pdfpages}
\usepackage[normalem]{ulem}

\usepackage{soul}
\usepackage{url}
\usepackage[ colorlinks = true,
linkcolor = red,
urlcolor  = red,
citecolor = blue,
anchorcolor = red]{hyperref}

\begin{document}

\title{Thermopower probing emergent local moments in magic-angle twisted bilayer graphene}
%Evidence of enhanced thermopower from emergent local moments in flatbands of magic-angle twisted bilayer graphene
\author{Ayan Ghosh$^{1}$\footnote{equally contributed}, Souvik Chakraborty$^{1}$\footnote{equally contributed}, Ranit Dutta$^{1}$\footnote{equally contributed}, Adhip Agarwala$^{2}$, K. Watanabe$^{3}$,T. Taniguchi$^{3}$, Sumilan Banerjee$^{1}$, Nandini Trivedi$^{4}$, Subroto Mukerjee$^{1}$ and Anindya Das$^{1}$\footnote{anindya@iisc.ac.in}}

\maketitle

\begin{affiliations}
\item Department of Physics, Indian Institute of Science, Bangalore, 560012, India.
\item Indian Institute of Technology Kanpur, Kalyanpur, Uttar Pradesh 208016, India
\item National Institute for Materials Science, 1-1 Namiki, Tsukuba 305-0044, Japan.
\item Department of Physics, The Ohio State University, Columbus, Ohio - 43210, USA.
\end{affiliations}

%%%%%%%%%%%%%%%%%%%%%%%%%%%%%%%%%%%%%%%%%%%%%
\noindent\textbf{%Recent experiments on magic-angle twisted bilayer graphene (MATBLG) indicate a coexistence of heavy~\cite{xie2019spectroscopic,wong2020cascade,saito2021isospin,rozen2021entropic,kerelsky2019maximized,jiang2019charge} and light~\cite{cao2018unconventional1,wu2021chern,zondiner2020cascade,lu2019superconductors} fermions, \textcolor{red}{similar to heavy-fermion physics~\cite{song2022magic,PhysRevLett.127.026401,PhysRevLett.131.026501,PhysRevLett.131.026502,PhysRevB.106.245129,PhysRevLett.127.266601}, where the strong correlations play a big role.}
\textcolor{black}{Recent experiments on magic-angle twisted bilayer graphene (MATBLG) have revealed the formation of flatbands, suggesting that correlation effects are likely to dominate in this system. Yet, a global transport measurement showing distinct signatures of strong correlations like local moments arising from the flatbands is missing}. Utilizing thermopower as a sensitive global transport probe for measuring entropy, we unveil the presence of emergent local moments through their impact on entropy. Remarkably, in addition to sign changes at the Dirac point ($\nu = 0$) and full band filling ($\nu = \pm 4$), the thermopower of MATBLG demonstrates additional sign changes at the location, $\nu_{cross} \sim \pm 1$, %Importantly, the location of $\nu_{cross}$ 
which do not vary with temperature from $5K$ to $\sim 60K$. \textcolor{black}{This is in contrast to sensitive temperature-dependent crossing points seen in our study on twisted bilayer graphene devices with weaker correlations.} \textcolor{black}{Further,} we have investigated the effect of magnetic field ($B$) on the thermopower, both $B_{\parallel}$ and $B_{\perp}$. Our results show a $30\%$ and $50\%$ reduction, respectively, \textcolor{black}{that is consistent with suppression seen in the layered oxide due to the partial polarization of the spin entropy.} The observed robust crossing points, together with suppression in a magnetic field, cannot be explained solely from the contributions of band fermions; instead, our data is consistent with the dominant contribution arising from the entropy of the emergent localized moments of a strongly correlated flatband.}

\section*{Introduction:}
The introduction of a relative twist angle between two or more van der Waals layers, resulting in the formation of a moir\'e superlattice, has opened a new field of exploration in condensed matter research, termed 'twistronics'\cite{cao2018correlated1,cao2018unconventional1,lu2019superconductors,liu2020tunable}. %cao2020tunable,su2023superconductivity
Within the family of twisted heterostructures, the magic-angle twisted bilayer graphene (MATBLG) with a twist angle ($\theta_M \sim 1.1^0$) is extensively studied~\cite{cao2018unconventional1,lu2019superconductors,cao2018correlated1,das2021symmetry,nuckolls_strongly_2020,stepanov2020untying,liu2021tuning,wong2020cascade,zondiner2020cascade}. %cao2020strange
The inter-layer hybridization between the rotated mono-layers of MATBLG plays a crucial role in forming isolated flat bands, resulting in the effective electronic kinetic energy being significantly smaller than the effective Coulomb interactions, thus enabling the realization of a rich phase diagram dominated by strong correlations~\cite{cao2018correlated1,kerelsky2019maximized,xie2019spectroscopic,choi2019electronic}. Emergent phenomena like superconductivity~\cite{cao2018unconventional1,yankowitz2019tuning,lu2019superconductors,doi:10.1126/science.abc2836,saito2020independent}, %codecido2019correlated,jaoui2022quantum
correlated Mott insulators~\cite{cao2018correlated1,lu2019superconductors}, Coulomb blockade in STM~\cite{xie2019spectroscopic,kerelsky2019maximized,choi2019electronic,choi2021correlation}, orbital ferromagnetism~\cite{sharpe2019emergent,lin2022spin}, anomalous Hall effects\cite{tseng2022anomalous}, quantized anomalous Hall effects~\cite{Serlin900}, nematicity~\cite{doi:10.1126/science.abc2836}, Chern insulators~\cite{wu2021chern,das2021symmetry,nuckolls_strongly_2020}, strange metal~\cite{cao2020strange}, Pomeranchuk effect~\cite{rozen2021entropic,saito2021isospin}, and giant thermopower~\cite{paul2022interaction} at low temperatures have been reported. These experimental features show a combination of properties: some are associated with itinerant electrons, while others relate to atomic orbital physics with localized moments. %combinatory properties, some associated with itinerant electrons and others associated with atomic orbital physics having localized moments.

\noindent To elucidate these experimental observations, a %appropriate 
comprehensive framework of heavy fermion physics has been invoked~\cite{PhysRevLett.127.026401,PhysRevLett.131.026502,song2022magic,PhysRevB.106.245129,PhysRevLett.127.266601}. %cualuguaru2024thermoelectric,PhysRevLett.131.026501,zhou2024kondo,wang2024molecular
In this framework, the more dispersive c-band and the flat f-band hybridize to form the emergent band of MATBLG with narrow bandwidth. In addition to the normalized bands, we expect that the flat-band of MATBLG would lead to the formation of emergent local moments due to strong correlations, with orbital, valley, and spin characteristics. \textcolor{black}{Although thermodynamic probes such as compressibility measurements\cite{rozen2021entropic,saito2021isospin} have indicated the presence of local moment fluctuations in MATBLG, and scanning tunneling microscope (STM) studies\cite{xie2019spectroscopic,kerelsky2019maximized,choi2019electronic,choi2021correlation,wong2020cascade} suggest that strong correlations dominate in MATBLG $-$ one of the key ingredients for exhibiting local moments, \textcolor{black}{a global transport measurement is still lacking to conclusively establish the presence of emergent local moments in MATBLG}.} Recent experiments on twisted bilayer~\cite{merino2024evidence} and twisted trilayer graphene~\cite{batlle2024cryo} using photo-thermoelectric effect have turned to heavy fermion theory~\cite{cualuguaru2024thermoelectric} to interpret the observed results, primarily originating from the lighter c-electrons, as opposed to f-electrons with a much shorter lifetime. \textcolor{black}{In this work, we use thermopower $-$ a global transport probe to measure a system's entropy, to study the presence of emergent local moments due to strong correlations in the flat-band of MATBLG. We believe this is a crucial step in understanding thermopower and its dichotomy with electrical transport.}

\noindent\textcolor{black}{We have conducted comprehensive thermopower studies, varying the filling factor ($\nu$) and temperature ($T$), on two types of twisted bilayer graphene (TBLG)-based devices: (i) type 1 $-$ MATBLG device with expected $U/W \gtrsim 1$, and (ii) type 2 $-$ TBLG devices with expected $U/W \lesssim 1$, where $U$ and $W$ represent the interaction strength and bandwidth, respectively. Except for low temperatures ($T < 7K$), the thermoelectric voltage ($V_{Th}$) of the MATBLG for both the conduction and valence bands remains symmetric with respect to the Dirac point (DP) with opposite signs. Aside from the anticipated sign changes in $V_{Th}$ at the DP ($\nu = 0$) and full band filling ($\nu = \pm 4$), additional sign changes or crossing occur around $\nu \sim \pm 1$. %for positive and negative filling with respect to the charge-neutral Dirac point. 
Our significant observations are as follows: \\
(1) The positions of the three crossing points $-$ at the Dirac point and the additional crossing points, $\nu_{cross} \sim \pm 1$ $-$ remain constant from $T \sim 5K$ to $60K$ for MATBLG with $U/W \gtrsim 1$. This is in contrast to the temperature-sensitive behavior of the three crossing points for TBLG devices with $U/W \lesssim 1$, for which three crossings merge into a single one around $30K$. \\
(2) Using non-interacting band electrons and weak-coupling interactions with Hartree-Fock (HF), we find that theoretical models based on the single-particle picture predict that the additional crossing points ($\nu_{cross}$) are highly temperature-dependent, and fail to capture the robustness of these crossings observed experimentally for MATBLG. \\
(3) Our theoretical models, based on dynamical mean field theory (DMFT) and a minimal model of the atomic limit with strong correlation ($W \lesssim U$), successfully capture the robustness of the additional crossing points $\nu_{cross}$ to temperature changes. \\
To validate the presence of strong correlation, we investigate the effect of a magnetic field ($B$) on $V_{Th}$ for MATBLG. We find that with $B_{\parallel}$ and $B_{\perp}$, $V_{Th}$ shows a $30\%$ and $50\%$ reduction, respectively, with minimal change in resistance. The response to magnetic field, along with the robust $\nu_{cross}$, suggests that $V_{Th}$ in MATBLG predominantly arises from the emergent local moment degrees of freedom (spin/valley), highlighting the importance of strong correlations in flat bands.}
\begin{figure*}[htbp]
\centerline{\includegraphics[width=1.0\textwidth]{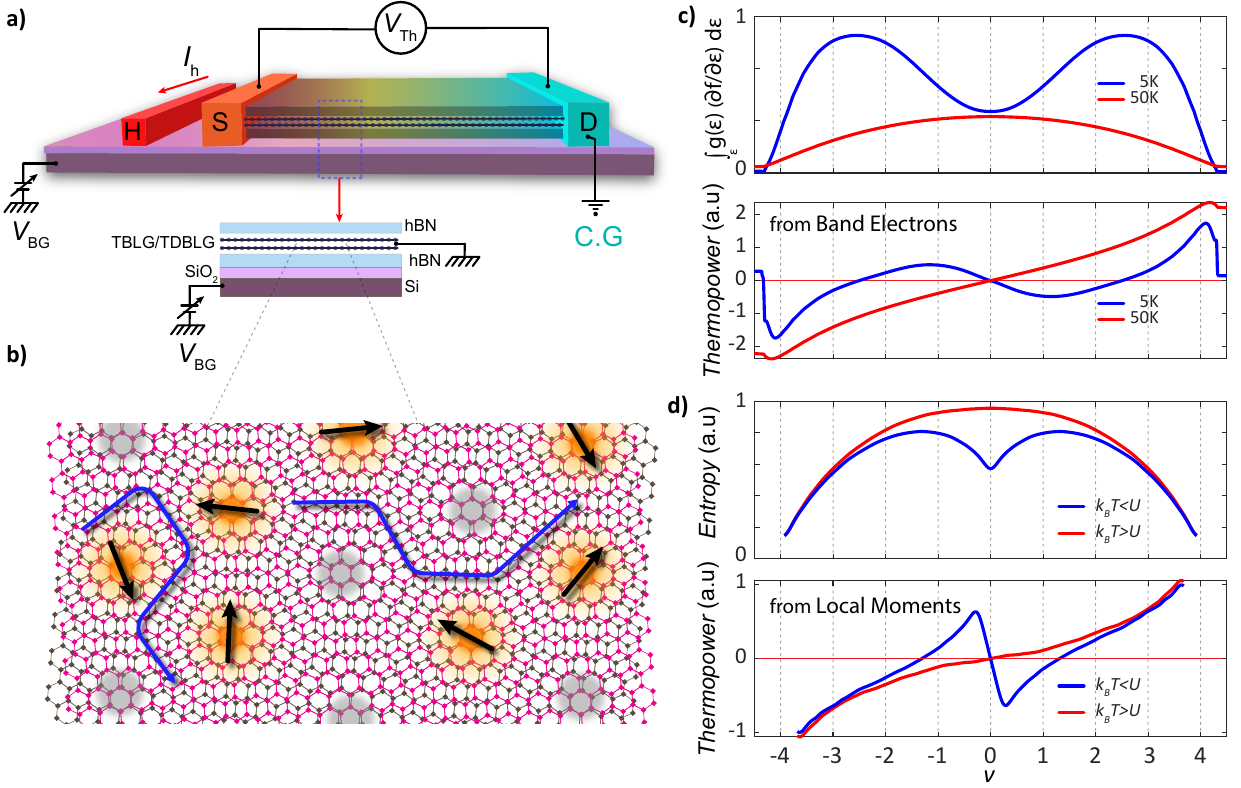}}
\caption{\label{mf1}\textbf{Device schematic and thermopower from band electrons versus strongly correlated electrons.} \textbf{a}, Schematic of the measurement setup. The devices comprise hBN-encapsulated twisted bilayer graphene (TBLG) on a $Si/SiO_2$ substrate. For the $V_{Th}$ measurement, an isolated gold heater line is positioned parallel to one side of the stack. The bottom inset shows the side view of the heterostructure. \textbf{b}, A schematic of TBLG with AA and AB stacking regions. In the strongly correlated regime, the majority of the carriers are localized at the AA sites with local moments as indicated by the black arrow, while the mobile itinerant carriers primarily move through the AB regions as indicated by the blue lines. %The f-electrons localize at the AA sites, while c-electrons primarily move through the AB regions. The f and c - bands hybridize and give rise to the normalized flat band of MATBLG. In the strongly correlated regime, 
At a given partial filling, $\nu$, some sites are occupied (yellow filled regions) while others remain empty (grey regions), and various ways to occupy the AA sites 
%arrange this occupation, 
give rise to the dominant contribution to the thermopower via configurational entropy together with the local moment's degrees of freedom (spin/valley character). \textbf{c}, In contrast, the contribution of thermopower from the non-interacting band electrons arises from the particle-hole asymmetry of a band. %{\color{red} For (c) and (d), is it possible to indicate in a figure heading that c is the thermopower from band electrons and d from local moments} 
Top panel - the effective density of states ($\int g(\varepsilon)(-df/d\varepsilon)d\varepsilon$) due to thermal broadening, where $g(\varepsilon)$ is the density of states %continuum model 
- DOS. Bottom panel - the expected thermopower from band electrons using the semi-classical equation at different temperatures. %is found to be highly sensitive to temperature. 
\textbf{d}, Top panel - the
entropy per unit moire unit cell ($S^M_{en}$) from atomic limit with $\nu$ for $k_B T > U$ and $k_B T < U$, where $U$ is the onsite Coulomb repulsion. Bottom panel - the thermopower, $S_{th} = -\frac{1}{e} \times \frac{\partial S^M_{en}}{\partial \nu}$ with $\nu$.} %For $k_B T < U$, thermopower exhibits additional sign changes around $\nu \sim \pm 1.3$ (blue line).
\label{Figure1}
\end{figure*} 

%%%%%%%%%%%%%%%%%%%%%%%%%%%%%%%%%%%%%%%%%
%Diffusive versus configurational entropic thermopower
\textcolor{black}{\subsection{Thermopower of band electrons versus strongly correlated electrons of a flat-band:} Figure~\ref{Figure1}b illustrates a schematic of twisted bilayer graphene featuring AA and AB stacking regions. The f-electrons localize at the AA sites~\cite{PhysRevLett.127.026401,li2023topological,song2022magic,cualuguaru2024thermoelectric,PhysRevLett.131.026501,lau2023topological}, while c-electrons primarily move through the AB regions~\cite{li2023topological,PhysRevLett.131.026501,lau2023topological}, as depicted in the schematic. These f- and c-electrons hybridize, resulting in the normalized flat-band of MATBLG. In the strongly correlated regime ($W \lesssim U$) of a flat-band, the thermopower ($S_{th}$) and the total entropy ($S_{en}$) are related by the Heikes limit\cite{Mukerjee_PRB_2005,PhysRevB.13.647} ($S_{th} =  -\frac{1}{e} \times \frac{\partial S_{en}}{\partial N}$), where $N$ is the total number of particles. The entropy can be considered a combination of configurational ($S_{conf}$) and spin ($S_{spin}$) contributions\cite{Mukerjee_PRB_2005}. The $S_{conf}$ can be envisioned as follows: Since the AA sites approximately contain $\sim 95\%$ of the carriers~\cite{song2022magic,PhysRevLett.131.026501}, at any filling ($\nu < \pm 4$), the $S_{conf}$ represents the number of ways the AA sites can be filled with carriers (Figure~\ref{Figure1}b), and its behavior with $\nu$ is illustrated in Figure~\ref{Figure1}d (top panel) for two regimes, $k_B T < U$ (blue line) and $k_B T > U$ (red line). The $S_{spin}$ arises from the local moment's degrees of freedom (spin/valley). The corresponding thermopower~\cite{Mukerjee_PRB_2005,peterson2010kelvin} is depicted in Figure~\ref{Figure1}d (bottom panel). As long as $k_B T < U$, thermopower exhibits additional sign changes around $\nu \sim \pm 1.3$.}

\noindent In the regime where $U \approx 0$, the thermopower of a flat band can be represented by non-interacting band electrons, for which $S_{th}$ is expected to be proportional to the derivative of the density of states (DOS) $-$ $g(\varepsilon)$. However, as temperature increases, one must consider the impact of the effective DOS ($\int g(\varepsilon)(-df/d\varepsilon)d\varepsilon$) due to thermal broadening. This effect is represented by the blue and red lines in the top panel of Figure~\ref{Figure1}c for $5K$ and $50K$, respectively. The calculated thermopower using a semi-classical description with a DOS following the continuum model\cite{} is displayed in Figure~\ref{Figure1}c (bottom panel) and is highly sensitive to temperature. If thermal broadening becomes comparable to half of the bandwidth ($3.5 k_B T \sim W/2$), the additional sign changes merge with the Dirac point.

%The mobile c-electrons contribute to the diffusive component of thermopower, proportional to the derivative of the density of states (DOS - $g(\varepsilon)$). However, as temperature increases, one should consider the impact of effective DOS ($\int g(\varepsilon)(-df/d\varepsilon)d\varepsilon$) due to thermal broadening, (represented by the blue and red lines in the top panel of Figure~\ref{Figure1}c  for $5K$ and $50K$, respectively) on the diffusive thermopower. The calculated thermopower using the semi-classical equation with a continuum model is displayed in Figure~\ref{Figure1}c (bottom panel), which is highly sensitive to temperature. If thermal broadening becomes comparable to half of the bandwidth ($3.5 k_B T \sim W/2$), the additional sign changes merge with the DP. %The total thermopower comprises contributions from both the entropic and diffusive components. 

\subsection{Device and measurement setup:} Figure~\ref{Figure1}a illustrates the schematic of our thermopower measurement setup. The devices comprise of hBN-encapsulated twisted bilayer graphene with $Si/SiO_2$ or graphite back-gated. Further details are provided in the supplementary information (SI-1 and Methods section). For the thermopower measurement, an isolated gold heater line, depicted in Figure~\ref{Figure1}a, is positioned parallel to one side of the stack. The thermoelectric voltage, $V_{Th}$ measurement employs a well-established $2\omega$ lock-in technique at $\omega = 7 Hz$~\cite{zuev2009thermoelectric,PhysRevB.80.081413,nam2010thermoelectric,wang2011enhanced,duan2016high,ghahari2016enhanced,mahapatra2020misorientation} in linear regime (see SI-2), whereas we have employed Johnson noise thermometry\cite{paul2022interaction} to measure the temperature difference ($\Delta T$) to extract the thermopower (see SI-Fig. 5). However, except low temperatures ($<10 K$), we have only concentrated on the thermoelectric voltage, $V_{Th}$, since extracting \(\Delta T\) using noise thermometry is challenging at elevated temperatures due to high background equilibrium thermal noise. %and presented throughout the manuscript. 
We have carried out measurements on four devices: MATBLG ($1.05^\circ$), near MATBLG ($0.95^\circ$), twisted double bilayer graphene (TDBLG $\sim 1.1^\circ$) with $SiO_2$ back-gated and $\sim 1.2^\circ$ TBLG with graphite back-gated. The advantages of our geometry (Figure~\ref{Figure1}a, SI-1) are discussed in SI-2, and further details can be found in our previous work~\cite{paul2022interaction,ghosh2023evidence}, where $V_{Th}$ is investigated at lower temperatures, specifically below $<10K$.

%In the following sections, we will present how the thermo-electric voltage, $V_{Th}$ evolves with the temperatures. We will consider the different temperature regimes in comparison with the energy scales of bandwidth ($W$) of the flatband, the interaction strength ($U$), and the energy gap between the dispersing lower and upper bands ($\Delta_{g}$). The experimentally determined typical energy scales for MATBLG are approximately $W \sim 10$ meV\cite{?}, $U \sim 20$ meV\cite{?}, and $\Delta_{g} \sim 30$ meV\cite{?}.

\begin{figure*}[htbp]
\centerline{\includegraphics[width=1.0\textwidth]{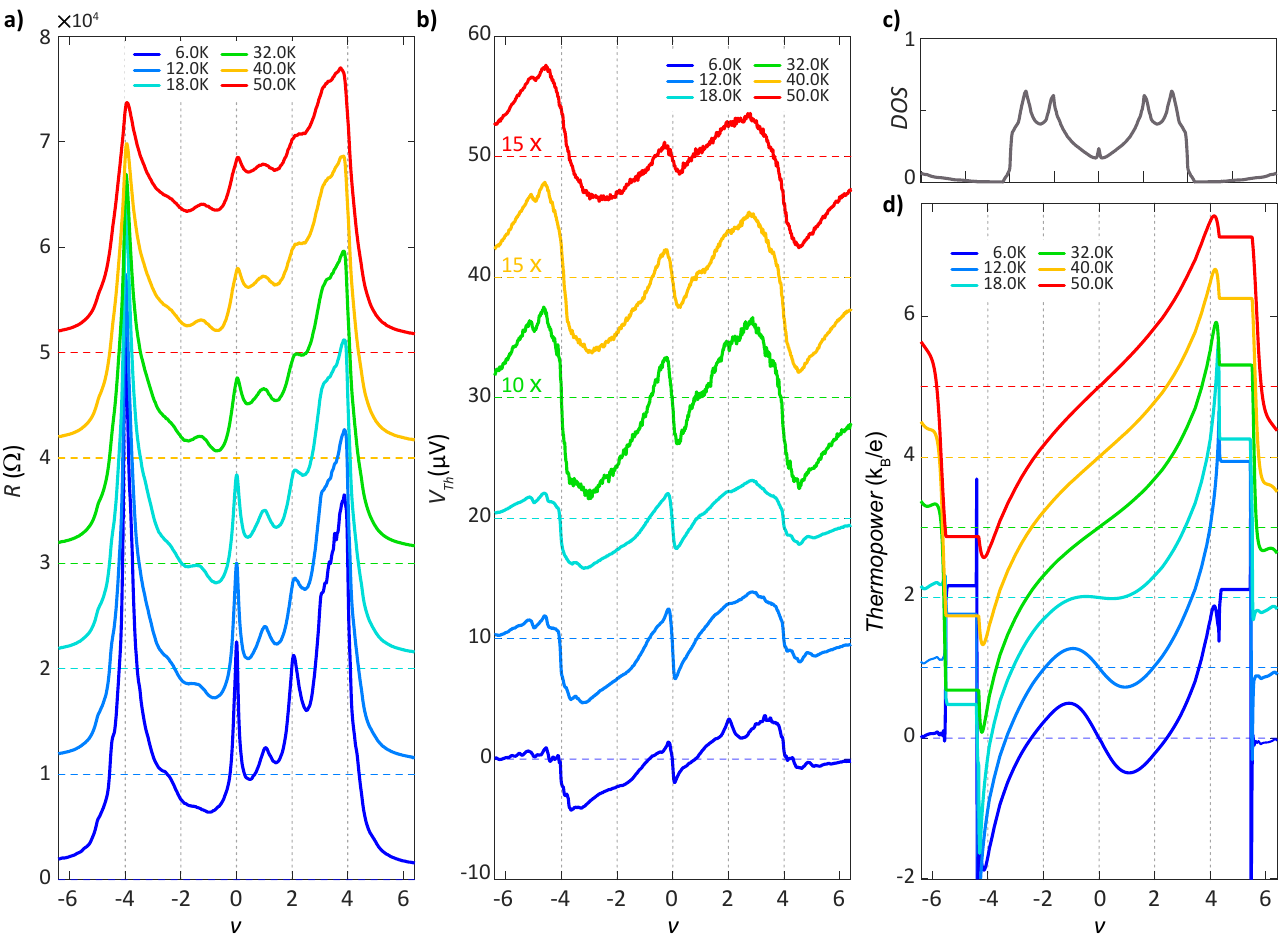}}
\caption{\label{mf1}\textbf{Low-temperature evolution of %resistance 
$R$ and %thermopower 
$V_{Th}$ of MATBLG, contrasted with the expected thermopower of band fermions.} \textbf{a}, $R$ as a function of $\nu$ plotted for several $T$. For visual clarity, each data set is offset by $10$ K$\Omega$ along the y-axis with increasing $T$. In general, $R(T)$ shows metallic behavior except at the DP ($\nu=0$), and at $\nu=\pm4$ when the bands are fully filled, where $R(T)$ shows insulating behaviour. Additional peaks in $R(\nu)$ emerge at positive integer filling ($\nu=1,2,3$), i.e. for the conduction band, but are weaker for the valence band. 
%With $T$, the $R$ shows metallic behavior except at DP and $\nu=\pm4$. 
 \textbf{b} $V_{Th}(\nu)$ for several $T$ with an offset of $10\mu V$. $V_{Th}$ is symmetric for the conduction and valence bands with opposite signs (except at $6K$), which starkly contrasts the asymmetry of resistance data. The sign of $V_{Th}$ changes at DP and $\nu=\pm4$ with additional points at $\nu_{cross} \approx\pm1$. It is significant to note that the location of $\nu_{cross}$ barely changes with increasing $T$. \textbf{c} Non-interacting single-particle density of states (DOS) of flat-bands using a continuum model\cite{zondiner2020cascade} with the inclusion of parabolic dispersive bands\cite{das2021symmetry}. \textbf{d} Thermopower {in shifted plot} calculated semi-classically ($S_{SC}$) using the DOS in (c). Note that at low $T$, $S_{SC}$ shows crossing between $0<\mid\nu\mid<4$, which is strongly dependent on $T$, and vanishes beyond $18K$. $S_{SC}$ at higher $T$ becomes positive (negative) for the conduction (valence) band caused by the thermal broadening ($-df/d\varepsilon$).}
\label{Figure2}
\end{figure*}

\noindent In the upcoming sections, we discuss our measurements of the temperature-dependent evolution of the thermoelectric voltage ($V_{Th}$). We explore different temperature ranges in relation to the key energy scales of the system, including the bandwidth ($W$) of the flat band, the interaction strength ($U$), and the energy gap between the dispersive lower and upper bands ($\Delta_{g}$). \textcolor{black}{We compare the results from two types of devices: (i) $U/W \gtrsim 1$ $-$ $SiO_2$ back-gated $1.05^\circ$ MATBLG %and $0.95^\circ$ near-MATBLG,
and (ii) $U/W \lesssim 1$ $-$ $1.1^\circ$ TDBLG and $1.2^\circ$ TBLG with graphite gating. This categorization is based on their resistance response to filling and temperature, as well as their distinct thermopower responses. The stronger correlation in $SiO_2$ back-gated MATBLG is evidenced by the more pronounced resistance peaks at integer fillings, which persist beyond 100K, in contrast to the weaker resistance peaks in graphite-gated TBLG, which vanish within 50K (see Figure. \ref{Figure2},\ref{Figure5} and SI-Fig. 9). This is further supported by experimentally determined typical energy scales for the $SiO_2$ back-gated MATBLG (via STM\cite{xie2019spectroscopic}), which are approximately $W \sim 20$ meV, $U \sim 20-30$ meV\cite{xie2019spectroscopic,wong2020cascade}, and $\Delta_{g} \sim 30-40$ meV\cite{cao2018correlated1}, suggesting that $SiO_2$ back-gated MATBLG device are in the strong correlation regime.}

%%%%%%%%%%%%%%%%%%%%%%%%%%%%%%%%%%%%%%%%%%%%%%%%%%%%%%%%%%%%%%%%%
%%%%%%%%%%%%%%%%%%%%%%%%%%%%%%%%%%%%%%%%%%%%%
%%%%%%% figure begins
%%%%%%% figure ends

%%%%%%%%%%%%%%%%%%%%%%%%%%%%%%%%%%%%%%%%%%%%%
%%%%%%%%%%%%%%%%%%%%%%%%%

%%%%%%%%%%%%%%%%%%%%%%%%%%%%%%%%%%%%%%%%%%%%%
\subsection{Anomalous thermopower response of MATBLG at low temperatures:} %Temperature evolution of resistance and thermopower response of MATBLG: 
In this section, we showcase the data within a temperature range that is both lesser and comparable to $W$, yet lower than $U$ and $\Delta_{g}$. In Figure~\ref{Figure2}a, the resistance ($R$) as a function of filling ($\nu$) is plotted for several temperatures ($T$) between $6K$-$50K$. \textcolor{black}{The $\nu = 4n/n_{s}$ is the moire filling factor, where $n$ and $n_s$ are the carrier density induced by the gate voltage and the carrier density required to fully fill the flat band (4 electrons/holes per moire unit cell), respectively.} Insulating resistance peaks can be observed at the DP ($\nu=0$) as well as the band full-filling points ($\nu=\pm4$). Additional prominent peaks emerge at positive integer filling ($\nu=1,2,3$), i.e. for the conduction band, which survives even up to $100K$. In contrast, such peak features at integer filling remain very weak for the valence band. %Further, with $T$, the resistance mostly shows metallic behaviour (from $10K$ to $\sim 120 K$) throughout the flat band except at the primary Dirac point and full-filling points. 

\noindent In Figure~\ref{Figure2}b, thermoelectric voltage ($V_{Th}$) with $\nu$ is plotted for same $T$ values between $6K$-$50K$. $V_{Th}$ shows sign changes at the primary DP and full-filling points. Additionally, we observe sign changes near $\nu\approx\pm 0.95$. It is to be noted that the position of the crossing point at $\nu\approx\pm 0.95$ barely evolves with increasing $T$ from $5K-50K$, which is an order of magnitude change in temperature. At low temperatures, a peak in $V_{Th}$ appears around $\nu=2$. %and will be discussed later. %in the discussion section. 
However, at temperatures above $8K$, $V_{Th}$ appears very symmetric between the valence and conduction band with opposite signs, which starkly contrasts the asymmetric nature of the resistance data in Figure~\ref{Figure2}a. Specifically, $V_{Th}$ violates the Mott relation~\cite{PhysRev.181.1336} ($dlnR/d\nu$) at integer fillings, as depicted in Figure~\ref{Figure5}c (top three panels) and in SI-Fig. 27. The $dlnR/d\nu$, according to Mott's description of thermopower~\cite{PhysRev.181.1336}, anticipates sign changes at resistance peaks and dips, illustrated by the blue curves in Figure~\ref{Figure5}c. Even up to $T \sim 90$K, six sign changes are observed for the conduction band, distinct from the sign changes at the DP and band full-filling. This starkly contrasts with the measured $V_{Th}$ data presented in Figure~\ref{Figure2}b. \textcolor{black}{The Mott relation is violated in all our devices within certain density and temperature ranges, as shown in SI-15. Various possible origins for this violation are discussed in SI-16. Among these, the presence of interaction is one of the origins; however, it does not alone elucidate the physics of local moments.}

%We will establish that this discrepancy can be attributed to strong correlations manifested in the emergence of local moments, leading to markedly different signatures in resistance and thermopower.

%%%%%%%%%%%%%%%%%%%%%%%%%%%%%%%%%%%%%%%%%%%%
\begin{figure*}[ht!]
\centerline{\includegraphics[width=1.0\textwidth]{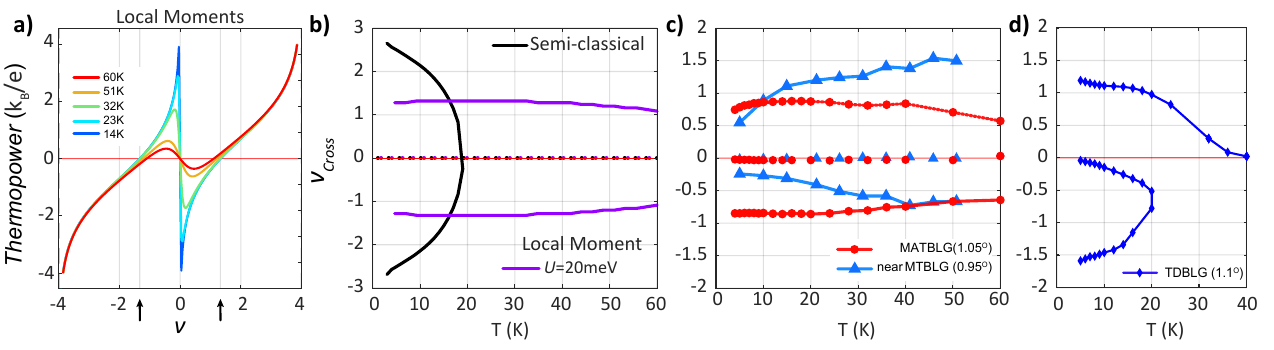}}
\caption{\label{mf3.5}\textbf{Expected thermopower from the local moments and comparison with the experiment:}
\textbf{a,} Theoretically calculated thermopower in strongly correlated regime using atomic limit\cite{PhysRevB.13.647,Mukerjee_PRB_2005,peterson2010kelvin} %from the configurational entropy of local moments~\cite{webb1962thermoelectricity} 
at different temperatures with $U = 20$meV. The additional crossings, $\nu_{cross}$ around $\pm 1.3$, are indicated by vertical black arrows. \textbf{b,} The extracted $\nu_{cross}$ from Figure~\ref{Figure3}a as a function of temperature is shown in a solid violet line. For contrast, $\nu_{cross}$ from non-interacting band electrons using semi-classical theory is shown in black line. %which remains sensitive with $T$ and merges to DP within $20K$.
%{{Temperature evolution of the entropic thermopower from the local moments in comparison with the experiment.} \textbf{a} Theoretical $\nu_{cross}$ from the entropy of the local moments (blue line) using atomic orbital physics or Heikes limit~\cite{webb1962thermoelectricity} for onsite Coulomb repulsion, $U = 20$ meV. $\nu_{cross}$ from the band electrons (black line) using semi-classical theory, %which remains sensitive with $T$ and merges to DP within $20K$.
\textbf{c,} Evolution of the crossing points (within $0<\mid\nu\mid<4$) with $T$ for MATBLG (red circles) and near MATBLG (blue triangles). Note that for the strongly correlated MATBLG with flatter bands, $\nu_{cross}$ remains 
almost constant, resembling the thermopower from the local moments in Figure~\ref{Figure3}b. Though, for the near MATBLG device with not as flatbands, there is greater variation in $\nu_{cross}$.
\textbf{d,} Twisted double bilayer graphene (TDBLG) with weaker correlation exhibits notable $\nu_{cross}$ variations with $T$: in the valence band, $\nu_{cross}$ shifts to lower $\nu$ and disappears by $20K$, while in the conduction band, $\nu_{cross}$ approaches $\nu=0$ around $35K$, and qualitatively resembles of black solid line {i.e. non-interacting band electrons} in Figure~\ref{Figure3}b.}
\label{Figure3}
\end{figure*} 
%%%%%%%%%%%%%%%%%%%%%%%%%%%%%%%%%%%%%%%%%%%%

%%%%%%%%%%%%%%%%%%%%%%%%%%%%%%%%%%%%%%%%%%%%%%%%

\noindent To get a sense of the effect of the band on temperature-dependent $V_{Th}$ (non-interacting), we have calculated the thermopower using a semi-classical approach (details in Methods and SI-8). The non-interacting single-particle density of states of the flat band calculated using the continuum model\cite{zondiner2020cascade} ($W \sim 10$meV) with the inclusion of parabolic dispersing\cite{das2021symmetry} lower and upper bands is shown in Figure~\ref{Figure2}c. Semi-classically, the thermopower can be approximated as; \\$S_{SC}=-(1/Te)[\int (\epsilon-\mu)g(\varepsilon)(-df/d\varepsilon)d\varepsilon]/[\int g(\varepsilon)(-df/d\varepsilon)d\varepsilon]$, where $e$, $T$, $\mu$, $g(\varepsilon)$ and $-df/d\varepsilon$ are, respectively, the electronic charge, temperature, chemical potential, density of states and derivative of Fermi function. $S_{SC}$ has been calculated with self-consistently solved $\mu$. At low temperatures, $S_{SC}$ also shows a crossing between $0<\mid\nu\mid<4$, which is strongly tunable with $T$ and vanishes beyond $18K$ as shown in Figure~\ref{Figure2}d. At elevated temperatures, $S_{SC}$ transitions to positive (negative) values for the conduction (valence) band. This arises due to thermal broadening (full width at half maximum of $-df/d\varepsilon \sim 3.5k_B T$), treating the flat-bands as a single band, and the charge neutrality point (CNP) behaves resembling a transition point from electron to hole-like for a half-filled single band. \textcolor{black}{In SI-8, calculations were further extended to varying bandwidths up to $40$ meV and using different types of DOS, from continuum to saw-tooth. The results show that the crossing points in thermopower from non-interacting band electrons is highly sensitive to temperature, although the merging to a single crossing point depends on the magnitude of the bandwidth. Additionally, calculations were performed in the presence of interaction in the weak-coupling regime using Hartree-Fock (HF), as shown in SI-Fig. 16, and the outcome remains similar to non-interacting band electrons. Comparing Figure~\ref{Figure2}b with Figure~\ref{Figure2}d, it is evident that single-particle band electrons fail to capture the measured $V_{Th}$. In the next section, we will discuss how thermopower from local moments qualitatively captures our experimental data.}

\subsection{Thermopower from the local moments and comparison with the experiment:} \textcolor{black}{In order to capture the effect of correlation, we have considered various theoretical models (SI-10 and SI-11) with increasing correlation effects. Among these, the dynamical mean-field theory (DMFT) at intermediate interaction strength $U/W\lesssim 1$ captures the robust crossing in thermopower as shown in SI-Fig. 17. To see the effect of strong correlation~\cite{PhysRevB.108.075101} in thermopower, we employ a minimal model of atomic limit, the limiting case of extremely strong interaction, and the bandwidth approaching zero, $t\ll k_\mathrm{B}T\ll U$.} %To find the contribution of thermopower originating from the strong correlations~\cite{PhysRevB.108.075101} or local moments, we employ a minimal model of the atomic limit of 
In the atomic limit, the thermopower is known as Heikes limit\cite{PhysRevB.13.647,Mukerjee_PRB_2005} ($S_{th} =  -\frac{1}{e} \times \frac{\partial S^M_{en}}{\partial \nu}$, where $S^M_{en}$ - entropy per moire unit cell), and further can be derived from the Kelvin formula~\cite{peterson2010kelvin} (details in SI-11). %given by {$\frac{-dS_{con}}{e d\nu}$}. 
In the regime of strong correlation, %($W \leq U$), 
a reduction in entropy occurs when the band is half-filled (with 4 out of 8 orbitals filled), as depicted by the blue line in Figure~\ref{Figure1}d (top panel). This reduction leads to additional crossing points in the thermopower at $\nu_{cross} \sim \pm 1.3$, as observed in our theoretical calculations for $U = 20$ meV~\cite{xie2019spectroscopic} shown in Figure~\ref{Figure3}a. \textcolor{black}{The temperature independence of the crossing points in the atomic limit for varying $U$ from 15 to 40 meV is demonstrated in SI-Fig. 19c.
% , where robust crossing points persist above 60K. 
% Further, to consider the finite bandwidth of the flat bands, the calculations are done with multi-orbitals, which results in the robust crossing around $\sim \pm 1.1$, close to the experimentally observed one. 
To incorporate finite hopping in our calculation, thermopower is computed using exact diagonalization for a small cluster of moire sites via the Kelvin formula, showing the robust crossing (SI-Fig. 22). As shown below, among the various theoretical models, the simplest minimal model of atomic limit explains the most striking features of our experiment.} %. Additionally, thermopower calculations using DMFT with finite bandwidth in the intermediate to strong correlation regime (SI-Fig. 17) also capture robust crossing points in thermopower, in contrast to non-interacting band electrons. 

%Detailed descriptions of our theoretical calculations are provided in SI-4 as well as in form of Extended Data Fig. 9 (for different interaction scales).
%These models are discussed in detail in the later sections. We show that the simplest toy model based on atomic limit can explain the most striking features 

\noindent The solid violet line in Figure~\ref{Figure3}b illustrates the evolution of $\nu_{cross}$ with temperature, for thermopower from strong correlations using atomic limit in comparison with the non-interacting band electrons, $S_{SC}$ (solid black line). Figure~\ref{Figure3}b highlights that $\nu_{cross}$ from the strong correlations or local moments remains nearly constant, %within $60K$
contrasting with $\nu_{cross}$ from band electrons, which are sensitive to temperature. %The latter merges with the DP within $20K$. 
In Figure~\ref{Figure3}c, the evolution of experimentally measured $\nu_{cross}$ within $0<\mid\nu\mid<4$ is presented for MATBLG (red circles) and near MATBLG (blue triangles) devices, while Figure~\ref{Figure3}d shows the results for TDBLG and shown for graphite-gated TBLG in SI-Fig. 9b. Comparing Figure~\ref{Figure3}b with ~\ref{Figure3}c and SI-Fig. 9b, the MATBLG device closely resembles atomic limit physics with strong correlation, a trend also observed for the near MATBLG device, albeit with some variation (possibly arising from the larger bandwidth and effect from the higher dispersive band). %as well as some level of screening from the metal top gate). %and effect from the higher dispersive band. 
However, the TDBLG and graphite-gated TBLG devices qualitatively align closely with the thermopower from single-particle band electrons (or weaker correlation effect), which is not unexpected, considering that for TDBLG with $1.1^\circ$, correlation effects are weaker at zero displacement field %\cite{liu2020tunable,cao2020tunable} 
as well as expected for the screened TBLG device. The raw data of $V_{Th}$ related to all the devices are shown in SI-3.

%%%%%%%%%%%%%%%%%%%%%%%%%%%%%%%%%%%%%%%%%%%%%
%%%%%%% figure begins
\begin{figure*}[ht!]
\centerline{\includegraphics[width=1\textwidth]{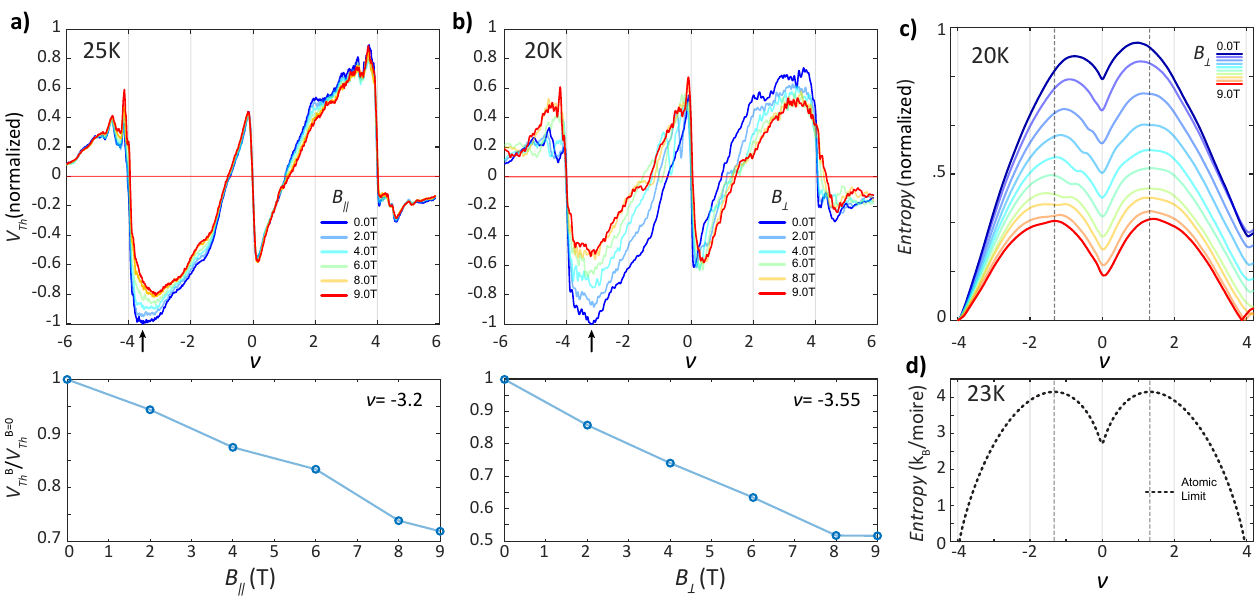}}
\caption{\label{mf3}\textbf{Magnetic field dependence of thermoelectric voltage.} \textbf{a}, $V_{Th}$ with $\nu$ for different $B_{\mid\mid}$ at $25K$. Here, the $V_{Th}$ is normalized to $V_{Th}$ measured at $B_{\mid\mid}=0T$. The reduction of $V_{Th}$ can be observed with increasing $B_{\mid\mid}$, and the effect being more pronounced for the valence band and plotted at $\nu=-3.2$ (vertical black arrow), which shows $30\%$ reduction (bottom panel). \textbf{b}, Evolution of normalized $V_{Th}$ with $B_\perp$ at {$20K$}. In contrast to $B_{\mid\mid}$, a larger decrease of the $V_{Th}$ signal is observed with increasing $B_\perp$, and plotted for $\nu=-3.55$ (vertical black arrow), where $50\%$ reduction can be seen (bottom panel). \textbf{c} Normalized entropy %(in arbitrary units) 
estimated from the integral of the of $V_{Th}$ with $\nu$ at $20K$ from $B_\perp$ data. The peak position of the entropy saturates to $\nu \sim \pm1.3$ (vertical dashed black lines) with higher $B_\perp$. The shape of the entropy resembles the theory %Heikes limit~\cite{webb1962thermoelectricity} 
from atomic limit shown in \textbf{d} for zero magnetic field.}
\label{Figure4}
\end{figure*} 
%%%%%%% figure ends
%%%%%%%%%%%%%%%%%%%%%%%%%%%%%%%%%%%%%%%%%%%%%
%%%%%%%%%%%%%%%%%%%%%%%%%%%%%%%%%%%%%%%%%%%%%
\subsection{Magnetic field dependence of thermopower:} %and evidence for local moments in MATBLG :} 
To establish that the primary contribution to $V_{Th}$ arises from the local moments of the flat-band%atomic orbital physics of f-electron
~\cite{PhysRevLett.127.026401,song2022magic,zhou2024kondo,PhysRevLett.131.026501,PhysRevLett.131.026502,cualuguaru2024thermoelectric}, it is necessary to demonstrate the influence of local moment degrees of freedom such as spin/valley with a magnetic field. For instance, the polarization of these degrees of freedom would lead to decreased entropy, resulting in a reduction of $V_{Th}$. To achieve the polarization of local moments, we investigate the impact of $B_{\parallel}$ and $B_{\perp}$.

\noindent Figure~\ref{Figure4}a-top panel shows $V_{Th}$ with $\nu$ for different $B_{\parallel}$ at $25K$. Here, the $V_{Th}$ at a given field is normalized to $B_{\parallel}=0T$. %With increasing $B_{\parallel}$ we observe no change in the positions of $\nu_{cross}$ near $\mid\nu\mid\approx1$. However, 
With increasing $B_{\parallel}$, in the range between a quarter ($\mid\nu\mid=1$) and full-filling ($\mid\nu\mid=4$) of both valence and conduction band, a clear decrease of %the magnitude of 
$V_{Th}$ is observed. The effect is more pronounced in the valence band and quantified by plotting its relative magnitude with $B_{\parallel}$ in Figure~\ref{Figure4}a-bottom panel for $\nu = -3.2$, and $30\%$ reduction can be seen. The reduction in $V_{Th}$ with $B_{\parallel}$ is in stark contrast to the resistance response (see SI-Fig. 10b), where no noticeable change in $R$ with $B_{\parallel}$ can be observed. Figure~\ref{Figure4}b (top panel) illustrates the evolution of normalized $V_{Th}$ with $B_{\perp}$ at $20K$. In contrast to $B_{\parallel}$, a more substantial decrease in the $V_{Th}$ signal is observed with increasing $B_{\perp}$, quantified in Figure~\ref{Figure4}b (bottom panel) for $\nu = -3.5$, where a $50\%$ reduction is evident. It is worth noting that at $20K$, no clear signature of Landau level (L.L) formation was observed with $B_{\perp}$ in the resistance data (see SI-Fig. 10f). %except very close to the DP
Additionally, L.Ls are expected to exhibit oscillations in $V_{Th}$~\cite{PhysRevB.29.1939}, which is not seen in Figure~\ref{Figure4}b due to thermal broadening except for a weak modulation near DP. Similarly, no signature of the Hofstadter-butterfly effect\cite{yu2022correlated} was seen in our data at $20K$. The larger reduction with $B_{\perp}$ compared to $B_{\parallel}$ may result from the polarization of both spin~\cite{wang2003spin} and valley. %sharpe2019emergent,serlin2020intrinsic,dutta2024electric
The valley can be coupled with $B_{\perp}$ through the orbital magnetic moments~\cite{lin2022spin,tseng2022anomalous,PhysRevB.102.041402,PhysRevLett.125.236804}, however, needs further theoretical exploration to understand its influence on thermopower. \textcolor{black}{In Figure~\ref{Figure4}c, we have plotted the integral of $V_{Th}$ to $\nu$ to see how the relative magnitude of entropy ($S_{en}$) reduces with increasing magnetic field. One can notice that the shape of Figure~\ref{Figure4}c resembles the theoretically calculated entropy from the atomic limit in Figure~\ref{Figure4}d (details in SI-Fig. 19). The study with the magnetic field is performed with a fixed heater current, keeping other parameters fixed. 
%and all other configurations remained unchanged, indicating tha
The measurements are done at a constant \(\Delta T\) (as in Ref. \cite{paul2022interaction}), hence the relative change in $V_{Th}$ is expected to be the same as in the thermopower. The exact magnitude of the entropy is shown in SI-Fig. 25 at $6K$ in units of \(k_B/\text{moire}\), and an estimation is shown for $ 20K$.}

\noindent \textcolor{black}{Note that the reduction in $V_{Th}$ with $B$ is consistent with the experiment on strongly correlated oxides \cite{wang2003spin} and seen in compressibility measurements\cite{rozen2021entropic,saito2021isospin} for MATBLG. Our results can be captured by a minimal model of atomic limit\cite{Mukerjee_PRB_2005}. However, almost no reduction is observed within $-1 < \nu < 1$ in $B_\parallel$ in contrast to $B_{\perp}$, which could arise from the influence of a finite bandwidth and from orbital effects that are not considered in our calculations. This requires higher-order corrections and is beyond the scope of our present calculation. It should be noted that the application of a magnetic field can also reduce $V_{Th}$ for non-interacting band electrons; %via polarization; 
however, as shown in SI-Fig. 15, at $20K$ non-interacting electrons show barely any change between $0T$ and $10T$. Though the reduction of \(V_{Th}\) with \(B\) is indicative of local moments, that feature alone is not conclusive evidence of local moments; instead, the combined effect of \(B\) and the robust crossing points in \(V_{Th}\) with temperature %behavior of the crossing points in \(V_{Th}\) with temperature (as observed in our experiment, where three crossing points remain at fixed positions from $5K$ to $60K$ for MATBLG) 
sheds light on the presence of local moments and strong correlation.}

%In Figure~\ref{Figure4}b (top panel), the shift of $\nu_{cross}$ to a higher value with $B_{\perp}$ is evident. This shift is further illustrated as entropy in Figure~\ref{Figure4}e (in arbitrary units), estimated from the integral of $V_{Th}$ to $\nu$ ($V_{Th}\approx\frac{-dS}{e d\nu}$, {where $S$ is entropy}). The shape of the entropy and the $\nu_{cross}$ approaching $\nu \sim \mid 1.3 \mid$ (peak positions of solid red line in Figure~\ref{Figure4}c) at $B_{\perp} \sim 8-9 T$ resembles the theoretically calculated entropy from atomic orbital physics (dashed black line in Figure~\ref{Figure4}c)~\cite{webb1962thermoelectricity,peterson2010kelvin}. We speculate that the shift of $\nu_{cross}$ towards $1.3$ with $B_{\perp}$ could be due to the localization of carriers induced by $B_{\perp}$ (via orbital effect), bringing the system more closer to resembling atomic orbitals.

%and is shown in SI-?, known as the Heikes limit~\cite{webb1962thermoelectricity}, where $V_{Th}\approx-\frac{\mu}{Te} \approx \frac{-dS}{e d\nu}$~\cite{peterson2010kelvin}.

%%%%%%%%%%%%%%%%%%%%%%%%%%%%%%%%%%%%%%%%%%%%%
%%%%%%%%%%%%%%%%%%%%%%%%%%%%%%%%%%%%%%%%%%%%%
%%% figure one is here 
\begin{figure*}[htbp]
\centerline{\includegraphics[width=1.0\textwidth]{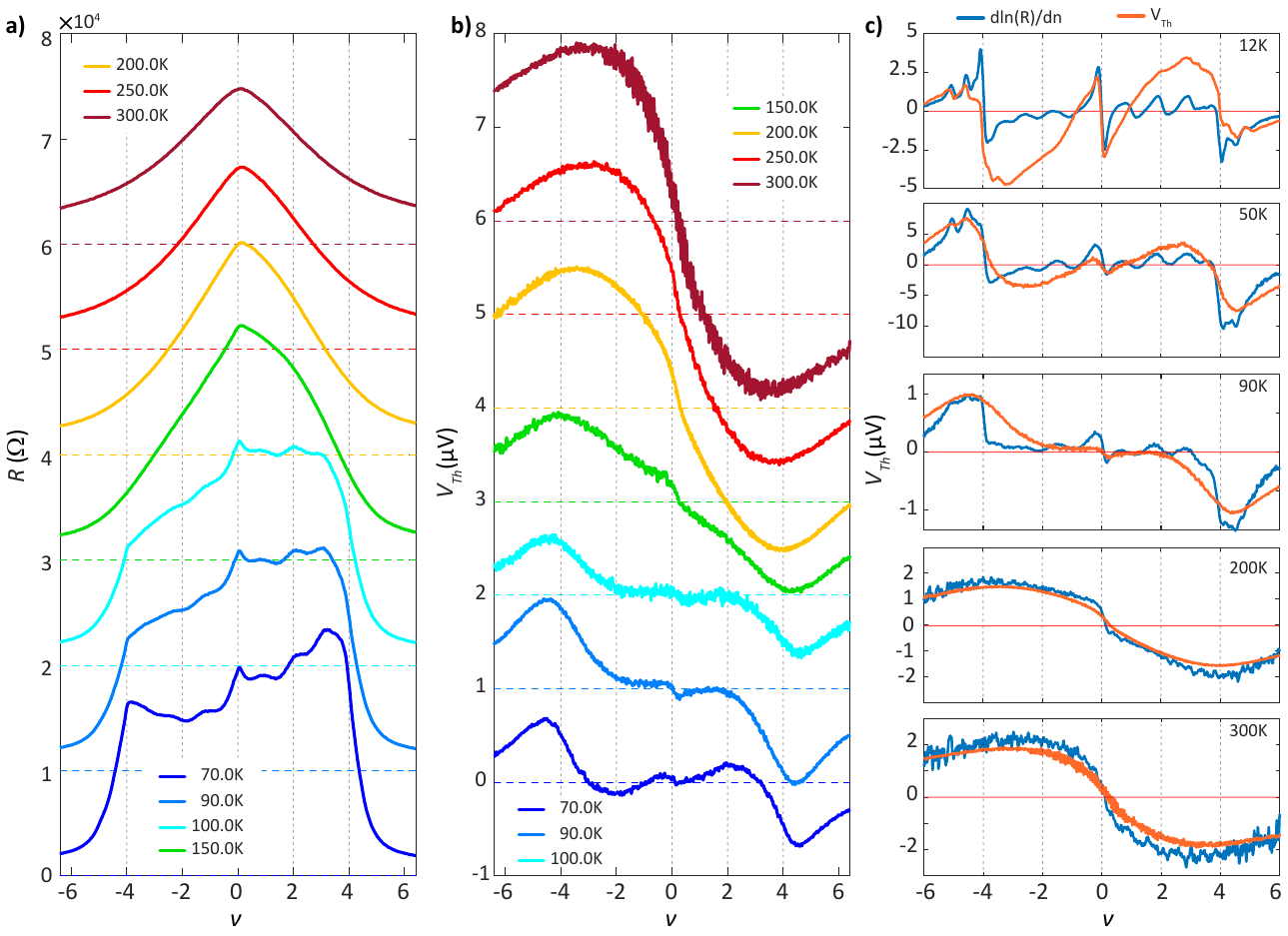}}
\caption{\label{mf2}\textbf{High-temperature response of %resistance 
$R$ and %thermopower 
$V_{Th}$ of MATBLG, and comparison with Mott formula.}  \textbf{a}, Evolution of $R$ with $\nu$ for $T \sim 70-300 K$. %In the temperature range $70 - 100K$, features of the flat band are seen as large resistance peaks at $\nu=\pm4$ and smaller kinks at integer $\nu$. 
Beyond $100K$, the flat band features fade and behave %revert to the underlying 
graphene-like resistance spectrum. %scan as a function of doping. 
\textbf{b},  Evolution of $V_{Th}$ with $\nu$ within the temperature range $70-300 K$. %Above $70K$, no crossings as a function of $\nu$ are observed inside the flat band except at the Dirac point. 
For $T \sim 80-100K$, the $V_{Th}$ remains flat and close to zero within the doping range $\mid\nu\mid\approx 2$. For $T$ beyond $120K$, $V_{Th}$ remains negative (positive) throughout the conduction (valence) band, similar to graphene. \textbf{c}, Validation of Mott,  measured $V_{Th}$ and derivative of resistance ($dlnR/d\nu$) of MATBLG at several $T$. The $dlnR/d\nu$ in light blue lines and $V_{Th}$ in orange curves. %Except for the trivial crossing points at $\nu=0,\pm4$, showing only one crossing point for each band. 
The violation of the Mott formula persists up to $120K$, beyond which $V_{Th}$ starts to fall in good agreement with Mott, as evident from $200K$ and $300K$ data.}
\label{Figure5}
\end{figure*}
%%%%%%%%%%%% figure ends

\subsection{High-temperature response of thermopower of MATBLG:} %resistance $R$ and %thermopower $V_{Th}$ of MATBLG, and comparison with Mott formula: 
After discussing the $V_{Th}$ in the intermediate temperature ($5K-60K$), in this section, we present the results for a higher temperature range of $70K-300$ K, such that we can study $V_{Th}$ for the non-interactive regime ($k_{B} T \gtrsim U$) as well as the effect of upper and lower dispersive bands ($k_{B} T \gtrsim \Delta_g$).
%After establishing the dominant contribution of thermopower coming from the local moments in the temperature range, $5-60K$ ($\approx W$, but $<U$ and $<\Delta_g$), in this section, we present the results for a higher temperature range of $70-300$ K, such that we can study $V_{Th}$ for the non-interactive regime ($k_{B} T \geq U$) as well as the effect of upper and lower dispersive bands ($k_{B} T \geq \Delta_g$).

\noindent Figure~\ref{Figure5}a illustrates the evolution of $R$ with $T$ in the range of $70-300$ K as a function of $\nu$. At $70$ K, flat band features persist, evident in resistance peaks at $\nu=\pm4$ and kinks at integer values. The asymmetry between valence and conduction bands is noticeable. As temperature increases beyond {$120$K}, flat band features fade, and the resistance spectrum resembles a graphene-like behavior with a single Dirac peak at the CNP. The spectrum becomes more symmetric with rising temperature. In Figure~\ref{Figure5}b, the evolution of $V_{Th}$ with $T$ is depicted for $\nu$ in the range of $70-300$ K. Above $70$ K, no crossings occur inside the flat band, except at the Dirac point. Between $80-120$ K, $V_{Th}$ remains nearly flat and close to zero within $\mid\nu\mid\approx2$. Beyond $120$ K, the response of $V_{Th}$ is consistently negative (positive) throughout the conduction (valence) band, akin to the thermopower response observed in graphene~\cite{zuev2009thermoelectric,wei2009anomalous,PhysRevLett.116.136802}. The comparison between the measured $V_{Th}$ and derivative of resistance ($dlnR/d\nu$) of MATBLG versus $\nu$ is shown in Figure~\ref{Figure5}c and details can be found in SI-15. %for the higher temperature range ($150 - 300K$) is shown by the bottom two panels in Figure~\ref{Figure5}c, and shown in details for various temperatures in Extended Data Fig. 5. 
Beyond $120 K$, $V_{Th}$ starts to fall in good agreement with Mott, as evident %from $300K$ data 
in Figure~\ref{Figure5}c-bottom panels. 

\noindent In this paragraph, we summarize the observed thermopower across different temperature ranges. As previously mentioned, the key energy scales include the bandwidth ($W$) of the flat band, the strength of interaction ($U$), and the energy gap ($\Delta_{g}$) between the dispersing lower and upper bands (illustrated schematically in SI- Fig. 23). (i) When $W < U < \Delta_g < k_BT$, the contributions from higher dispersing bands dominate, resulting in positive thermopower for $\nu<0$ and negative for $\nu>0$ at high temperatures, as depicted for $T > 120K$ in Figure~\ref{Figure5}b. This behavior is characteristic of graphene-based systems due to electron-hole symmetry, which manifests as a crossing at $\nu=0$ (Dirac point). (ii) When $W < U < k_BT \approx \Delta_g$, the effect of the flat band %(in a non-interacting regime) 
starts contributing along with higher dispersing bands. However, these contributions exhibit opposite signs, leading to nearly flat, close-to-zero $V_{Th}$, as observed for $T \sim 80 - 120K$ in Figure~\ref{Figure5}b. (iii) In the scenario where $W < U\approx k_BT <\Delta_g$, the influence of the flat band dominates, overshadowing contributions from higher dispersing bands. This %non-interacting 
regime, illustrated by the solid red line in Figure~\ref{Figure1}d (bottom panel), with single crossing at $\nu=0$ with positive values for $\nu>0$ and negative for $\nu<0$. %exhibits maximum entropy at $\nu=0$, resulting in a single thermopower crossing at $\nu=0$ with positive values for $\nu>0$ and negative for $\nu<0$, as expected (refer to solid red line in Figure~\ref{Figure1}d - bottom panel and Extended Data Fig. 9). 
This behavior is observed within a narrow temperature range around $70 K$, as depicted in Figure~\ref{Figure5}b. (iv) When $ W \leq k_BT \leq U$, the strong correlation starts dominate, leading to three prominent crossings ($\nu = 0, \nu \pm 1$), as illustrated in Figure~\ref{Figure2}b. %dominance of interaction dictates the entropy, leading to three prominent crossings ($\nu = 0, \nu \pm 1$), as illustrated in Figure~\ref{Figure2}b. Extended Data Fig. 8 
SI-Fig. 24 provides a pictorial summary of the thermopower response of MATBLG across various temperature ranges, comparing it with theoretical expectations.

\subsection{Conclusion:} \textcolor{black}{In summary, our thermopower response is consistent with the effects of strong correlations in MATBLG. However, there are still puzzles that require an understanding of orbital and valley effects to gain a clearer picture of the system and to design novel methods to differentiate these contributions. Such insights will pave the way for the future development of thermoelectric devices that harness the unique characteristics of flatbands in twisted heterostructures.}

\newpage
\newpage
\newpage
\section*{Acknowledgements}
A.D. thanks to Prof. Dmitri K. Efetov, Prof. Andrei Bernevig and Prof. Manish Jain for the useful discussions. The authors are grateful to Arup Kumar Paul for initial help in measurements, and Ujjal Roy for numerous discussions on the fabrication of twisted heterostructures and measurements. A.D. thanks the Department of Science and Technology (DST) and Science and Engineering Research Board (SERB), India, for financial support (SP/SERB-22-0387) and acknowledges the Swarnajayanti Fellowship of the DST/SJF/PSA-03/2018-19. A.D. also thanks CEFIPRA project SP/IFCP-22-0005. Growing the hBN crystals received support from the Japan Society for the Promotion of Science (KAKENHI grant nos. 19H05790, 20H00354 and 21H05233) to K.W. and T.T. N.T. thanks Indian Institite of Science, Bengaluru for their hospitality during her sabbatical.

%%%%%%%%%%%%%%%%%%%%%%%%%%%%%%%%%%%%%%%%%%%%%

\section*{Author contributions}
A.G. contributed to developing measuring codes and carried out all the measurements, data acquisition, and analysis. S.C. and R.D. contributed to device fabrication. A.A. developed atomic orbital calculation. A.A., S.B., S.M., and N.T. contributed to the development of the theory, and N.T. contributed to the understanding of data. K.W. and T.T. synthesized the hBN single crystals. A.D. contributed to conceiving the idea and designing the experiment, data interpretation, and analysis. All the authors contributed to the data interpretation and writing the manuscript.

%%%%%%%%%%%%%%%%%%%%%%%%%%%%%%%%%%%%%%%%%%%%%
\section*{Competing financial interests}
The authors declare no competing interests.

\section*{Data and materials availability:}
The data presented in the manuscript are available from the corresponding author upon request. 

\pagebreak
\section*{References}
\bibliography{ref}

\pagebreak
\section*{Materials and Methods}
\subsection{Device fabrication and measurement scheme:}
We have employed the modified "tear and stack" technique to fabricate the twisted heterostructures, and details are mentioned in the supplementary information (SI). We have used four devices: three of them are controlled by Si/SiO$_2$ gating, whereas the graphite gating controls one TBLG device. All the contacts are made of  $Cr(2nm)$, $Pd (10nm)$ and $Au(70nm)$. For all the devices, the heater line, which is electrically isolated from the devices, is used to create the temperature gradient. The thermoelectric voltage is measured using standard V$_{2\omega}$ technique, and the temperature gradient is measured using Johnson noise thermometry. The details can be found in the SI.

\subsection{Semiclassical calculation:}
The semi-classical description of Boltzmann theory to calculate thermopower for the band electrons is given by the following relation: considering energy-independent relaxation approximation,
\begin{equation}
\label{semieq}
S_{SC}(\mu)=-\frac{1}{Te}\frac{\int_{-\infty}^\infty (\epsilon-\mu)g(\varepsilon)(-df/d\varepsilon)d\varepsilon}{\int_{-\infty}^\infty g(\varepsilon)(-df/d\varepsilon)d\varepsilon},
\end{equation}
where $g(\epsilon)$ refers to the density of states (DOS) as a function of energy ($\epsilon$), $f$ is the fermi function, $e$, being the electronic charge and $\mu$ chemical potential. Note that in Eq.~\ref{semieq}, $df/d\epsilon$ has a finite width beyond which $df/d\epsilon\rightarrow0$, thus eliminating the necessity to integrate over $-\infty<\epsilon<\infty$. For all practical purposes, integral over the low-energy flat band and some portion of the dispersive band accurately depicts the semi-classical behavior of thermopower. The self-consistently solved $\mu$ is used to calculate the thermopower. The details can be found in the SI.

\subsection{Thermopower from atomic limit:} In this regime, the flatbands and interactions therein are expected to dominate. We work in the atomic limit where Hamiltonian is given by $
H = \sum_{\alpha} \Big( U n_{\alpha \uparrow}{n_{\alpha \downarrow}} + \epsilon_\alpha n_\alpha \Big)$
where $\alpha=\{1,\ldots,4\}$ labels the four orbitals and $U$ is the Hubbard interaction. Evaluating the partition function $Z$ and the corresponding Helmholtz free energy is given by $F = - k_B T \log Z$
and the corresponding entropy is given by$
S_{en} = - \Big( \frac{\partial F}{\partial T} \Big)
$. The thermopower is then given by $
S_{th} = -\frac{1}{e}\frac{\partial S_{en}}{\partial \nu}$
where $e$ is electric charge. It is to be noted that $\mu$ is solved self-consistently to fix the filling $\nu$. The details are described in the SI. 

\newpage
\thispagestyle{empty}
\mbox{}


\section*{Supplementary material}

**This supplementary information contains the following details:**

**1. Device fabrication, characterization and twist angle determination:**

**2. Thermopower measurement scheme: (i)** $V_{2\omega}$ **for thermoelectric voltage, (ii) noise thermometry for temperature gradient, (iii) linear response regime:**

**3. Resistance and thermoelectric responses of MATBLG** ($1.05^0$ - $SiO_2$ **back-gated), near MATBLG** ($0.95^0$ - $SiO_2$ **back-gated), TDBLG** ($1.1^0$ - $SiO_2$ **back-gated) and TBLG** ($1.2^0$ **- Graphite back-gated):**

**4. Thermoelectric voltage of MATBLG device** ($1.05^0$) **with in-plane and perpendicular magnetic fields:**

**5. Theoretical modeling using different methods to understand the thermoelectric response of TBLG-based devices: (i) non-interacting calculation using semi-classical Boltzmann description, (ii) Hartree-Fock mean field theory for Dirac revivals, (iii) dynamical mean-field theory (DMFT), and (iv) strongly correlated atomic limit:**

**6. Local moments and their experimental signatures in various measurement techniques:**

**7. Thermopower and local moments:**

**8. Thermopower from non-interacting electrons using semi-classical description with different bandwidths (10-40 meV); different kinds of density of states (DOS) - continuum, sawtooth, asymmetric; thermopower for non-interacting electrons with in-plane magnetic field:**

**9. Thermopower in Hartree Fock mean-field theory for Dirac revivals in MATBLG:**

**10. Thermopower in dynamical mean-field theory (DMFT) for MATBLG:**

**11. Thermopower in the atomic limit: (i) single orbital -** $SU(2)$**, (ii) multi-orbitals -** $SU(8)$**, (iii) atomic limit with finite bandwidth, (iv) cluster calculation with extended Hubbard model, (v) effect of higher energy dispersive bands and understanding of** $V_{Th}$ **at different temperature regime, (vi) effect of in-plane magnetic field in the atomic limit:**

**12. Different origins of entropy for non-interacting band electrons to strongly correlated electrons, and relation between entropy and thermopower:**

**13. Estimation of entropy from the measured thermopower:**

**14. Entropy of non-interacting itinerant electrons:**

**15. Comparison of measured** $V_{Th}$ **with Mott formula:**

**16. Various origins of the violation of Mott formula:**

**SI-1: Device fabrication, its electrical response with gate voltages and twist angle determination.**

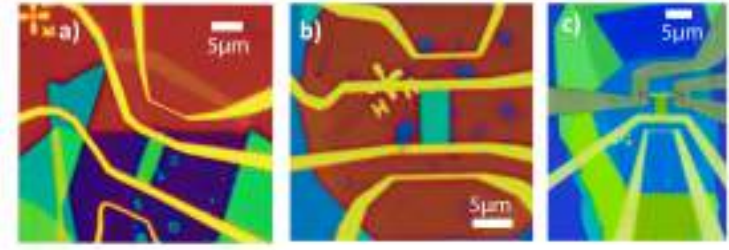

**SI-Fig. 1: Optical images of the devices: (a)**, **(b)** and **(c)** are the optical images of MATBLG, TDBLG, and graphite-gated TBLG, respectively.

We have employed the modified "tear and stack" [1] technique to fabricate the twisted heterostructures. First, on Si/SiO$_2$ substrates, hBN and graphene are separately exfoliated. Then, suitable flakes are sorted for fabrication. Suitable single larger graphene flakes are then pre-cut to form into two, to be picked up sequentially. hBN, with thickness between $\sim 20nm - 30nm$ are commonly used for encapsulation. A film of transparent poly propyl carbonate (PPC) on top of a hemispherical PDMS stamp is used to pick up the top hBN layer. Using this PDMS-PPC-hBN stamp, we pick up the pre-cut graphene (mono/bi) layers, one after the other, while introducing a rotation between subsequent pick-ups to introduce the desired twist angle. Finally, another hBN flake is picked up, which encapsulates the TBLG/TDBLG layers. We have also fabricated a graphite gated device, for which an additional graphite flake is picked up after the bottom hBN. The whole stack is then placed on a clean Si/SiO$_2$ substrate. During the whole stack-making process, a temperature at or below 55$^o$ C is maintained to reduce the chances of thermal relaxation of the twist angle. The final landing on Si/SiO$_2$ substrate is also done at 75$^o$ C for the aforementioned reason.

To make the 1D edge contacts, we spin-coat with 495A4-950A4 PMMA bilayer, which acts as a negative resist. Unlike the shorter duration of usual 180$^o$C baking, each PMMA layer was baked at 80-100$^o$C for 15 minutes to avoid thermal relaxation of the introduced twist angle. Then, the contacts, including a thin heater line, are defined using standard e-beam lithography (EBL), with the dose optimized for low baking temperature of PMMA. Following the lithography, contacts are etched with CHF$_3$-$O_2$ plasma followed by thermal deposition of $Cr(2nm)$, $Pd(10nm)$ and $Au(70nm)$. The width of source and drain contacts on all devices are $\sim 1\mu m$, while the heater lines are $\sim 200 - 300nm$ in width. The heater lines are separated by around $3\mu m$ gap from the source contacts. The Si/SiO$_2$ substrate acts as a back gate for three devices, whereas one TBLG was graphite-gated. Finally, the channel geometry is mapped out by EBL, and the rest of the graphene domains are etched out to define the channel geometry. Optical images of the representative TBLG devices are shown in SI-Fig. 1. The typical dimensions for our devices with channel length $\sim 5 - 8\mu m$ and width $\sim 1.5 - 2\mu m$.

The graphite-gated TBLG is screened compared to the $SiO_2$ back-gated MATBLG, as evidenced by their resistance responses to temperature. As shown in Fig. 2a, Fig. 5a of the manuscript, and SI-Fig. 9, the resistance peaks at the integer fillings due to correlation vanish within 50K for the graphite-gated TBLG (1.2°) (SI-Fig. 9a), whereas these resistance peaks persist up to 100K for the $SiO_2$ back-gated MATBLG (1.05°) (Fig. 5a in main manuscript).

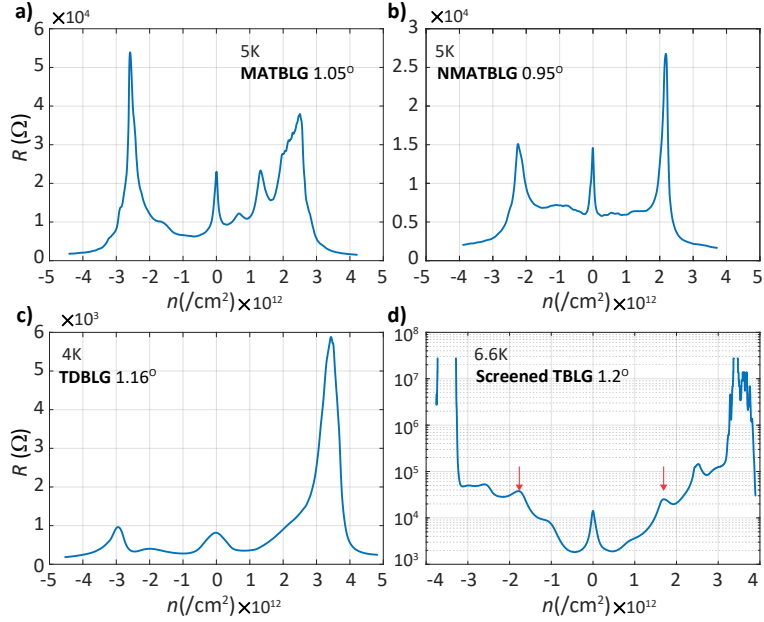

**SI-Fig. 2:** **Resistance response as a function of density ($n$): (a), (b), (c)** and **(d)**, Resistance as a function of density for MATBLG (at $5K$), near-MATBLG (at $5K$), TDBLG (at $4K$) and graphite back-gated TBLG ($6K$), respectively. For all four devices, the Moire superlattice gap opening can be observed as secondary resistance peaks that emerge around $2 \times 10^{12} cm^{-2} <| n |< 4 \times 10^{12} cm^{-2}$ alongside the primary Dirac peak at zero density.

**Twist-angle determination:** For all four devices, we have used the full-filling ($\nu = \pm 4$) carrier density to measure the twist angle ($\theta$). At full-filling, the moire superlattice has 4 electrons/holes per unit cell. The twisted superlattice unit cell area is given by $A \sim \sqrt{3}a^2/2\theta^2$, where $a = 0.246 nm$ is the lattice constant of monolayer graphene. At the full-filling, the superlattice density $n_s$ is given by $n_s = 4/A = 8\theta^2/\sqrt{3}a^2$. Using this expression, the $\theta$ is determined from the densities at $\nu = \pm 4$. The resistance peak of the MATBLG symmetrically occurring on either side of the primary Dirac point can be used to locate the full-filling density. For this device for $\nu = \pm 4$ the densities are $n \sim \pm 2.58 \times 10^{12} cm^{-2}$ (SI-Fig. 2a), which translate into a twist angle of $\sim 1.05^0$ i.e. the magic angle. The superlattice peaks of the near MATBLG occurs at densities $n \sim \pm 2.15 \times 10^{12} cm^{-2}$ (SI-Fig. 2b), which results in a twist angle of $\sim 0.95^0$. The secondary Dirac peaks of TDBLG are not precisely symmetric. Thus the average position of full-filling density is $n \sim \pm 3.19 \times 10^{12} cm^{-2}$ (SI-Fig. 2c), which translate into a twist angle of $\sim 1.1^0$. Similarly, for the graphite back-gated TBLG device, the half-filling ($\nu \pm 2$) density (marked by red arrows in SI-Fig. 2d) occurs at $n \sim \pm 1.7 \times 10^{12} cm^{-2}$ (SI-Fig. 2d) and correspond to $\sim 1.2^0$ twist angle.

**SI-2: Thermopower ($S_{th}$) measurement scheme:**

**Thermoelectric voltage ($V_{Th}$) measurement:** A simplified $V_{Th}$ measurement scheme is shown in SI-Fig. 3a. We inject ac ($I_h$) through the heater line at 7 Hz (via a $1k\Omega$ resistance connected in series to the heater line acting as the ballast resistance). The $V_{Th}$ is measured across the source and drain contacts as the second harmonic (14 Hz) Y-component ($V_{2\omega}$) signal generated across the sample, which is measured via

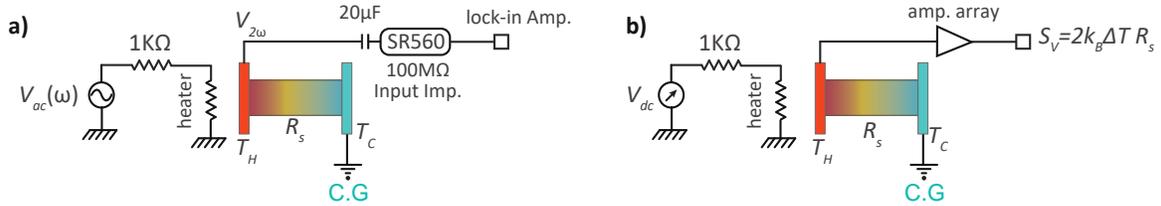

**SI-Fig. 3: Thermoelectric Voltage ($V_{Th}$) and temperature gradient measurement scheme.** **(a)** A current is passed through the heater to create the temperature gradient across the sample. The thermoelectric voltage was measured using standard $V_{2\omega}$ method. **(b)** Johnson noise thermometry to measure the temperature difference between the source and drain.

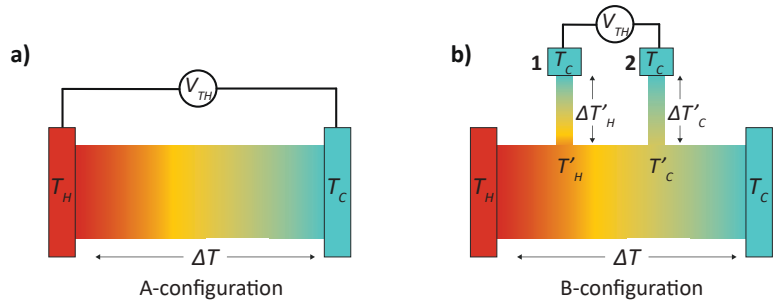

**SI-Fig. 4: Temperature profile and thermoelectric voltage measurement in different geometry:** **(a)** and **(b)**, respectively, show the temperature profile across the measurement leads for 2-probe, A-configuration and 4-probe, B-configuration.

5

the lock-in using a 100-gain amplifier (SR-560). The drain probe (i.e., the farthest device contact from the heater) is directly bonded to the cold finger; this thermally anchors the drain to the MC plate of the dilution refrigerator, which serves as a cold ground (C.G).

For $V_{Th}$ measurement, one has to be mindful of stray components of thermoelectric voltage emerging from temperature gradients between the measurement probes and sample which can potentially render the $V_{Th}$ measurement inaccurate. In following segment, we discuss various inaccuracies that may or may not influence the $V_{Th}$ measurement depending on the measurement scheme.

In SI-Fig 4, in a cartoon diagram, we highlight the temperature gradients (as color gradient) across the sample as well as the probes that comprise the measured $V_{Th}$ for a 2-probe as well as a 4-probe geometry. While a 4-probe geometry is ubiquitous for negating contact-resistance effects for $R_{xx}$ measurement, the same is not entirely true for measuring the longitudinal component of thermoelectric voltage $V_{Th}^{xx}$. Therefore, it is not accurate to label $V_{Th}$ in 2-probe or 4-probe geometry, rather we will address them as $V_{Th}$ in A-configuration or B-configuration, respectively. However, in practice, a B-configuration geometry is arguably complicated for $V_{Th}$ measurement, as many stray components of temperature gradients across the measurement probes also get included in the measured $V_{Th}$. Referring to SI-Fig 4b, (because of the large metal effectively acting as a temperature bath), the metal leads at contacts 1 and 2 essentially maintain the bath temperature $T_C$ or very close to that. However, the thin probes connecting the channel to the metal leads do experience temperature variations. For probe 1 (2), the temperature difference is $\Delta T'_H = T'_H - T_C$ ($\Delta T'_C = T'_C - T_C$), $T'_H$ ($T'_C$) being local the temperature of the channel at the position of probe 1 (2). These will introduce stray components of thermoelectric voltage in series to the measurement circuit. Hence the measured $V_{Th}$ will have the following components-

$$V_{Th}^{4-probe} = -V_{Th}^{\Delta T'_H} + V_{Th}^{T'_H - T'_C} + V_{Th}^{\Delta T'_C} \tag{1}$$

Where, $-V_{Th}^{\Delta T'_H}$ comes from temperature gradient across probe 1 (note the $-ve$ sign is considered since, because of the temperature gradient direction, $V_{Th}^{\Delta T'_H}$ will oppose rest of the thermoelectric components). Similarly, $V_{Th}^{\Delta T'_C}$ and $V_{Th}^{T'_H - T'_C}$ are the components coming from probe 2 and the channel (between probe 1 and 2), respectively. Even in a simplified assumption that both the probes have an identical material composition and geometry (i.e, identical thermoelectric properties), $V_{Th}^{\Delta T'_H} \neq V_{Th}^{\Delta T'C}$ since $\Delta T'_H \neq \Delta T'_C$, which depends on various details like the temperature profile created by the heater and, the thermal conductivity of the metals/substrates etc. Thus unlike $R_{xx}$ measurement where we nullify contact effects (by open circuit voltage measurement), $V_{Th}^{4-probe}$ will have non-negligible influences of the probes. In practice, these probe contributions can be significant as they rarely are identical (due to strain or being of a different composition than the channel or because of the inaccuracies of fabrication etc). On the other hand, none of these aforementioned components are present in A-configuration geometry. Note that the interfaces generally are also a contributor to thermoelectric voltage in both A-configuration and B-configuration geometry, but between the probes and metal leads the interfaces are effectively one-dimensional, thus temperature drop across the interfaces ($\Delta T$) can be neglected, hence having negligible contribution in thermoelectric voltage. In our measurement, we have utilized A-configuration for measuring $V_{Th}$ for the four devices.

**Temperature gradient measurement scheme with noise thermometry:** We have employed Johnson noise thermometry to measure the temperature difference, $\Delta T$, at lower temperatures below $10K$ as shown

in SI-Fig. 3b. For this measurement, equivalent dc current is sent through the heater line to create the temperature gradient across the sample, and thermal noise is measured using a LC-resonant circuit. The details can be found in our earlier work [2, 3]. In the absence of any heater current, the sample temperature maintains equilibrium with the MC plate temperature of the dilution refrigerator via cold ground. In this case, measured thermal noise is given by $S_V{}^0 = 4k_BT_cR_s$, where $T_c$ is the temperature of the MC plate of the dilution refrigerator and $R_s$ refers to the sample resistance between the cold ground and the hot contact. In the presence of finite heater current, the thermal noise will rise to $S_V{}^{Av} = 4k_BT_{Av}R_s$ due to the average temperature $T_{Av}$ of the device being higher than $T_c$. Due to the applied temperature gradient, the excess thermal noise is given by $\Delta S_V = S_V{}^{Av} - S_V{}^0 = 2k_B\Delta T R_s$ [2, 3]. A detailed derivation of this expression is provided in our earlier works [2, 3]. Note that, at temperatures beyond $10K$, $\Delta T$ measurement by noise thermometry is challenging. Extracting the excess noise over the already high background equilibrium thermal noise is challenging, although not impossible, as it requires noise data acquisition over a significantly extended time. In the main manuscript, since we only concentrate on the sign changes and relative amplitude of the thermopower data, we present only the thermoelectric voltage $V_{Th}$ except, the data at $6K$ is converted to thermopower and shown in SI-Fig. 5c.

**Measuring thermoelectric voltage in linear response regime:** In the top panel of SI-Fig. 5a and SI-Fig. 5d, we show the measured $V_{2\omega}$ signal versus back gate voltage ($V_{bg}$) at a few different applied heater currents for MATBLG device, at temperature $6K$ and $40K$, respectively. SI-Fig 5a, 5d-bottom panels show the same thermoelectric voltages normalized with respective heater current squared ($I_h^2$). All the normalized $V_{2\omega}/I_h^2$ values overlap well, since the Joule heating, and hence, the temperature difference is proportional to $I_h^2$. This shows the $V_{2\omega}$ measurements have been performed in the linear response regime. Further, the extracted $\Delta T$ from the thermal noise measurement is shown as a function of $I_h^2$ in SI-Fig. 5b. Within the maximum heater current range, a clear linear trend (dashed red line) can be observed between $\Delta T$ and $I_h^2$, and $\Delta T << T$. Finally, SI-Fig. 5c shows the thermopower calculated from $V_{2\omega}$ signal and $\Delta T$ data, and around $6K$ its magnitude is of the order of $\sim 8\mu V/K$. We have measured all the thermoelectric voltages for different temperatures in the linear regime by making sure that the normalized thermoelectric voltages ($V_{2\omega}/I_h^2$) collapses to a single curve similar to SI-Fig. 5a-bottom panel. SI-Fig. 5e shows the same linearity check performed in TDBLG device at $40K$. To further demonstrate the linear regime transport, SI-Fig.6 shows the resistance response of the MATBLG device at $40K$ with and without heater current. We observe the resistance remains almost identical in both scenarios. SI-Fig. 5 d, e thermopower linearity and, SI-Fig.6 demonstrates the linear regime of our experiment at higher temperatures, which means the temperature difference between the hot and cold end is much smaller than the bath temperature (mean temperature). This also reaffirms that the applied temperature gradient does not smear out any features of our experiment due to local heating.

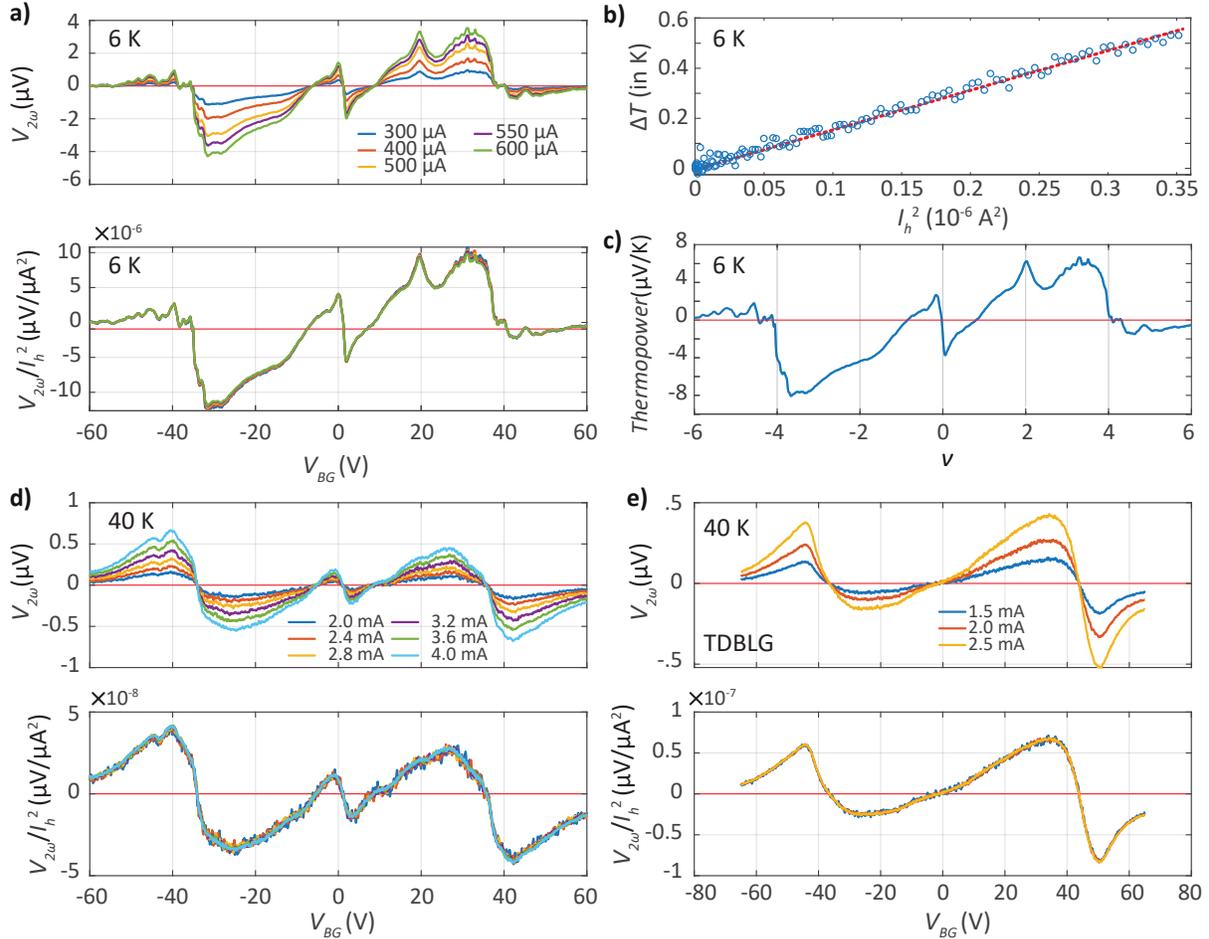

**SI-Fig. 5: The linear regime of thermopower measurement: (a)** $V_{2\omega}$ signal versus $V_{bg}$ for different applied heater currents ($I_h$), respectively, at 6K for MATBLG device (top panel). $V_{2\omega}$ signal normalized by the square of respective heater currents ($I_h^2$) versus $V_{bg}$ (bottom panel). **(b)** Measured temperature difference ($\Delta T$) as a function of $I_h^2$ at 6K at backgate voltage $V_{bg} = 7.6V$ **(c)** Thermopower as a function of filling $\nu$ derived from measured $V_{2\omega}$ signal and $\Delta T$. **(d), (e)** (Top panels) $V_{2\omega}$ signal versus $V_{bg}$ for different applied heater currents ($I_h$), at 40K for MATBLG and TDBLG device, respectively. (Bottom panels) $V_{2\omega}$ signal normalized by the square of respective heater currents ($I_h^2$) versus $V_{bg}$.

8

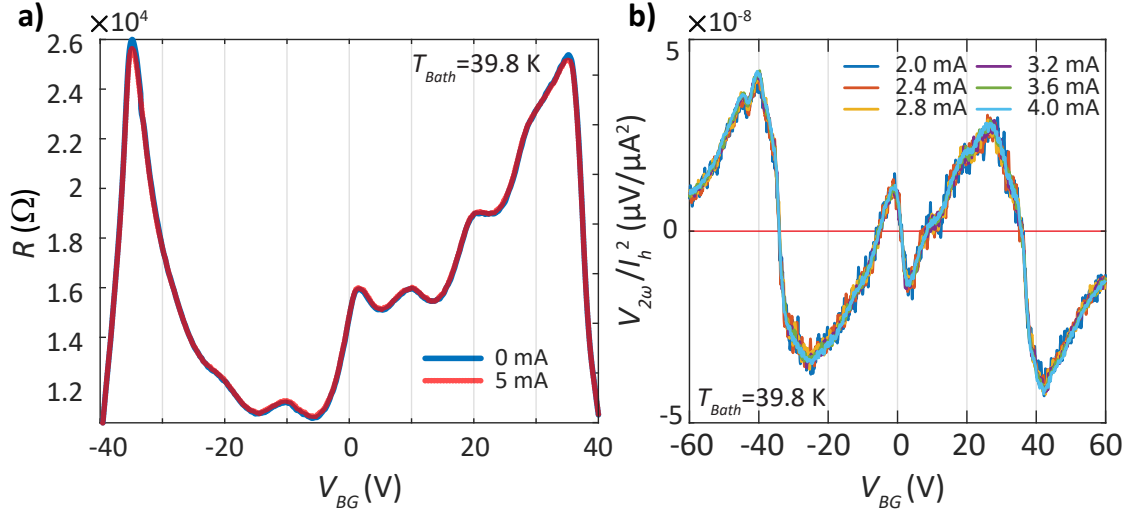

**SI-Fig. 6: Resistance and thermoelectric voltage at different heater currents: (a)** Resistance at two different heater currents at $40K$ as a function of gate voltage. The resistance response remains identical with or without the presence of heater current. **(b)** $V_{2\omega}$ signal normalized by the square of respective heater currents ($I_h^2$) versus $V_{bg}$ at $40K$, shows all the normalized thermopower overlap indicating linear regime.

**SI-3: Resistance and thermoelectric response of near MATBLG ($0.95^o$), TDBLG ($1.1^o$) and graphite gated TBLG device ($1.2^o$):**

In this section, we summarise results from near MATBLG ($0.95^o$), TDBLG ($1.1^o$), and graphite gated TBLG ($1.2^o$) devices. In SI-Fig. 7a, we present the resistance ($R$) response for near MATBLG at various temperature ranges (from $5K$-$100K$). $R$ remains featureless at integer fillings, except for the Dirac point and $\nu = \pm 4$. However, the weaker resistance peaks at the positive integer fillings appear at higher temperatures ($20K - 50K$), as can be seen in SI-Fig. 7a and SI-Fig. 28 (derivative of resistance). This suggests that the effect of correlation in near MATBLG device could be weaker compared to the MATBLG ($1.05^o$) device. The $V_{Th}$ in SI-Fig. 7b resembles that of MATBLG ($1.05^o$), exhibiting similar three crossing points near $\nu = 0$ and $\nu \approx \pm 1$ within $-4 < \nu < +4$. However, in comparison to MATBLG, the additional crossing points near $\nu \approx \pm 1$ show some variation with $T$ and persist only up to $50K$ (this could be due to the larger bandwidth as compared to MATBLG ($1.05^o$) device). Beyond $100K$, $V_{Th}$ of the device, as well as $R$, begins to resemble a graphene-like spectrum.

In contrast, the $V_{Th}$ of TDBLG in SI-Fig. 8b has a distinct temperature dependence. Notably, $V_{Th}$ remains asymmetric for the conduction and valence bands at lower $T$, and the three crossing points merge to a single crossing point beyond $20K$. The most notable departure from the MATBLG ($1.05^o$) can be seen beyond $20K$. At these temperatures, $V_{Th}$ remains completely positive (negative) for the whole span of the conduction (valence) band within $-4 < \nu < 4$. This resembles qualitatively the expected thermopower from the non-interacting itinerant electrons (which can be described by a semi-classical model (as well as the Hartree-Fock model at larger temperatures) as shown in the main manuscript, and will be discussed later SI-Fig.11e,(SI-Fig.16). The resistance in TDBLG (see SI-Fig. 8a) at all temperatures shows large

9

asymmetry. The asymmetry is more pronounced at higher temperatures.

The resistance and $V_{Th}$ response of the graphite-gated TBLG ($1.2^o$) device is shown in SI-Fig. 9. This device shows resistance peaks at the integer fillings. The resistance peaks coming from the flat band are weaker compared to the MATBLG ($1.05^o$) device, as shown in Fig. 2a and 5a of the manuscript. Further, peaks at integer fillings hardly can be seen beyond $30K$ in contrast to MATBLG ($1.05^o$) device, where they survive beyond $100K$. This suggests that the graphite-gated TBLG ($1.2^o$) device is weakly correlated compared to the $SiO_2$-gated MATBLG ($1.05^o$) device. The $V_{Th}$ shows three prominent crossing points around $\nu = 0$ and $\nu \approx \pm 1$ within $-4 < \nu < +4$ at lower temperatures. However, beyond $30K$, these three crossing points become a single crossing point, similar to the TDBLG device, and qualitatively follow the expected thermopower from the non-interacting itinerant electrons or weakly correlated system. Note that an asymmetry between the conduction and valence band emerges at higher temperatures ($T > 30K$), for which the single crossing point of the $V_{Th}$ occurs at a relative offset from the Dirac point towards the electron doping or conduction band, and seen for the TDBLG device in Fig. 3d of the manuscript. This offset could be arising from the band structure effect, which plays the dominant role in the thermopower response for the weakly correlated system. The offset in $V_{Th}$ can be captured by invoking an asymmetric non-interacting density of states (DOS) for the conduction and valence bands using a semi-classical description, as shown later (see SI-Fig.14). Such asymmetric DOS is already reported in theoretical calculations based on relaxed moire structure[4].

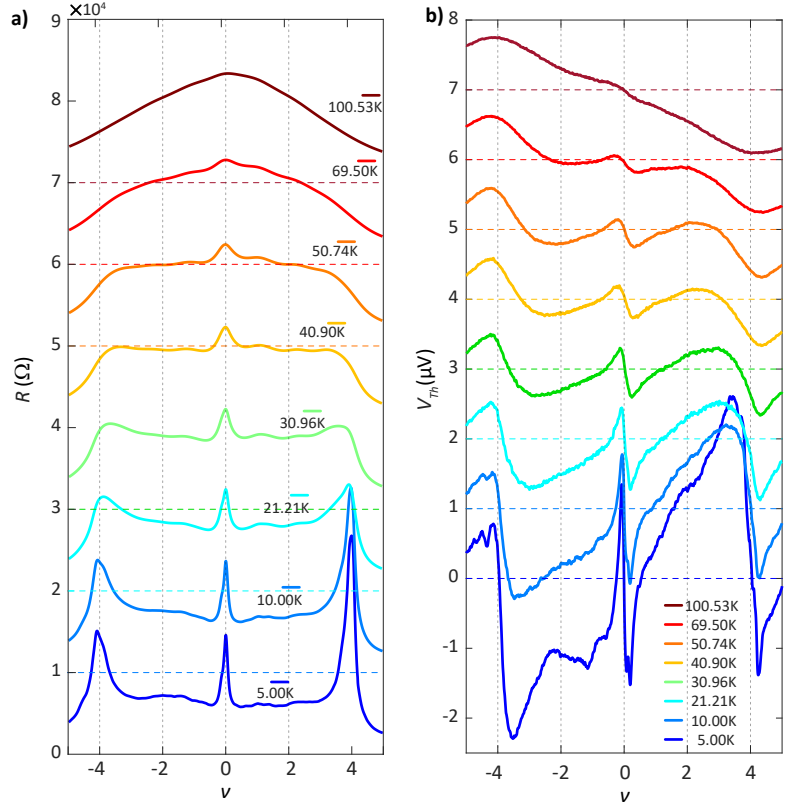

**SI-Fig. 7: Resistance and $V_{Th}$ response for near MATBLG ($0.95^o$) device:(a)**, Resistance response as a function of $\nu$ at different temperatures, depicted in a shifted plot. The horizontal lines represent the zero resistance reference at each temperature. **(b)**, $V_{Th}$ with $\nu$ at different $T$. The $V_{Th}$ exhibit crossing points near $\nu \approx \pm 1$ at lower $T$ alongside at the Dirac point and $\nu = \pm 4$. However, unlike MATBLG ($1.05^o$), the $\nu_{cross}$ for this device shows some variation with $T$ and three crossing points within $-4 < \nu < +4$ survive till $50K$. At higher temperatures above $100K$, a graphene-like spectrum in $V_{Th}$ is obtained.

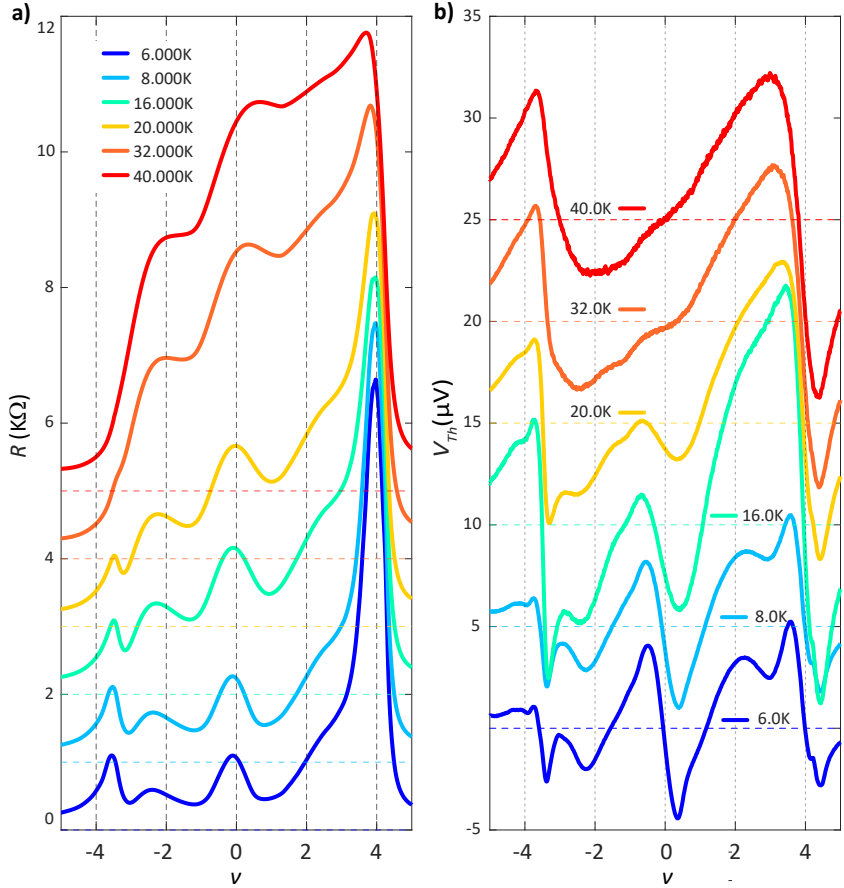

**SI-Fig. 8: Resistance and $V_{Th}$ response for TDBLG ($1.1^o$) device: (a)**, Resistance response as a function of $\nu$ at different temperatures, depicted in a shifted plot. The horizontal lines represent the zero resistance reference at each temperature. **(b)**, $V_{Th}$ with $\nu$ at different $T$. In contrast to MATBLG ($1.05^o$), the additional crossing points in the conduction and valency bands vanish at a much lower temperature, around $20K$. In other words, the three crossing points merge to a single crossing point within $-4 < \nu < +4$. The most distinctive difference is, above $20K$, $V_{Th}$ remain completely positive (negative) in the conduction (valence) band, which is opposite to how $V_{Th}$ behaves for MATBLG device.

12

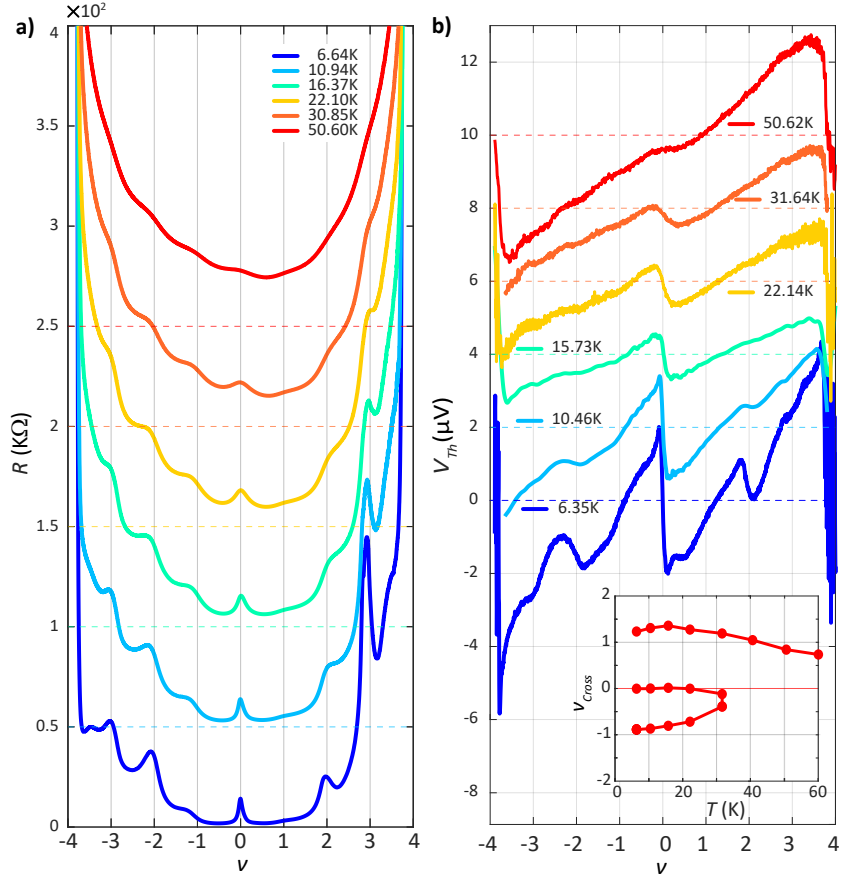

**SI-Fig. 9: Resistance and $V_{Th}$ response for graphite-gated TBLG ($1.2^o$) device: (a)**, Resistance response as a function of $\nu$ at different temperatures, depicted in a shifted plot. The horizontal lines represent the zero resistance reference at each temperature. **(b)**, $V_{Th}$ with $\nu$ at different $T$. The crossing points vary quite significantly with temperature. The three crossing points become a single crossing point above $30K$, after which the electron and hole bands, respectively, exhibit positive and negative thermopower. The $V_{Th}$ response of this device is similar to TDBLG, but quite distinct from the MATBLG device ($1.05^o$) as shown in the manuscript. The inset shows the crossing point evolution with temperature.

13

## SI-4: Magnetic field dependent thermoelectric voltage of MATBLG device:

In the main manuscript, we have discussed the decrement of the normalized $V_{Th}$ signal with the application of $B_\parallel$ at $25K$ (also presented in SI-Fig10a). Particularly, in the valence band, we observed a decrement of $30\%$. However, no noticeable change can be observed in the resistance response, as shown in SI-Fig10b. This is also evident from $dln(R)/dn$ from the resistance data (shown in the inset of SI-Fig10b), as well as the 2D plot of the resistance with $\nu$ and $B_\parallel$, which remain unaltered with $B_\parallel$ in SI-Fig10c. In SI-Fig10d, we show the $V_{Th}$ with $B_\parallel$ at $5K$, and the significant reduction of $V_{Th}$ is clearly visible. In SI-Fig10 e, f, g and h, we present the data with $B_\perp$. The strong dependence of $V_{Th}$ with $B_\perp$ is shown in the main manuscript (and also in SI-Fig10e) at $20K$. The corresponding resistances are shown in SI-Fig10f with $dln(R)/dn$ in the inset. The signature of the Landau levels formation can be felt as oscillations in $V_{Th}$ with $\nu$ within $\nu \pm 1$. However, away from the Dirac point, the signature of Landau levels can hardly be seen in SI-Fig10e and SI-Fig10g (2D plot). The Mott-formula, i.e., $dln(R)/dn$ (SI-Fig10f inset), also fails to capture any such decrement with $B_\perp$ seen in SI-Fig10e. In SI-Fig10h, the $V_{Th}$ evolution with $B_\perp$ is shown at $10K$, and reduction of $\sim 40\%$ in $V_{Th}$ at $\nu \simeq -3$ can be seen. Note that all the measurements with $B$ were conducted with a fixed heater current, and all other configurations remained unchanged, indicating that the measurements were done at a constant $\Delta T$ (as established in our earlier work [2]). Therefore, the relative change in thermoelectric voltage is expected to be the same as in thermopower.

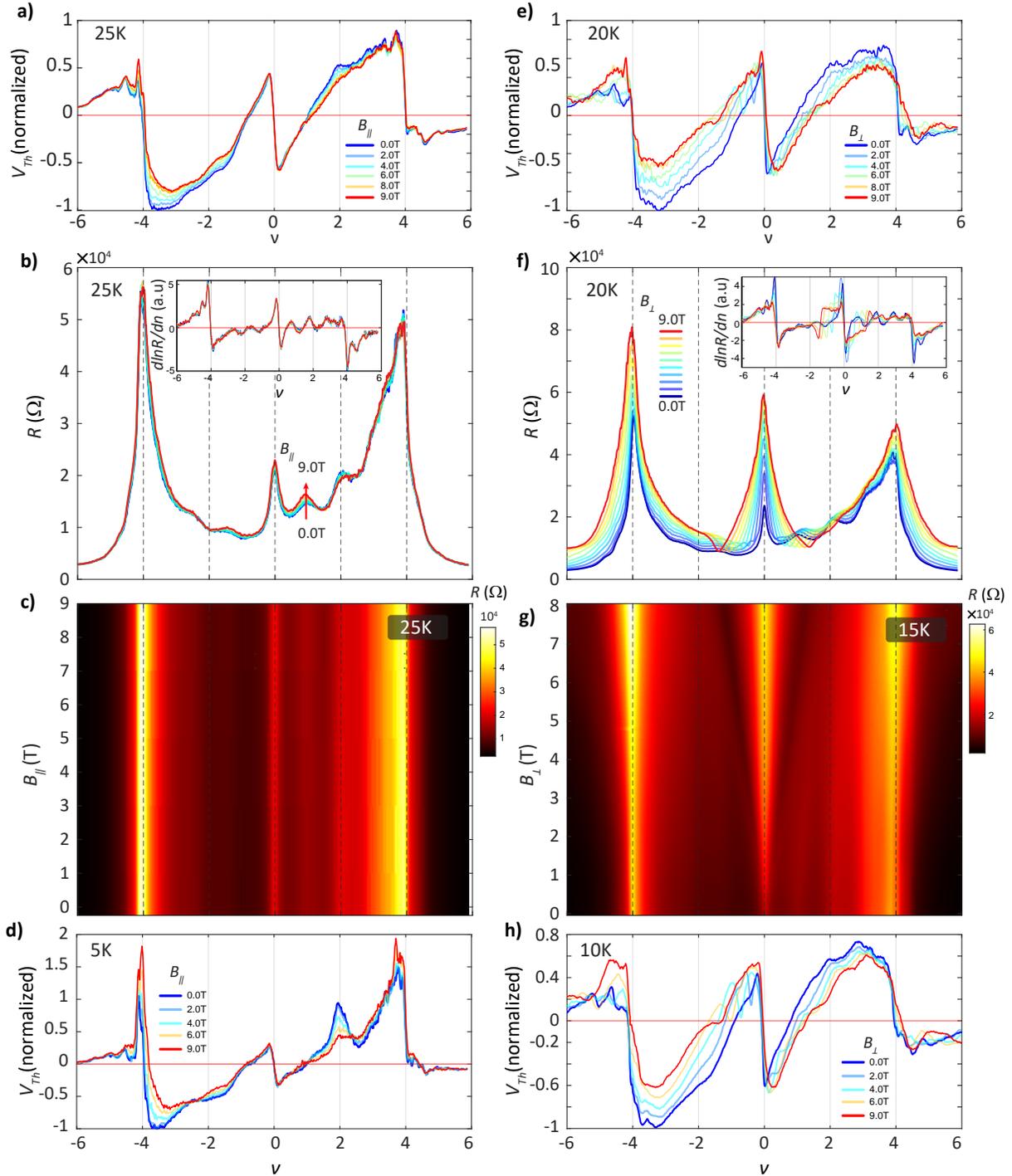

**SI-Fig. 10: Thermoelectric voltage and resistance with with in-plane and out-of-plane magnetic fields:** (a) $V_{Th}$ with $\nu$ for different $B_{||}$ at $25K$. Here for figures **(a),(d),(e)** and **(h)**, the $V_{Th}$ is normalized to measured data at $B_{||} = 0T$. (b) Resistance as a function of filling ($\nu$) at different in-plane magnetic fields ($B_{||}$) at $25K$. Inset shows the $dln(R)/dn$ at corresponding $B_{||}$. (c) 2D plot of Resistance as a function of $\nu$ and in-plane magnetic field at $25K$. (d) normalized $V_{Th}$ with $\nu$ for different $B_{||}$ at $5K$ (e) Evolution of normalized $V_{Th}$ with $B_\perp$ at $20K$. (f) Resistance as a function of $\nu$ at different out of plane magnetic fields at $20K$. Inset shows the $dln(R)/dn$ at corresponding $B_\perp$. (g) 2D plot of Resistance as a function of $\nu$ and out-of-plane magnetic field at $15K$. (h) Evolution of normalized $V_{Th}$ with $B_\perp$ at $10K$.

**SI-5: Various theoretical modeling to understand the thermoelectric response of TBLG:**

In order to understand the thermoelectric response of four devices, we will discuss various theoretical approaches in this section. Let us summarize our experimental findings. In our experiment, we have studied four TBLG-based systems: $1.05^0$ MATBLG ($SiO_2$ gating), $0.95^0$ near MATBLG ($SiO_2$ gating), $1.1^0$ TDBLG ($SiO_2$ gating) and $1.2^0$ TBLG (graphite back-gating). We have carried out thermoelectric voltage measurements from $5K$ to $300K$ and, most importantly, extracted the crossing points in thermopower as a function of temperature. At lower temperatures like $5-10K$, all these devices show three crossing points within $-4 < \nu < +4$, including at the Dirac point. Remarkably, for the $1.05^0$ MATBLG, the three crossing points remain robust or at the same position up to $\sim 60K$, whereas they show some variation for the $0.95^0$ near MATBLG device, but three crossing points survive up to $\sim 50K$. In contrast, for the $1.1^0$ TDBLG and $1.2^0$ TBLG (graphite back-gated) devices, the three crossing points remain sensitive to temperature and merge to a single crossing point above $20-30K$. Further, the thermoelectric voltage of the $1.05^0$ MATBLG device shows a reduction of $30\%$ with the application of an in-plane magnetic field of $B_{\parallel} = 9T$.

To understand these responses, we have used the following theoretical models: (i) non-interacting calculation using semi-classical Boltzmann description, (ii) a mean-field theory based on Hartree Fock (HF) approximation, (iii) dynamical mean-field theory (DMFT) at intermediate interaction strength $U/W \lesssim 1$, and (iv) the atomic limit, the limiting case of extremely strong interaction and the bandwidth approaching zero, $t \ll k_\mathrm{B}T \ll U$. These models are discussed in detail in the later sections. We show that the simplest toy model based on atomic limit can explain the most striking features of the experimentally observed thermoelectric response (temperature dependence of crossing points and in-plane magnetic field dependence) for the MATBLG device. This indicates that thermopower captures the effect of the many-body correlation, resulting in emergent local moment physics in the MATBLG device. However, for TDBLG and screened TBLG devices ($U/W < 1$, where $U$ is onsite Coulomb repulsion and $W$ is bandwidth), the thermopower shows weaker many-body correlation or local moments physics.

**SI-6: Local moments and their experimental signature in various measurement techniques:**

The local moments are expected to emerge in a strongly correlated system where the many-body correlation dominates. These systems are often theoretically described by the Hubbard-type models, where at large $U$ (on-site Coulomb repulsion) compared to the hoping term, double occupancy is suppressed, and each site has either a local spin whose magnitude grows toward $1/2$ with increasing $U$ or no spin (empty site). The motion of the electrons is strongly hindered due to the no-double-occupancy constraint. These relatively immobile electrons with spin $1/2$ moments are known as 'local moments', whose density depends on the filling fraction ($\nu$) or, in other words, how many sites per unit cell are filled. Thus, a large $U$ or strong correlation is required to have the local moments, and its existence can be verified by measuring the magnetization of these local moments. The magnetization of the local moments will be simply the filling fraction times the magnetization of a single spin $1/2$, which is nothing but $\nu \times \tanh(B/k_B T)$, where $B$ is the magnetic field, this can be further reduced to $\nu \times B/k_B T$ for smaller magnetic field or more precisely $B/k_B T < 1$, where $T$ is the temperature. In this limiting case, the magnetic susceptibility will be $\chi \sim \nu/k_B T$, known as Curie susceptibility. In contrast, for non-interacting itinerant electrons or mobile carriers, the magnetic susceptibility is typically expected to be independent of temperature, known as Pauli susceptibility. Thus,

measuring Curie susceptibility will be the smoking gun evidence for the existence of local moments. However, this is very challenging to measure in mesoscale devices with micron-size dimensions. In the absence of direct evidence for local moments, we discuss the following possible experimental signatures in various measurement techniques:

**(i) Scanning tunneling microscope (STM):** STM is an excellent probe for looking at the many-body correlation and is extensively used to study the magic-angle twisted bilayer graphene systems [5, 6]. Using STM, the non-interacting bandwidth (W) and the onsite Coulomb repulsion (U) are measured and reported of the order of $W \sim 20$ meV and $U \sim 20 - 30$ meV, respectively. STM measurements [6, 7, 8] also indicate spectral weight transfer to higher energies, revealing upper and lower Hubbard-band-like features, especially near integer fillings. Thus, these studies confirm that the interaction dominates in magic-angle twisted bilayer graphene systems indicating the strongly correlated nature of the carriers, which is one of the ingredients for exhibiting local moment behaviour. However, STM does not directly capture the signature of local moments since the tips are spin-unpolarized.

**(ii) Compressibility measurements:** Compressibility is another powerful technique for probing the many-body correlations. The measurement of thermodynamic quantities, like entropy, can shed light on the presence of local moments. For example, the local moments of electrons will give rise to spin entropy per particle of $k_B \ln 2$. However, in the extremely correlated atomic or completely flat band limit [9, 10] with spin, valley and sublattice degrees of freedom in magic-angle twisted bilayer graphene, one can expect the local-moment entropy per moire unit cell to be higher, e.g., $k_B \ln 4$ or $k_B \ln 8$. Most importantly, the application of an in-plane magnetic field will reduce the spin-entropy due to the polarization, which has been discussed theoretically in [11]. Ref. [12, 13] have used a compressibility probe to extract the entropy of magic-angle twisted bilayer graphene and found the entropy to be as large as of the order of $k_B$ per moire. Most importantly, a reduction of entropy with the variation of the in-plane magnetic field is seen. Around 10K, with $B_\parallel = 10$ T, almost $50\%$ reduction is reported in [12, 13], indicating a signature of loosely coupled local moments. It should be noted that for non-interacting itinerant electrons with a small bandwidth of $W \sim 20$ meV, one can also have large entropy of the order of $k_B$ per Moire at the similar temperature range $(10 - 20$ K), which is reported in Ref [12]. We also calculate and show the large entropy in the non-interacting limit in a later section (see SI-Fig.26). However, the entropy of the non-interacting itinerant electrons remains insensitive to the in-plane magnetic field (shown in the later section at SI-Fig. 15). Therefore, large entropy and its reduction with the in-plane magnetic field is the signature of the local moments as shown in Ref [11, 12].

**SI-7: Thermopower and local moments:**

Thermopower ($S_{th}$) is a unique probe used in literature to study the local moments in correlated systems like oxides [14]. The magnitude and sign of the thermopower (or crossing points), and its temperature and magnetic field dependence can reveal many interesting physics.

Let's start with non-interacting itinerant electrons, for which the thermopower can be understood using semi-classical Boltzmann transport theory, which is discussed in SI-8. Here, we briefly summarize it. At lower temperatures, thermopower is proportional to the particle-hole asymmetry or, in other words, to the

derivative of the density of states (DOS) $g(\varepsilon)$. For twisted bilayer graphene, based on the non-interacting DOS, three sign changes or the crossing points are expected within $-4 < \nu < 4$: at the Dirac point ($\nu = 0$) and around the van-hove singularities of the conduction and valence band. However, with increasing temperature, the effects of the sharp band structure features get reduced, and if the thermal broadening becomes comparable to half of the bandwidth ($3.5 k_B T \sim W/2$), the additional sign changes (around the van-hove singularities) merge with the Dirac point with a resultant single crossing. In this condition, the thermopower ($S_{th}$) and the total entropy ($S_{en}$) are related by the Mott-Heikes limit [11] as $S_{th} \approx \partial S_{en}/\partial N$, where $N$ is the total number of particles. The position of the crossing points for non-interacting itinerant electrons is sensitive to temperature and discussed in detail in the section of 'non-interacting calculation' (SI-8) with different bandwidths varying from 10 meV to 40 meV.

Now, let us discuss the other extreme limit of kinetic energy approaching smaller ($W \ll k_B T, U$) so that Coulomb energy completely dominates over the kinetic energy. For this scenario, known as the atomic limit, the thermopower ($S_{th}$) and the entropy ($S_{en}$) remain connected with $S_{th} \approx \partial S_{en}/\partial N$ at any temperature[11]. Therefore, one can easily understand the thermopower from the entropy. Let's consider the simplest case with one orbital (with spin degeneracy) per site. For $k_B T > U$, the entropy has one maxima at half-filling, and minima at zero and full-filling, and, therefore, will have one crossing in thermopower (at half-filling), similar to non-interacting itinerant electrons. This is not surprising since double occupancy per site is allowed for $k_B T > U$. However, for $k_B T < U$, the many-body correlation dominates, resulting in local moments formation (single occupancy or empty site), and the entropy can be categorized into real-space configurational (rearrangement of the carriers) and spin entropy from the local moments. The resultant entropy will have a dumbbell shape: two maxima, and three minima at zero, full-filling, and half-filling[11]. The minima at the half-filling arise due to reduced configurational entropy coming from many-body correlations. Thus, the thermopower will have three crossing points, whose position remains insensitive to temperature as long as the $k_B T < U$ condition is satisfied. This will be discussed in detail in the atomic limit section (SI-11).

Note that for both the above cases, a non-interacting band with finite bandwidth ($\sim 10 - 40$ meV) or atomic limit with bandwidth approaching zero, the magnitude of the thermopower is as high as $\sim k_B/e$; thus, the magnitude alone is not enough to distinguish the strongly correlated local moment regime in MATBLG. The crossing points of thermopower and their temperature dependence are unique diagnostics for studying strongly correlated systems. Further, the application of an in-plane magnetic field will reduce the spin entropy, resulting in the thermopower reduction, which further supports the local moment formation, as seen in correlated oxides [14]. Here, we have briefly summarized the expected thermopower for the two liming cases: a non-interacting band with finite bandwidth and an atomic limit with $U \gg W$. However, in reality, the twisted bilayer graphene is in the intermediate coupling regime where $U \lesssim W$, albeit strongly correlated. In the later sections, we discuss other theoretical approximations like a Hartree-Fock (HF) mean-field theory, dynamical mean-field theory (DMFT) and atomic limit with multi-orbitals to theoretically understand the main features of the thermopower for the realistic twisted bilayer graphene systems. However, capturing every aspect of the experimental data with variations of filling, temperature and magnetic field for a strongly correlated system like MATBLG is extremely challenging and out of the scope of this experimental work.

# SI-8: Thermopower from non-interacting electrons using semi-classical description:

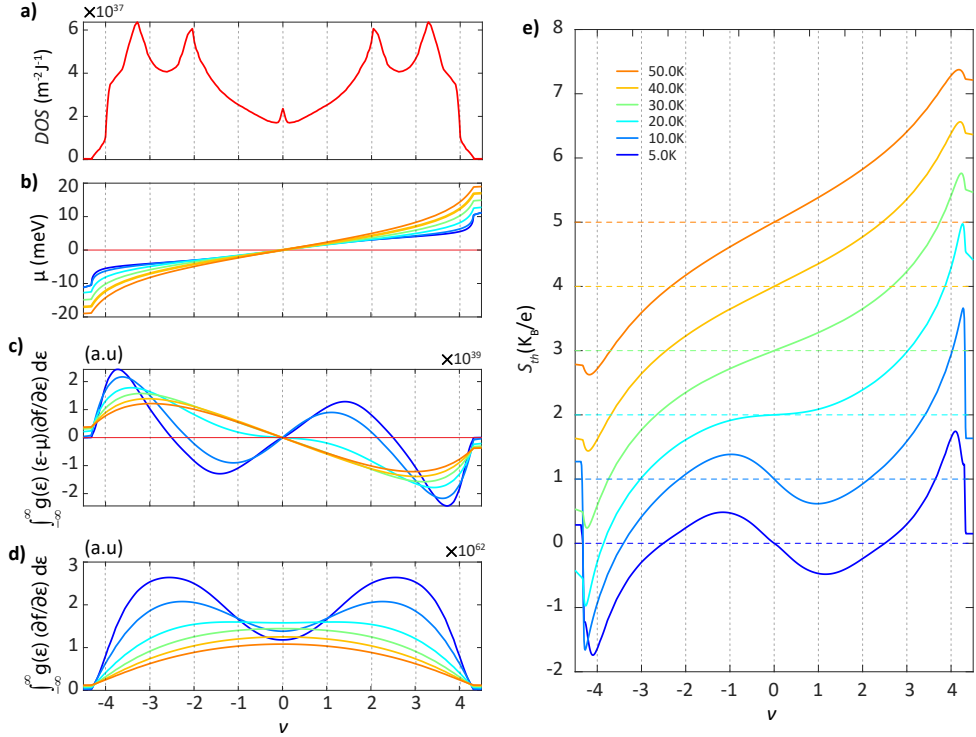

**SI-Fig. 11: Theoretical calculations of thermopower for non-interacting electrons using semi-classical equation: (a)** Continuum model DOS [15] as a function of filling ($\nu$) corresponding to twisted bilayer with twist-angle $1.05^0$. **(b)** Self-consistently solved chemical potential ($\mu$) as a function of $\nu$ at different temperatures. **(c)** and **(d)**, respectively, show the numerator and denominator of Eq. 2 as a function of $\nu$ at different temperatures. **(e)** Semi-classical thermopower with $\nu$ at different temperatures plotted with an offset of 1 $k_B/e$. The dashed horizontal lines correspond to the zero line references for respective temperatures.

The thermopower for the non-interacting itinerant electrons can be described using the semi-classical description (considering energy-independent relaxation approximation):

$$S_{SC}(\mu) = -\frac{1}{Te}\frac{\int_{-\infty}^{\infty}(\epsilon-\mu)g(\varepsilon)(-df/d\varepsilon)d\varepsilon}{\int_{-\infty}^{\infty}g(\varepsilon)(-df/d\varepsilon)d\varepsilon}, \quad (2)$$

where $g(\epsilon)$ refers to the Density of states (DOS) as a function of energy ($\epsilon$), $f$ is the fermi function, $e$, being the electronic charge. The integral runs over all possible energy states between $-\infty$ to $+\infty$. Thus, $S_{SC}(\mu)$ can be theoretically calculated if one is provided with the DOS ($g(\varepsilon)$) as a function of energy ($\varepsilon$). For any temperature, the chemical potential ($\mu(n)$) is self-consistently calculated. Note that in Eq. 2, $df/d\varepsilon$ has a finite width beyond which $df/d\varepsilon \to 0$, thus eliminating the necessity to integrate over $-\infty < \varepsilon < \infty$. For all practical purposes, integral over the low-energy flat band and some portion of the dispersive band accurately depicts the semi-classical behavior of thermopower. Here, we have taken single-particle continuum DOS [15] (as shown in SI-Fig. 11a for twisted bilayer graphene with a twist-angle $\theta = 1.05^o$ and bandwidth, $W \sim 10meV$. A gap of $5meV$ is considered between the low-energy flat band and the

higher-energy dispersive band. For DOS of the dispersive band, we have used the band dispersion by Das, I., Lu, X., Herzog-Arbeitman, J. et al. [16]. In SI-Fig. 11, we present the self-consistently solved $\mu$, the denominator, and numerator of Eq. 2, (i.e $\int_{-\infty}^{\infty} g(\varepsilon)(df/d\varepsilon)d\varepsilon$ and $\int_{-\infty}^{\infty}(\varepsilon-\mu)g(\varepsilon)(df/d\varepsilon)$ ) and finally the solved $S_{SC}$ for few temperatures (5K-50K in shift plot).

**Thermopower for non-interacting electrons with different bandwidths:** The thermopower is calculated by taking DOS with larger bandwidths of 20-40meV. For this purpose, we have used the same flat band DOS [15] (appropriately normalized) as in the previous section with larger bandwidths of 20 and 40meV, respectively. In SI-Fig. 12d, we plot the additional crossing points as a function of temperature for $10meV$, $20meV$, and $40meV$ bandwidths. The additional crossing points remain sensitive to temperature, and as expected, the critical temperature at which the additional crossing points merge to the Dirac point depends on the bandwidth.

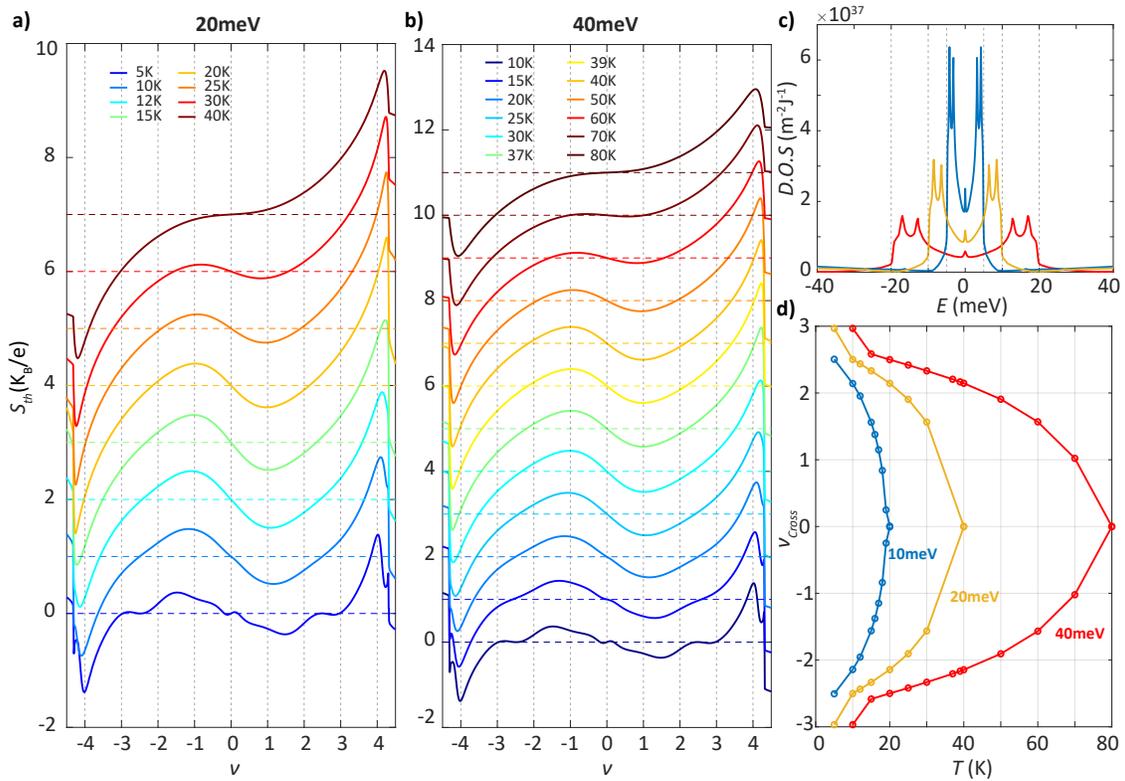

**SI-Fig. 12:** **Theoretical calculations of thermopower for differnt bandwidths: (a),(b)** The calculated thermopower for 20meV and 40meV bandwidth, respectively with $\nu$ at different temperatures. The dashed horizontal lines correspond to the zero line references for respective temperatures. **(c)** and **(d)**, respectively, show the DOS and crossing points versus temperature behavior for 10, 20, and 40meV bandwidth of the flat band.

The qualitative behavior of thermopower from non-interacting electrons does not change significantly on the nature of DOS. For example, a similar thermopower response is calculated for a saw-tooth DOS. We have used saw-tooth DOS with 10, 20, and 40meV bandwidth (as seen in SI-Fig 13 a,b,c). For all cases, we observed similar thermopower behavior with similar crossing point evolution (see SI-Fig 13 d) compared to the previously mentioned DOS based on the continuum model.

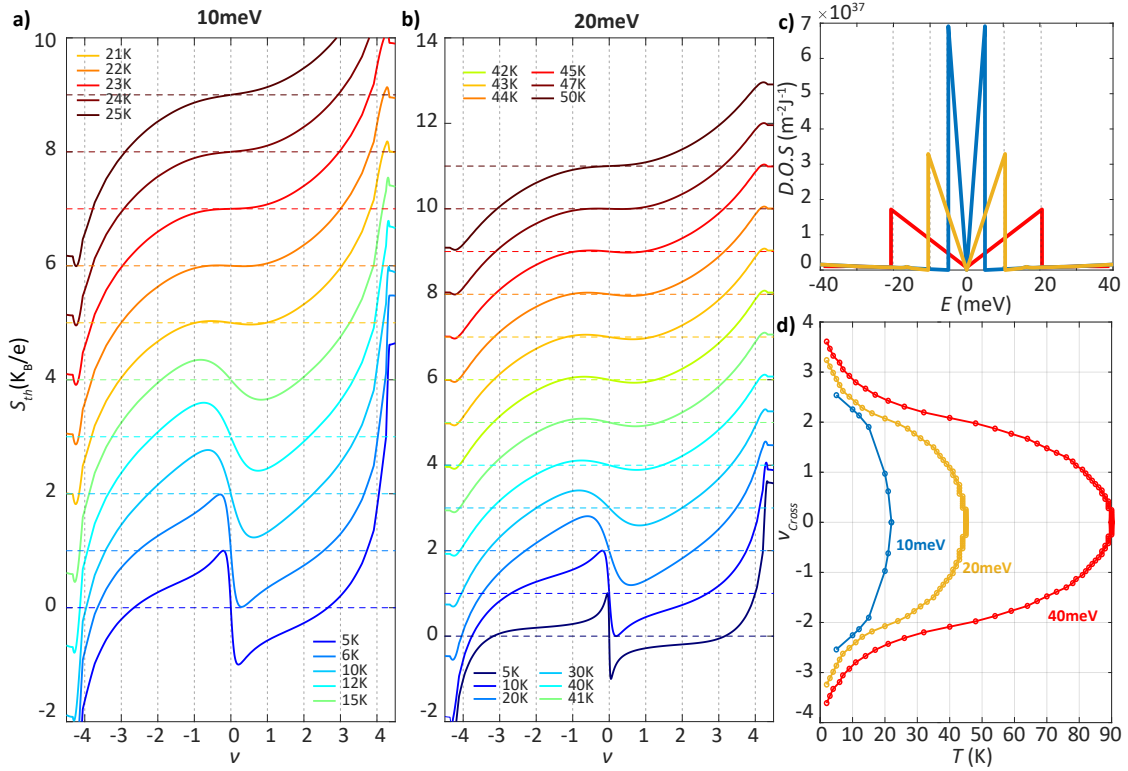

**SI-Fig. 13: Thermopower using Saw-tooth DOS: (a),(b)** Semi-classical thermopower calculated for 10meV and 20meV bandwidth, respectively, with $\nu$ at different temperatures. The dashed horizontal lines correspond to the zero line references for respective temperatures. **(c)** and **(d)**, respectively, show the saw-tooth DOS and crossing point versus temperature behavior for 10, 20, and 40meV bandwidth of the flat band.

21

**Thermopower for asymmetric DOS using semi-classical description:** SI-Fig. 14 shows the thermopower for asymmetric DOS at different temperatures. As expected from low to high temperatures, the three crossing points within $-4 < \nu < 4$ evolve into a single crossing point. However, as the temperature rises, the crossing point is offset with respect to the Dirac point, which arises due to asymmetric DOS. Such asymmetric DOS has already been reported in theoretical calculations based on relaxed moire structure [4]. The temperature response of thermopower for our graphite-gated TBLG device shows good qualitative agreement with this theoretical asymmetric DOS model.

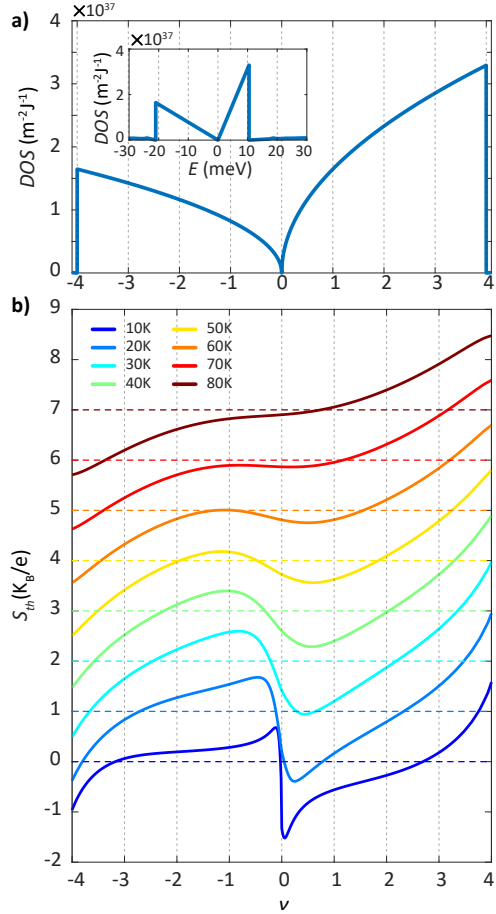

**SI-Fig. 14: Thermopower using asymmetric saw-tooth DOS: (a)** Asymmetric saw tooth DOS as a function of filling ($\nu$). **(b)** Semi-classical thermopower with $\nu$ at different temperatures plotted with an offset of 1 $k_B/e$ unit. The dashed horizontal lines correspond to the zero line references for respective temperatures.

**In plane magnetic field dependence of thermopower for non-interacting electrons:** In the absence of in-plane magnetic field, because of the spin degeneracy, the total DOS can be thought of as $DOS(\epsilon) = DOS_\uparrow(\epsilon) + DOS_\downarrow(\epsilon) = 2 \times DOS_{\uparrow/\downarrow}(\epsilon)$, where $DOS_\uparrow(\epsilon) = DOS_\downarrow(\epsilon)$ are the DOS for spin up and down states. In the presence of an in-plane magnetic field ($B$), spin degeneracy will be lifted. As a result, the effective DOS can be thought of as a combination of two identical DOS that respectively are shifted up and down in energy by $g\mu_B B$ (where $\mu_B$ is Bohr Magneton and $g$ is the Landé g-factor) for spin up and down bands. The effective DOS thus will have a form $DOS(\epsilon) = DOS_\uparrow(\epsilon - g\mu_B B/2) + DOS_\downarrow(\epsilon + g\mu_B B/2)$. Thus with increasing magnetic field as the spin up and spin down states move apart in energy the effective DOS will be altered, which intern will manifest as magnetic field dependence of the semi-classical thermopower. Taking $g = 2$ and $B = 10T$, the relative shift in energy is $g\mu_B B = 1.2$meV. The evolution of the thermopower with the magnetic field up to $10T$ at $20K$ is shown for the sawtooth DOS (see SI-Fig 15 a,c) as well as continuum model DOS (see SI-Fig 15 b,d). We observe no noticeable dependence of the thermopower on magnetic field, at least within $-3.5 < \nu < 3.5$ (see SI-Fig 15 c,d). However, near the band edge $|\nu| \approx 4$ thermopower shows a change (see SI-Fig 15 c,d), and this is expected as the only significant changes of the DOS happen near the band edge (see SI-Fig 15 a,b).

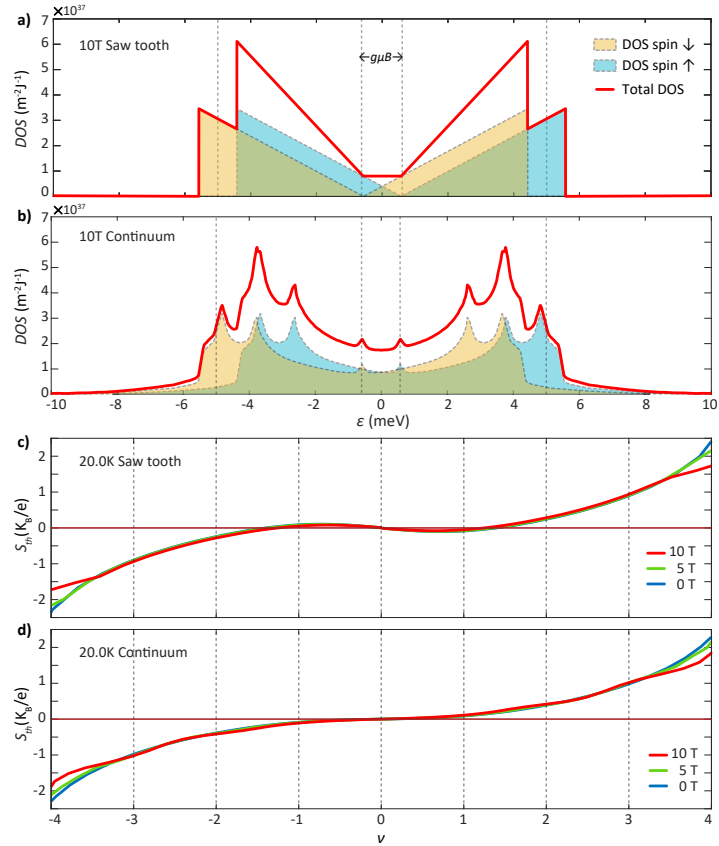

**SI-Fig. 15: Thermopower of non-interacting electrons with in-plane magnetic field: (a),(b)** Sawtooth and continuum DOS with energy at $10T$ in-plane magnetic field. The spin up and spin down state DOS are shown in yellow and blue shaded plots (respectively shifted by $\pm g\mu B/2$ along the X-axis), and the effective total DOS is shown as a solid red line. **(c),(d)** Calculated thermopower with filling at $20K$ for $B = 0, 5$ and $10T$ for saw-tooth and continuum DOS, respectively.

23

## SI-9: Thermopower in Hartree-Fock mean-field theory for Dirac revivals in MATBLG:

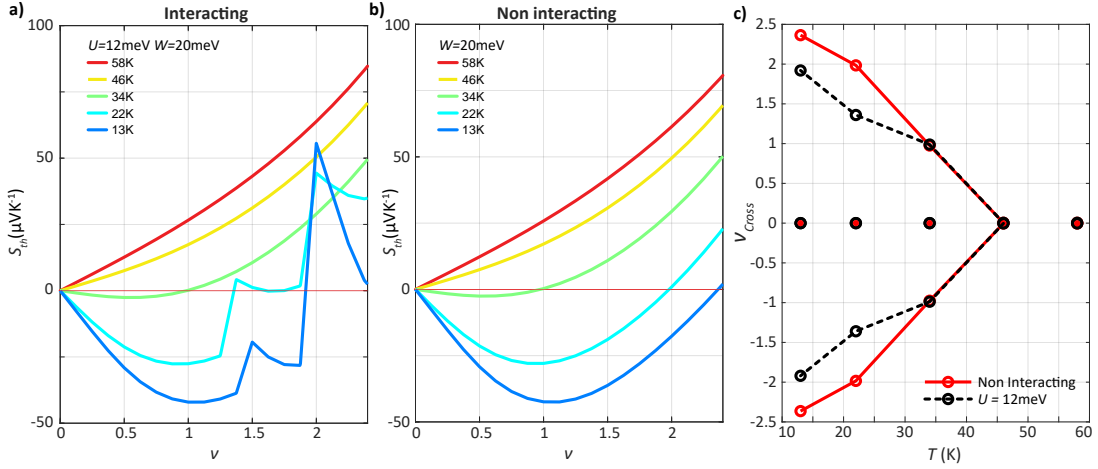

**SI-Fig. 16: Thermopower in Hartree-Fock theory: (a),(b)** Thermopower calculated using Hartree-Fock (HF) method with interaction $U = 12$ meV, compared with the non-interacting result with the same bandwidth ($W = 20$meV). **(c)** Crossing point versus temperature for HF thermopower compared with the non-interacting limit.

To contrast with our experimentally observed features of thermopower in MATBLG, namely the robust crossing point, captured within strong-coupling theories, like the dynamical mean field theory (DMFT) SI-10 and the atomic limit SI-11, here we discuss thermopower in a simplified weak-coupling Hartree-Fock (HF) theory for a minimal model [2, 15] for MATBLG with four fermionic flavours. The flavours describe the spin and valley degrees of freedom that interact via a local Coulomb interaction. HF theory is a 'static' mean field theory that can capture temperature and interaction dependent band-structure renormalizations for weak interaction, such as shift in the position of the bands and renormalization of bandwidths, e.g., due to low-temperature symmtry breaking. Various kinds of HF theories have been extensively used in refs[2, 17, 18, 19] to understand interaction effects in MATBLG. As discussed in our earlier work [2], and also in ref. [15], HF theory, even at a minimal level, leads to a cascade of Stoner-like Dirac revival transitions as a function of filling $\nu$ in MATBLG due to spontaneous spin and/or valley polarizations for temperature $T \leq 20K$. The Dirac revivals give rise to characteristic peaks with large thermpopower around integer fillings due to emergent low-energy particle-hole asymmetry. This is consistent with other measurements like STM [5, 6] and compressibility measurements [15]. However, inevitably, at a temperature above the symmetry breaking transition temperature $\sim 20K$, where the order parameter vanishes, the thermopower response in the HF and noninteracting theory (see SI-Fig 16) become identical. Thus, the HF theory only captures the interaction physics at a lower temperature and cannot capture the strong correlation regime with local moments at intermediate and high temperatures.

As is Ref.[15], we use the following Hamiltonian for the MTBLG

$$\mathcal{H} = \sum_{\mathbf{k},\alpha,b} (\epsilon_{\mathbf{k}\alpha b} - \mu) c^\dagger_{\mathbf{k}\alpha b} c_{\mathbf{k}\alpha b} + \frac{U}{A} \sum_{\alpha<\beta} \int d^2 r c^\dagger_\alpha(\mathbf{r}) c^\dagger_\beta(\mathbf{r}) c_\beta(\mathbf{r}) c_\alpha(\mathbf{r}), \quad (3)$$

where $\alpha = 1, \ldots, 4$ are the spin-valley indices, $b$ is the moire' band index, $\mathbf{k}$ is the momentum within the first moire' Brillouin zone, $A$ is the area of the moire' unit cell, $\mu$ is the chemical potential; $c^\dagger_{\mathbf{k}\alpha b}$ and $c_{\mathbf{k}\alpha b}$ are the

electron creation and annihilation operators. In the above, we have considered a local Coulomb interaction $U$ [2, 15], with $c_\alpha(\mathbf{r}) = (1/\sqrt{N}) \sum_{\mathbf{k}b} e^{i\mathbf{k}\cdot\mathbf{r}} c_{\mathbf{k}\alpha b}$ the local electron annihilation operator at $\mathbf{r}$; $N$ is the number of moiré unit cells. We restrict the summation over the band index to the flat bands ($b = 1, 2$) around the moiré Dirac points. We perform Hartree-Fock approximation for the Hamiltonian (Eq.3) using the mean-field Hamiltonian $\mathcal{H}_{MF} = \sum_{\mathbf{k},\alpha,b} (\epsilon_{\mathbf{k}\alpha b} + \mu_\alpha - \mu) c^\dagger_{\mathbf{k}\alpha b} c_{\mathbf{k}\alpha b}$, where $\mu_\alpha$ are the variational parameters, and obtain the self-consistency equations

$$\mu_\alpha = U \sum_{\beta \neq \alpha} n_\beta, \tag{4a}$$

$$n_\alpha = \int_{-\infty}^{\infty} d\epsilon\, g(\epsilon) f(\epsilon + \mu_\alpha - \mu). \tag{4b}$$

Here $n_\alpha$ is the occupation of flavor $\alpha$ and $g(\epsilon) = (1/N) \sum_{\mathbf{k}b} \delta(\epsilon - \epsilon_{\mathbf{k}\alpha b})$ is the non-interacting DOS (per unit cell), discussed in SI-Sec. 8, which does not depend on the flavor index. $f(\epsilon) = 1/(e^{\epsilon/k_B T} + 1)$ is the Fermi function at temperature $T$. We numerically solve the above HF self-consistency equations iteratively to obtain $\mu_\alpha$ ($n_\alpha$) along with equation $\sum_\alpha n_\alpha = (\nu + 4)$ for $\mu$, where $\nu$ ($-4 \leq \nu \leq 4$) is the fiiling measured from moiré Dirac point. We find the Dirac revival transitions around the integer fillings $\nu = 1, 2, 3$ up to temperature $T \sim 0.1 - 0.15 W$ for an interaction $U = 0.6 W$. For a bandwidth of $W = 20$ meV, this translates to symmetry breaking transition temperatures $\sim 20 - 30$ K.

We use the following expressions for thermopower (Seebeck coefficient) $S_{th}$, obtained from Kubo formulae neglecting vertex corrections [20],

$$S_{th} = -\frac{k_B}{e} \frac{A_1}{A_0} \tag{5a}$$

where

$$A_m = \int_{-\infty}^{\infty} d\omega\, \mathcal{A}^2(\epsilon, \omega) \Phi(\epsilon) \left(-T \frac{\partial f(\omega)}{\partial \omega}\right) \left(\frac{\omega}{k_B T}\right)^m, \tag{5b}$$

$$\mathcal{A}(\epsilon, \omega) = -(1/\pi) \text{Im}(1/(\omega + \mu - \epsilon - i\Gamma_0)) \tag{5c}$$

$$\Phi(\epsilon) = (1/NA) \sum_{\mathbf{k}\alpha b} (\partial \tilde{\epsilon}_{\mathbf{k}\alpha b} / \partial k_x)^2 \delta(\epsilon - \tilde{\epsilon}_{\mathbf{k}\alpha b}) \tag{5d}$$

with $m = 1, 2$; $\mathcal{A}(\epsilon, \omega)$ is the spectral density and $\Phi(\epsilon)$ the transport DOS. Here $\Gamma_0 = 0.0005 W$ is small impurity broadening put in by hand to incorporate elastic scatterings, and $\tilde{\epsilon}_{\mathbf{k}\alpha b} = \epsilon_{\mathbf{k}\alpha b} + \mu_\alpha$ is the HF dispersion. For our calculations we have approximated $v^x_{\mathbf{k}\alpha b} = \partial \tilde{\epsilon}_{\mathbf{k}\alpha b}/\partial k_x$ by a constant velocity. This makes $S_{th}$ (Eq.5a) independent of the band velocity and gives the same result as the semiclassical Boltzmann formula for thermopower in Eq.(2).

As discussed in Refs.[2, 15], in the Dirac revival picture for $\nu > 0$, the population ($n_\alpha$, $\alpha = 1, \ldots, 4$) of flavors follows a sequence of Stoner-like transitions where one, two and three of the flavors get fully occupied ($n_\alpha = 2$) at $\nu = 1, 2$ and 3 ($n/n_s = 0.25, 0.50$ and $0.75$), respectively, resetting the fillings of the remaining flavors at $n_\alpha = 1$. Due to these resettings, the effective 'non-rigid' single-particle DOS undergoes successive reconstructions, as a function of both filling and temperature, leading to rather drastic PH asymmetry even at very low energies around the chemical potential ($\mu$), near each integer fillings. Thermopower being

a sensitive probe of the low-energy PH asymmetry, exhibits prominent peaks around, albeit not always exactly at, the integer fillings, at low temperature $T < 20 - 30$ K, as shown in SI-Fig 16(a). For comparison, we also plot the thermopower in the non-interacting case in SI-Fig 16(b). As evident in SI-Figs 16(a),(b), HF result for $S_{th}$ becomes identical to that of non-interacting systems above $T \sim 30$ K, where the ordering vanishes. Consequently, as demosntrated in SI-Fig 16 (c) the crossing point of HF thermopower tracks the non-interacting crossing point and is strongly temperature dependent, unlike in our experiments (see Figure. 2,3 of main manuscript) and in the strongly correlated regime captured by the DMFT and the atomic limit (see SI-10,11).

**SI-10: Thermopower in dynamical mean-field theory (DMFT) in MATBLG:**

As discussed in the preceding sections, thermopower obtained in the non-interacting model and weak-coupling HF theory for MATBLG cannot capture experimentally observed robust crossing point that is relatively insensitive to temperature over a large range ($\sim 20 - 60$ K) (Fig. 2 and Fig. 3 of the manuscript.) To address the putative strongly correlated regime realized in MATBLG, here we discuss the thermopower computed within a single-site dynamical mean field theory (DMFT) [21] in the minimal interacting model (Eq.(3)) that we employed for the HF theory of MATBLG in SI-Sec.9. Over the last two decades, the single-site DMFT and its cluster extensions have emerged as one of the most successful approximations to describe strongly correlated materials [21, 22]. DMFT, in its single-site implementations, maps the interacting model into an effective impurity problems in a self-consistent bath [21], and, as a result, captures non-trivial dynamical quantum correlations crucial to access strong-correlation phenomena, but ignores spatial fluctuations. As well-known from studies on one-band Hubbard model, for large interaction $U$ comparable or larger than the bandwidth $W$, i.e., $U \gg$ hopping scale $t$, DMFT naturally leads to local moment formation both in the Mott insulating phase at half filling, and in the metallic phase at half filling as well as away from half filling [21]. In particular for the paramagnetic Fermi liquid (FL) metallic phase at large $U$, the usual single-site DMFT gives rise to a spin susceptibility $\chi \sim 1/(T + T_{coh} + J)$ [21], with an FL coherence temperature $T_{coh} \sim t^2/U$ and an exchange scale $J \sim t^2/U$. Thus for intermediate and high temperature $t^2/U \lesssim T \lesssim U$, the susceptibility is Curie like, indicating the presence of local moments over a broad temperature range. The thermopower computed within DMFT for large $U \lesssim W$ shows robust crossing point as in our experiment, and as obtained in the atomic limit calculation SI-Sec.11. The presence of local moments is inevitable in the latter limit, and, thus, the DMFT results for $U \lesssim W$ correlates the robust crossing point with the strongly correlated local moment regime while smoothly connecting it to the extreme atomic limit of $U \gg W$.

To set up the DMFT computation involving the four moire$'$ flat bands, we envisage an effective hexagonal lattice model for the moire lattice with four flavours with $SU(4)$-symmetric on-site Hubbard interaction term, as in the HF calculations (SI-Sec.9), indexed by $\alpha = 1, \ldots, \mathcal{M}$ with $\mathcal{M} = 4$, such that there are four bands that can hold maximum eight electrons per triangular unit cell, i.e.

$$\mathcal{H} = -\sum_{ij,\alpha} t_{ij} c_{i\alpha}^\dagger c_{j\alpha} + U \sum_{i,\alpha<\gamma} n_{i\alpha} n_{i\gamma} \qquad (6)$$

Here $c_{i\alpha}$ is the electron operator for $i$-th hexagonal lattice site. The hoppings are in general complex and can be chosen to fit [17, 23, 24] the flat-band dispersions from band-structure calculation. However, the four-flavour model is insufficient [17, 25] to reproduce the topology of the Bloch wave functions while keeping all the low-energy symmetries of the continuum model [26]. Within our simplified DMFT approximation,

26

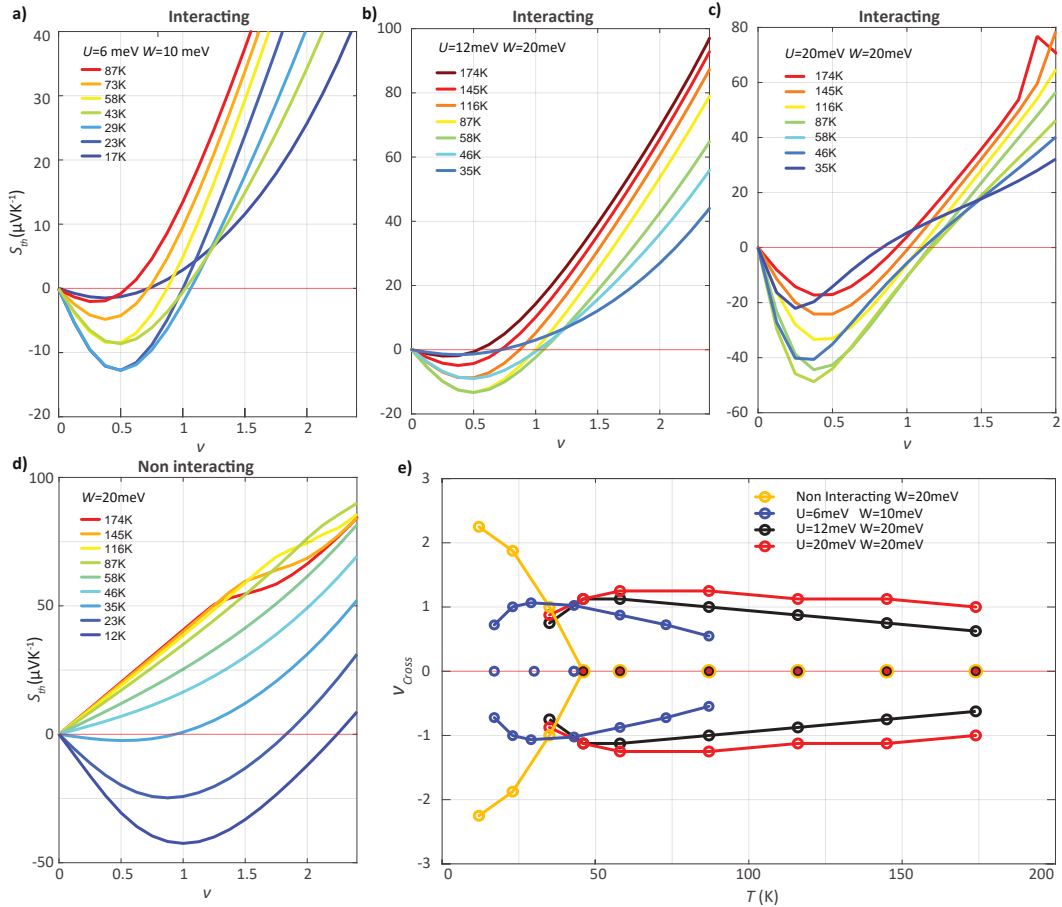

**SI-Fig. 17: DMFT calculation:** The thermopower calculated in DMFT is shown for **(a)** $U = 6$ meV, $W = 10$ meV, **(b)** $U = 12$ meV, $W = 20$ meV, and **(c)** $U = 20$ meV, $W = 20$ meV, at various temperatures, and contrasted with the non-intecrating results for **(d)** $U = 0$, $W = 20$ meV. **(e)** Temperature evolution of the crossing point extracted from panels **(a), (b), (c), (d)** for $U \lesssim W$ has only weak temperature dependence over a broad range, unlike the non-interacting limit, where the crossing point varies significantly with temperature and merges to a single crossing point at much lower temperature.

27

only the DOS of the low-energy bands enter and we take the DOS discussed in SI-Sec.8. Hence the tight-binding parameterization for a low-energy effective lattice model is not explicitly needed for our DMFT calculations.

In the DMFT approximation [21], and assuming a homogeneous state with flavour ($SU(4)$) symmetry, we reduce the model of Eq.(6) to an effective single-site impurity problem with imaginary-time action

$$S_{imp} = -\int_0^\beta d\tau d\tau' \sum_\alpha \mathcal{G}^{-1}(\tau - \tau')\bar{c}_\alpha(\tau)c_\alpha(\tau') + U \sum_{\alpha<\gamma} \int_0^\beta d\tau n_\alpha(\tau) n_\gamma(\tau) \tag{7}$$

Here $\beta = 1/k_B T$ and $(\bar{c}_\alpha, c_\alpha)$ are fermionic Grassmann variables with $n_\alpha = \bar{c}_\alpha c_\alpha$. The dynamical mean field $\mathcal{G}^{-1}(i\omega_n) = i\omega_n + \mu - \Delta(i\omega_n)$ for the Matsubara frequency $\omega_n = (2n+1)k_B T$, with $n$ being an integer, is determined by the hybridization function $\Delta(i\omega_n)$ which is self-consistently determined using the non-interacting lattice DOS as we discuss below. The chemical potential is fixed by the filling.

Since we need the real-frequency electronic Green's function $G(\omega) = G(i\omega_n \to \omega + i0^+)$ to compute dc transport coefficients, e.g. Seebeck coefficient, we use an approximate impurity solver, namely modified iterative perturbation theory (IPT) [21, 27] and its generalization for the multi-orbital case [28]. The latter has been benchmarked with numerically exact techniques for solving the single-impurity problem [28] and expected to work quite well for moderate interaction strengths $U \lesssim W$. Within IPT, we obtain the impurity self-energy as

$$\Sigma(\omega) = (\mathcal{M}-1)U\langle n \rangle + \frac{A(\mathcal{M}-1)\Sigma^{(2)}(\omega)}{1-B(\mathcal{M}-1)\Sigma^{(2)}(\omega)}, \tag{8}$$

and thus the impurity Green's function is obtained from the Dyson equation $G^{-1}(\omega) = \mathcal{G}^{-1}(\omega) - \Sigma(\omega)$. The first term in Eq.(8) is the Hartree self energy and $\Sigma^{(2)}(\tau) = U^2 \tilde{\mathcal{G}}^3(\tau)$ is the second-order self-energy obtained using the Hartree-corrected impurity Green's function $\tilde{\mathcal{G}}^{-1}(\omega) = \mathcal{G}^{-1}(\omega) - U\langle n \rangle$. The coefficients $A$ and $B$ are chosen to satisfy certain sum rules and known high-frequency behavior of impurity Green's function [28] and are given by

$$A = \frac{\langle n \rangle(1-\langle n \rangle)}{\langle n_0 \rangle(1-\langle n_0 \rangle)} + \frac{(\mathcal{M}-2)(\langle nn \rangle - \langle n \rangle^2))}{\langle n_0 \rangle(1-\langle n_0 \rangle)} \tag{9}$$

$$B = \frac{(1-(\mathcal{M}-1)U\langle n \rangle) + \mu_0 - \mu}{(\mathcal{M}-1)U^2 \langle n_0 \rangle(1-\langle n_0 \rangle)} \tag{10}$$

Here $\langle n \rangle = -(1/\pi)\int_{-\infty}^\infty d\omega f(\omega)\text{Im}G(\omega) = (\nu+4)/8$ is a the occupation of a single orbital. The density-density correlation function $\langle nn \rangle \approx -\int_{-\infty}^\infty d\omega f(\omega) \text{Im}(\Sigma G)/(\pi U(\mathcal{M}-1))$. We fix the 'pseudo' chemical potential $\mu_0$ from $\langle n_0 \rangle = -(1/\pi)\int_{-\infty}^\infty d\omega f(\omega)\text{Im}\tilde{\mathcal{G}}(\omega) = \langle n \rangle$ using the Hartree corrected Green function $\tilde{\mathcal{G}}^{-1}(\omega) = \omega + \mu_0 - \Delta(\omega) - U\langle n \rangle$ [29].

Once the self-energy [Eq.(8)] is obtained, the DMFT self-consistency condition is used to compute the local lattice Green's function

$$G(\omega) = \frac{1}{2\mathcal{M}} \int_{-\infty}^\infty d\epsilon \frac{g(\epsilon)}{\omega + \mu - \epsilon - \Sigma(\omega)}, \tag{11}$$

28

assumed to be the same as the impurity Green's function for a single flavour. The self-consistency loop is closed by obtaining the new hybridization function $\Delta(\omega) = \omega + \mu - \epsilon - \Sigma(\omega) - G^{-1}(\omega)$. The loop is iterated till convergence is reached.

The Seebeck coefficient [20] $S_{th}$ is obtained from the Kubo formula in Eq.(5), where

$$A_m = \int_{-\infty}^{\infty} d\omega d\epsilon \mathcal{A}^2(\epsilon, \omega) \Phi(\epsilon) \left(-T \frac{\partial f(\omega)}{\partial \omega}\right) (\beta\omega)^m \tag{12}$$

with $m = 0, 1$. $\mathcal{A}(\epsilon, \omega) = -(1/\pi)\mathrm{Im}(1/(\omega + \mu - \epsilon - \Sigma(\omega)))$ is obtained from DMFT self-energy and $\Phi(\epsilon) = (1/NA) \sum_{n=1,\mathbf{k}}^{4} (\partial\epsilon_{n\mathbf{k}}/\partial k_x)^2 \delta(\epsilon - \epsilon_{n\mathbf{k}})$ is the transport DOS. We again make a constant band velocity approximation, as in the HF calculation (see SI-Sec.9).

The results for thermopower calculated in DMFT are shown in SI-Fig. 17 for three values of $U \lesssim W$ and constratsed with the non-interacting case. The DMFT thermopower exhibits a robust weakly temperature dependent crossing point over a broad range of temperature, unlike the non-interacting results.

**SI-11: Thermopower from atomic limit:**

The Kubo formula for thermopower, which is formally exact, is given by [30]

$$S_{th} = \frac{1}{T} \frac{\int_0^\infty dt \int_0^\beta d\tau \langle \hat{J}_x^E(t - i\tau) \hat{J}_x(0) \rangle}{\int_0^\infty dt \int_0^\beta d\tau \langle \hat{J}_x(t - i\tau) \hat{J}_x(0) \rangle} - \frac{\mu(T)}{q_e T} \tag{13}$$

where $J_x^E$ and the $J_x$ are the energy and current density operators, $\mu$ is the chemical potential, and $T$ is the temperature. While for weakly interacting diffusive metals, this boils down to Mott's formula. For systems with interactions retaining strong correlations and transport coefficients in a single formalism is difficult. Thus approximate methods have been found useful particularly for systems in strong correlations [11, 30, 31, 32, 33].

In particular, for strong Hubbard interactions $U$, one can work in the limit where the hopping integral $t$ can be ignored when $t \ll U$, this leads to the atomic limit calculations [11]. Here the contribution of the first term comes directly through the transport, and second term which is also transport driven is albeit considered entropic contributions leading to the idea that thermopower often measures the entropy per carrier. While $\mu$ is essentially an entropy derivative with respect to density at a constant energy; a particularly useful related quantity is

$$S_{th} = -\frac{1}{e} \left(\frac{\partial S_{en}}{\partial N}\right)_{T,V} \tag{14}$$

where $S_{en}$ is the total entropy and $N$ is the total number of particles. The above equation goes under the name of Kelvin formula where the derivative is now done at a constant temperature. This follows from the so-called slow limit which is known to capture the many-body density of states enhancements even though it misses velocity and relaxation contributions[34, 35]. We use the Kelvin formula in the atomic limit to capture the essential experimental phenomenology. This is particularly suited since in the flat band limit, we are working where the bandwidth is much smaller than the Hubbard interactions.

**Entropy calculations using atomic limit:** We first consider a particular atomic limit which best captures the most prominent feature of our experimental data in MATBLG, namely three temperature-independent robust crossing points in thermopower. In this case, we take effective intra-flavour or 'intra-orbital' spin SU(2)-symmetric on-site Hubbard interaction without any inter-orbital interaction. The Hamiltonian in this atomic limit is given by

$$H = \sum_\alpha \left( U n_{\alpha\uparrow} n_{\alpha\downarrow} + \epsilon_\alpha n_\alpha \right) \quad (15)$$

where $\alpha = \{1,\ldots,4\}$ labels the four orbitals correspsponding to the two valley and two sublatices in the completely flatband limit [9, 10] and $U$ is the Hubbard interaction strength. We also keep on-site energies $\epsilon_\alpha$ as tuning parameters that can mimic the effects of finite energy dispersion breaking the degeneracy of the bands. Upon evaluating the partition function $Z$, we obtain the corresponding Helmholtz free energy $F = -k_B T \log Z$ and from that, the entropy $S_{en} = -(\partial F/\partial T)$.

The thermopower is then given by (using the Kelvin formula),

$$S_{th} = -\frac{1}{e}\frac{\partial S_{en}^M}{\partial \nu} \quad (16)$$

where $S_{en}^M$ is the entropy per moire unit cell, $e$ is electric charge[34]. It is to be noted that $\mu$ is solved self-consistently to fix the filling $\nu$. In strongly interacting systems, particularly when electronic bandwidth is comparatively smaller to temperature and/or interactions, interpretation of thermopower when analysed in atomic limit has met remarkable success. In particular, it has been useful to explain its non-monotonic behavior (as a function of temperature) in strongly correlated systems[11, 36, 37], their magnetic field dependence [38] and predict characteristic zero crossings of thermopower as a function of density where effective phase space gets modified due to strong interactions[39]. The recurring idea here is that at large temperatures, even in the presence of strong interactions, the thermopower is essentially how *thermodynamic entropy* changes with variations in carrier density [30]. Given bilayer graphene at magic angles has a flat band and comparatively strong interactions [9, 26, 40, 41] – we explore this limit of analysis in the context of the experimental data. Assuming $-\epsilon_1 = -\epsilon_2 = \epsilon_3 = \epsilon_4 = 0.2 meV$ we investigate the behavior of thermopower as a function of temperature and Hubbard interaction strength $U_\alpha = U$ which is kept the same for all the orbitals. Finite values of $\epsilon_1,\ldots,\epsilon_4$ is kept to model a finite bandwidth of the flat bands. The resulting behavior is shown in SI-Fig. 19.

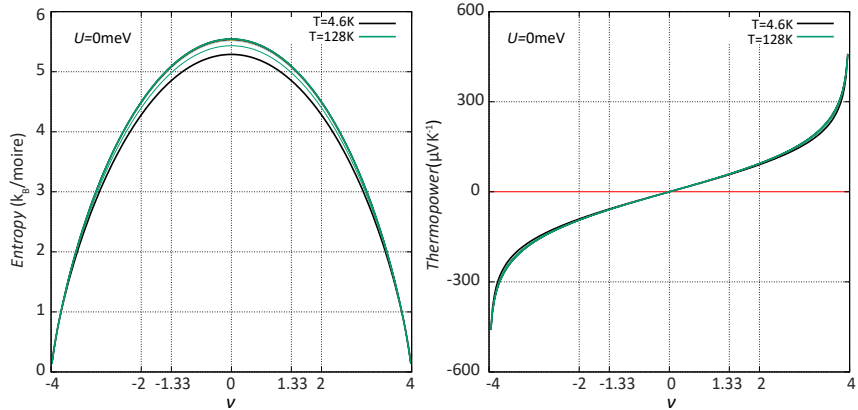

**SI-Fig. 18:** Behavior of entropy and thermopower for the four orbital atomic limit where the onsite energies are $-\epsilon_1 = -\epsilon_2 = \epsilon_3 = \epsilon_4 = 0.2 meV$, Hubbard interaction $U = 0 meV$ as a function of filling. Different curves correspond to different temperatures: 4.6K, 7K, 14K, 16K, 19K, 21K, 23K, 26K, 35K, 46K, 58K, 70K, 81K, 93K, 104K, 116K and 128K. Thermopower has a zero crossing at only at $\nu = 0$.

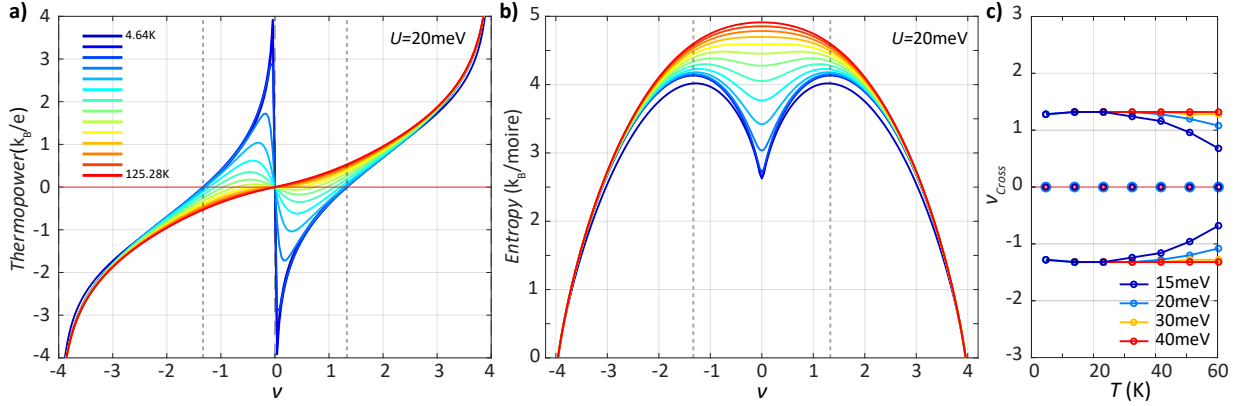

**SI-Fig. 19:** **(a),(b)** Behavior of thermopower and entropy for the four orbital atomic limit where the onsite energies are $-\epsilon_1 = -\epsilon_2 = \epsilon_3 = \epsilon_4 = 0.2 meV$, Hubbard interaction $U = 20 meV$ as a function of filling. Different curves correspond to different temperatures: 4.6K, 14K, 23K, 33K, 42K, 51K, 60K, 70K, 79K, 88K, 97K, 106K, 116K and 125K. At lower temperatures, thermopower has zero crossings only at $\nu = 0, \pm\frac{4}{3}$. **(c)** Thermopower zero-crossing points as a function of temperature for different interaction energies (U=20,30 and 40 meV).

While in the absence of any interactions, the thermopower shows a single crossing at $\nu = 0$ (see SI-Fig. 18) with increasing temperature. This single crossing happens here since the entropy maximizes exactly at $\nu = 0$. This is essentially the point where the phase space is maximum, given both the particles and holes are equal. However, as interactions are slowly increased, a characteristic undulation develops near $\nu = 0$, which creates two extra crossings, which remain stable even at relatively large high temperatures. These interaction dominated crossings develops at $\nu = \pm\frac{4}{3}$. Even for a single orbital model, given $N_A$ atoms and $N$ electrons in the system, the total number of states are [39]

$$g = \frac{N_A! 2^N}{N!(N_A - N)!} \quad (17)$$

In the thermodynamic limit, this gives the thermopower as $S_{th} \sim \frac{d\log g}{d\rho} \sim \log(2(1-\rho)/\rho)$, thus giving a zero crossing at $\rho = \frac{2}{3}$. This corresponds to for the four orbital model $\nu = 4(\rho - 1) = -\frac{4}{3}$. This is strikingly similar to the experimental observations at high temperatures, which seems to point out that even at high temperatures, strong correlation physics is at play in the twisted bilayer graphene system. A similar calculation taking into account doublons will lead to a crossing near $\nu = \frac{4}{3}$. Expanding the orbital energies such that $-\epsilon_1 = \epsilon_4 = 5 meV$ $-\epsilon_2 = \epsilon_3 = 1.7 meV$ and with a Hubbard interaction of around $25 meV$ we again find that the extra crossings appear for $T < 100K$ (see SI-Fig. 20) however now closer to $\nu \sim \pm 1.1$. This reflects that the atomic orbital limit captures the essential experimental phenomenology of thermopower in the magic angle twisted bilayer graphene.

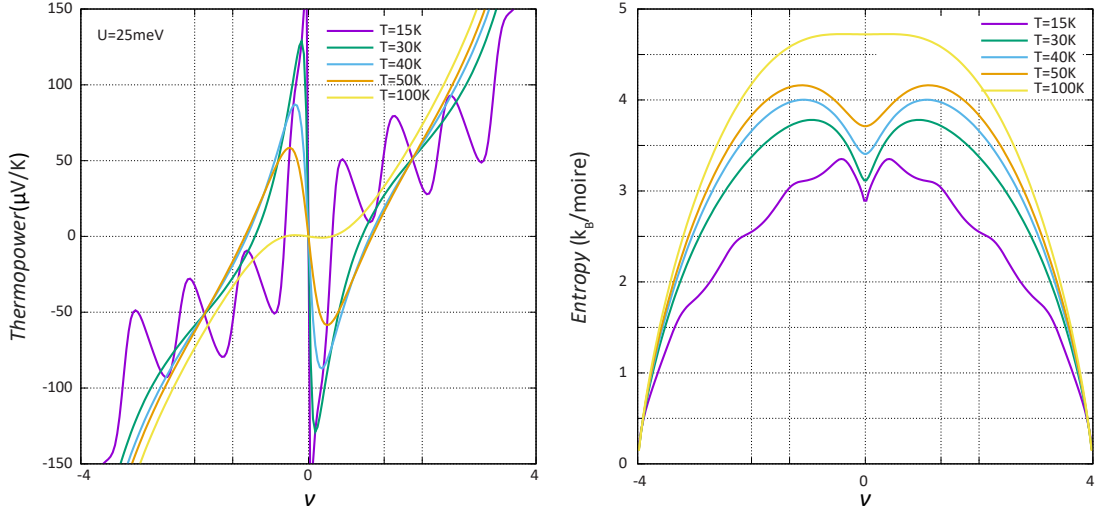

**SI-Fig. 20:** Behavior of thermopower and entropy for the four orbital atomic limit where the onsite energies are $-\epsilon_1 = \epsilon_4 = 5meV$ $-\epsilon_2 = \epsilon_3 = 1.7meV$, Hubbard interaction $U = 25meV$ as a function of filling.

**Atomic limit with multi orbitals:** In SI-Fig. 21 we show the results for the multi-orbital case with another set of on-site energies and $U$, where thermopower response shows three robust crossing points as long as orbital energies ($\epsilon$) are smaller than thermal broadening and $U$. However, at lower temperatures, thermopower shows more modulation. In our modelling of experimental thermopower in the atomic limit, we have considered intra-orbital interaction and ignored the inter-orbital Hubbard interactions, which reduces the on-site Hubbard Hamiltonian to four decoupled spin SU(2) terms. However, in the completely flat band limit [10, 17], one expects a fully SU(8)-symmetric Hubbard interaction involving inter-orbital interactions between spin, valley and sublattice flavors, i.e.

$$H_{\text{init}} = U(\sum_{i=1,\sigma}^{4} n_{i\sigma})^2 \tag{18}$$

Considering the the above SU(8) Hubbard interactions, we have calculated the thermopower and shown SI-Fig. 21, where one can see more modulation/crossing points in thermopower depending on the interaction strength. Thus, it suggests that our experimental data in Fig. 2b and Fig. 3c of the manuscript can be best explained under the SU(2) toy model. In Ref [12], using compressibility measurement, the extracted global entropy shows a dumbbell shape with prominent minima only at the Dirac point and some modulation at other integer fillings, indicating that inter-orbital interaction could be weaker. Moreover, such an effective SU(2) model can, in principle, originate at intermediate temperature from some even higher temperature symmetry broken 'parent' states, like valley-symmetry broken intervalley coherent (IVC) state, as indicated by early HF calculations [17]. The behavior of entropy and thermopower is shown in Fig. 21 with $U = 1$ in arbitrary unit. As is clear that SU(8) symmetric interactions predict multiple oscillations and crossings which are not observed in the experiments.

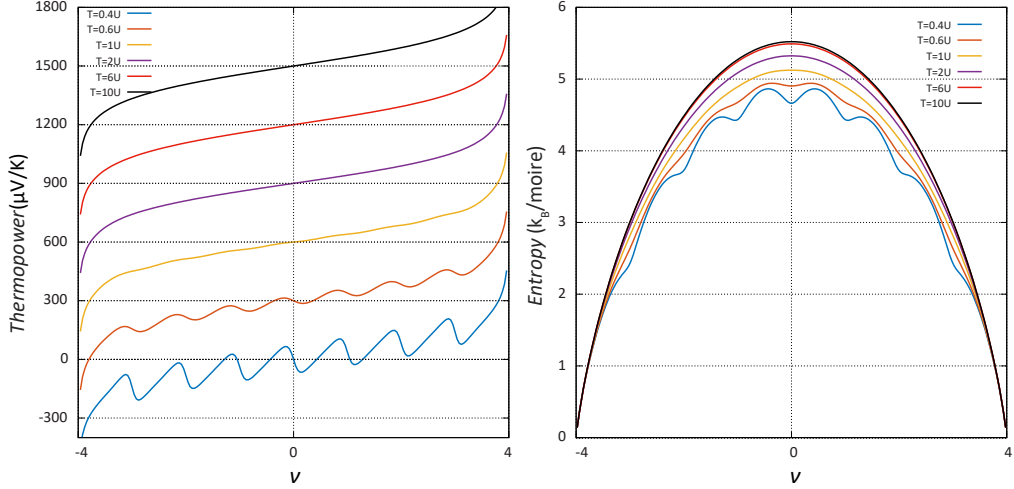

**SI-Fig. 21:** Behavior of thermopower and entropy for a SU(8) symmetric system for various values of temperature in units of $U$.

**Effects of finite bandwidth:** So far, we have discussed the thermopower calculated via the Kelvin formula [Eq.(14)] [34] in the atomic limit with bandwidth or hopping limiting to zero. One can arrive at the atomic limit for thermopower starting with a realistic Hamiltonian with both the hopping ($t$) and onsite Coulomb repulsion ($U$) term and treating hopping perturbatively, as discussed in Refs. [11, 34, 36, 39]. It can be shown that for $t \ll k_B T \ll U$, the calculation is reduced to the atomic limit, known as the Mott-Heikes (MH) formula, which only keeps the second term in Eq.(13), and corresponds closely to the Kelvin formula [Eq.(14)]. This atomic limit, $t \ll k_\mathrm{B} T \ll U$, though may not be strictly justified for any realistic TBLG device, including ours, still provides the important limiting case to understand the most prominent features of the thermopower data in our strongly correlated MATBLG devices in the intermediate temperature range ($10-60K$). It should be noted that even though the atomic limit could qualitatively capture the response for for the MATBLG in the intermediate temperature range, but it is expected to fail quantitatively in the limit of $k_B T \lesssim t \lesssim U$, which is a more realistic situation with respect to our experiments, given the bandwidth $W \simeq 20$ meV [5]. This situation is captured by our DMFT calculations discussed earlier in Sec.SI-10. Notwithstanding, the DMFT results nicely correlate with the thermopower from the atomic limit. Below, we discuss the exact diagonalization (ED) calculations of thermopower using Kelvin formula for a small cluster of moire sites that incorporates finite hopping. These calculations further corroborate the connection between strongly correlated regimes in the extreme atomic limit and away from strictly atomic limit.

**Cluster calculation with extended Hubbard model:** We now discuss exact diagonalization (ED) calculations of thermopower using kelvin formula for a hexagonal cluster of moire sites, as shown in SI-Fig.22(a). Here, the sites denote the AA stacking positions. This was introduced in [5] to describe the spectroscopic signatures of many-body correlations in magic-angle twisted bilayer graphene. The two orbitals per site correspond to the two flat bands. The spin flavors were not explicitly introduced and we continue to follow

the same notation. The Hamiltonian is given by

$$H = \sum_i [\epsilon(n_{1,i} - n_{2,i}) + U_1 n_{1,i} n_{2,i}] + \sum_{\langle ij \rangle} t(c^\dagger_{1,i} c_{1,j} + c^\dagger_{2,i} c_{2,j} + \text{h.c.}) \quad (19)$$

$$+ \sum_{\langle ij \rangle} [V_0(n_{1,i} n_{1,j} + n_{2,i} n_{2,j}) + V_1(n_{1,i} n_{2,j} + n_{2,i} n_{1,j})] \quad (20)$$

where $\langle i, j \rangle$ denotes the nearest neighbors and $1, 2$ labels the orbital index. $V_0$ and $V_1$ characterize the on-site and nearest neighbor interactions. The behavior of thermopower and entropy for different values of temperature is shown in SI-Fig.Fig. 22

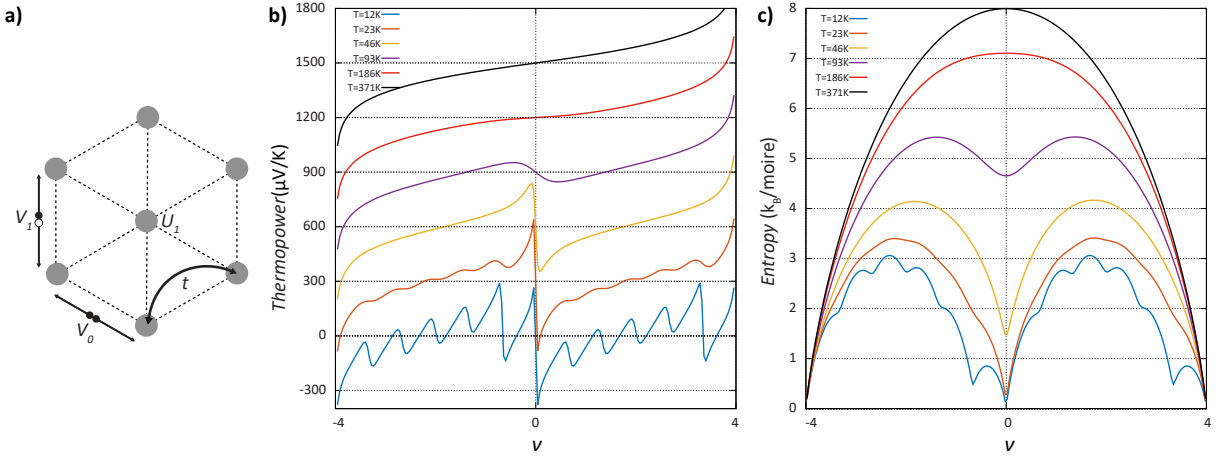

**SI-Fig. 22:** Behavior of thermopower and entropy for extended Hubbard model for, $\epsilon = 9 meV, t = 0.75 meV, U_1 = 30 meV, V_0 = 5 meV, V_1 = 3.8 meV$ for various values of temperature. The parameters has been taken from [5].

The parameter values used for different interaction and energy scales are exactly used as mentioned in, [5], which also studies the MATBLG system with $SiO_2$ gating. At lower temperatures $k_B T \lesssim V_0$ ($T < 15K$) thermopower show several crossings. However, beyond this temperature scale, we only observe three robust crossings. The behaviour of thermopower and corresponding entropy suggests that temperature scales beyond $V_0$ ($3.5 k_B T \sim 5 meV$) are only dominated by the onsite interaction scale $U_1$ between the two orbitals. This, in turn, could also be interpreted as the Hubbard interaction between the two opposite spins on the same site. This justifies why our single site SU(2) Hubbard model is able to qualitatively capture the essential dependence of the thermopower on temperature. In fact, at lower temperatures where the system becomes long-range correlated, we do not expect the thermopower dependence as predicted via atomic limit calculations to be valid, and there, indeed, long-range correlations leading to Dirac revival physics or related phenomena become crucial [1, 6, 15]. In SI-Fig. 22, and SI-Fig. 21 we have rescaled the density of particle filling from 0 to 2 per site to -4 to 4, in order to compare with experimental phenomenology.

**Understanding the response of $V_{Th}$ at different temperature regime:** The experimental data for the thermoelectric voltage, $V_{Th}$ is shown in Fig. 2 and Fig. 5 in the manuscript has lots of features apart from the evolution of three crossing points within $-4 < \nu < 4$. It shows a distinct evolution at the high-temperature

limit, where it can be described as arising from an effective graphene band. As the temperature is lowered to the scale of the bandwidth of the flat band, the behavior changes. In particular, from a single crossing where the sign of thermopower changes just at $\nu = 0$ to three crossings where additional crossing appears near $\nu \sim \pm 1$. In order to explain these responses one needs to consider the higher energy dispersive bands. We show below that the flat band, strong correlation, and effect of a higher dispersive band qualitatively explain our data.

It is useful to identify the different energy scales associated with the system. (i) The bandwidth of the flatband $\equiv W(\sim 10-20~meV)$, (ii) the Hubbard interaction strength ($U \sim 20~meV$), (iii) the gap between the dispersing lower and upper bands ($\Delta_g \sim 30~meV$). The different energy scales are schematically shown in SI-Fig. 23.

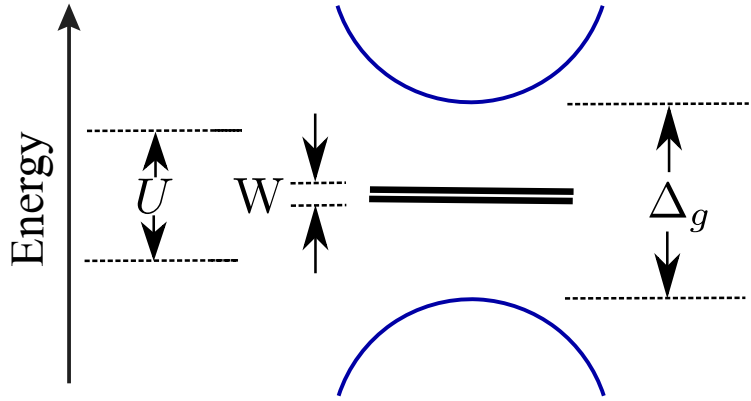

**SI-Fig. 23:** Schematic representation of different energy scales; $U$, $W$ and $\Delta_g$ represent onsite Hubbard interaction strength, the bandwidth of the flatband and the gap between the dispersive bands, respectively.

$T > \Delta_g$: The high-temperature thermopower is positive for $\nu < 0$ and negative for $\nu > 0$ (as depicted in the 1st row of Table 1 in SI-Fig. 24). This is a monolayer graphene-like characteristic of graphene-based systems where electron-hole symmetry renders one crossing at $\nu = 0$, and with respect to this Dirac point, hole (electron) based carriers lead to positive (negative) thermopower for $\nu < 0 (> 0)$ [42].

As the temperature falls below $\Delta_g$, the effect of the higher energy dispersive bands gets reduced, and slowly, the flat bands start dominating with the decrement of the temperature. For $U \lesssim k_B T < \Delta_g$, one can consider the non-interacting atomic limit for the flat band. This will give a single crossing in thermopower at $\nu = 0$ with negative thermopower for $\nu < 0$ and positive for $\nu > 0$ as shown by the 3rd row of Table 1. However, at a slightly higher temperature range, $U < k_B T \approx \Delta_g$, both the contributions from the higher dispersive bands (1st row of Table-1) and flat bands (3rd row of Table-1) are comparable (which are opposite in sign). As a result, it can produce almost zero thermoelectric voltage around $\nu = 0$ as shown by the 2nd row in Table 1.

$k_B T < U$: In this regime, many body correlation or strong correlation dominants, which can be understood

35

using a toy model based on the atomic limit and shows three prominent crossings in thermopower within $-4 < \nu < 4$ as shown in the last row of Table 1.

| Temperature scales | Theoretical $V_{Th}$ | Characteristic Behavior | Experimental $V_{Th}$ |
|---|---|---|---|
| $K_B T > \Delta_g$ | 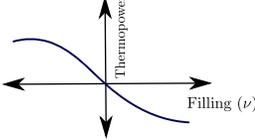 | >0 for v<0 and <0 for v>0 | 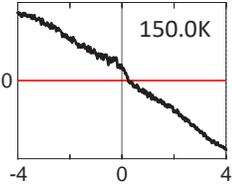 150.0K |
| $U < K_B T \approx \Delta_g$ | 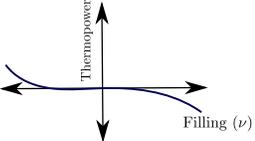 | ~0 in a range of v | 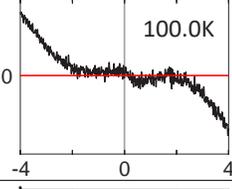 100.0K |
| $U \lesssim K_B T < \Delta_g$ | 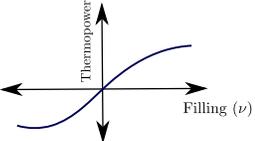 | <0 for v<0 and >0 for v>0 | 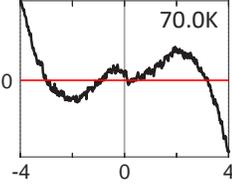 70.0K |
| $K_B T < U$ | 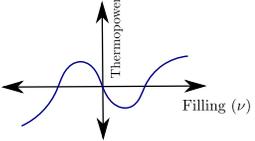 | Three prominent crossings | 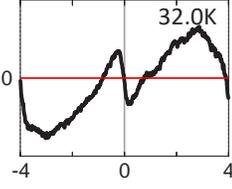 32.0K |

**SI-Fig. 24:** **Table 1:** Qualitative behavior of the thermopower in different temperature regimes. Expected theoretical $V_{Th}$ at different temperature regimes are shown schematically in the 2nd column. The 3rd column indicates the sign of $V_{Th}$ around $\nu = 0$. The last column is the experimentally measured $V_{Th}$.

$k_B T < W$: More precisely, $k_B T < t$, where $t$ is the hopping term[11]. In this limit, the atomic limit can not be used for thermopower. Since other terms related to the band also contribute significantly[11]. We have used HF, DFMT and ED calculations on small clusters to qualitatively capture the measured thermopower at low temperatures [see sections SI-9,10,11].

**Effect of in-plane magnetic field:** In the strongly correlated local moment regime, the application of an in-plane magnetic field will reduce the spin entropy, resulting in a reduction of thermopower, as observed experimentally for oxides [14], where around 40% reduction in thermopower is seen with $B_\parallel = 10T$ at $T = 10K$. Theoretically, the reduction is calculated in the atomic limit in Ref. [11] and is consistent with the experiment. In our experiment, we also observe around $\sim 30\%$ reduction in thermopower with $B_\parallel = 9T$ for the magic-angle twisted bilayer graphene, which is similar to the reduction observed in the entropy measurements in Ref [12]. However, both experiments[12, 13], hardly see any changes within $-1 < \nu + 1$, and our thermopower results show similar behavior. Note that non-interacting electrons

36

are not expected to show any significant changes in thermopower in the studied temperature and magnetic field range as shown in SI-Fig.15. Our thermopower results (crossing points, its temperature dependence, magnetic field dependence) are best explained with a simple toy model of atomic limit confirming the effect of many-body correlation/local moments in magic-angle twisted bilayer graphene. However, to understand all the features in our thermopower study quantitatively, further theoretical studies are required.

## SI-12: Different origins of entropy from non-interacting band electrons to strongly correlated electrons, and the relation between entropy and thermopower:

In this section, we will discuss the microscopic origins of entropy. We will consider three different regimes: (i) For the non-interacting band electrons with $k_B T \ll \varepsilon_f$ and $U \approx 0$, the thermopower ($S_{th} \sim \frac{k_B}{e} \times \frac{k_B T}{\varepsilon_f}$) is essentially the entropy per particle ($S_{en}/N \sim k_B \times \frac{k_B T}{\varepsilon_f}$), with $S_{en}$ representing the total entropy and $N$ the number of particles. In this limit, only a fraction of the total particles, of the order of $\sim \frac{k_B T}{\varepsilon_f}$, contributes to the total entropy, resulting in a much lower magnitude. The origin of this entropy can be understood through the $k$-space or momentum space picture. Here, the available $k$-states within the energy scale of $k_B T$ around the Fermi energy represent microstates with well-defined energy with spin degeneracy. In this limit, it is not feasible to separate the total entropy into real-space configurational entropy contributions, as the energy is not well-defined in real space. Similarly, the spin entropic part cannot be distinctly separated from the total entropy.

(ii) For $\varepsilon_f \ll W \ll k_B T$ with $U \approx 0$ (where $W$ is the bandwidth), all the carriers in the band will contribute to the entropy, and the entropy per particle will be of the order of $k_B$. This situation is similar to the flat-band limit, where both $k$-space and real-space microstates have well-defined energies. Therefore, one can consider the real-space configurational entropy. This can be qualitatively understood because all $k$-points contribute to the entropy. Since the total number of $k$-points in a band equals the number of unit cells in real space, the $k$-space and real-space pictures yield the same result. For $N$ particles distributed among $M$ sites (with $N < M$), the number of available microstates is given by $\Omega \sim {}^{2M}C_N$, where the factor of 2 accounts for the spin degeneracy. The entropy $S_{en}$ is then given by $S_{en} \sim k_B \ln \Omega$. Theoretically, separating the total entropy into configurational and spin entropy components is difficult. In this high-temperature limit, the thermopower and entropy are related as $S_{th} \sim -\frac{1}{e} \times \frac{\partial S_{en}}{\partial N}$, known as the Heikes limit[11]. However, the total entropy of this limit will differ from that associated with local moments formation when $U \gg k_B T \gg W$, because, in the latter case, there is no double occupancy of a site.

(iii) For the limit $U \gg k_B T \gg W$, the total entropy can be unambiguously separated into real-space configurational entropy and spin entropy. This is explained as follows: For $N$ particles distributed among $M$ sites (with $N < M$), the number of available microstates is given by $\sim {}^M C_N \times 2^N = \Omega_{conf} \times \Omega_{spin}$, where the first term arises from the real-space configuration and the second term from the spin contribution (2 possible states with up or down spin). The entropy $S_{en}$ is then given by $S_{en} \sim k_B \ln \Omega = k_B \ln \Omega_{conf} + k_B \ln \Omega_{spin} = S_{conf} + S_{spin}$, where $S_{spin} = N k_B \ln 2$. Note that the separation of configurational entropy and spin entropy is possible due to the formation of local moments, meaning there is no double occupancy due to the many-body correlation effect (large on-site Coulomb repulsion, $U$). In this limit, the thermopower and entropy are related by the Heikes limit as $S_{th} \sim -\frac{1}{e} \times \frac{\partial S_{en}}{\partial N}$[11].

The effect of an in-plane magnetic field ($B$) on entropy with large $U$ can be qualitatively understood by looking at how the spin entropy ($k_B ln 2$) gets reduced and eventually approaches zero at a larger field

($k_B ln 1$). In other words, the microstates will change from $\sim {}^M C_N \times 2^N$ at zero $B$ to $\sim {}^M C_N \times 1^N$ = ${}^M C_N$ at larger $B$. The effect of $B$ on spin entropy for the band electrons is tricky since, for the band electrons, one can not separate out spin entropy from the total entropy. However, one can roughly estimate the spin entropic contribution per particle, which will be of the order of $\sim k_B \times \frac{k_B T}{\varepsilon_f} \times \ln 2$, where $k_B \ln 2$ is the maximum spin entropy per particle. Due to polarization, the spin entropic part of band electrons will be reduced by applying an in-plane magnetic field, but its variation is significantly reduced by the factor $\frac{k_B T}{\varepsilon_f}$. However, for the limit of $k_B T > W$ with band/itinerant electrons ($U \approx 0$), one can have significant entropy change with $B$. Indeed, one would expect the microstates will change from $\sim {}^{2M} C_N$ at zero $B$ to $\sim {}^M C_N$ at larger $B$ due to spin polarization.

Note that the above discussions about entropy are at a qualitative level for the limiting cases; in reality, finite temperature broadening as well as finite $U$ will modify it. We have theoretically calculated the total entropy for different temperatures and $U$ in the SI-11 section using atomic limit-based calculation and shown the corresponding thermopower using Heikes limit[11]. Similarly, for the band electrons, using the semi-classical description, we have calculated the thermopower, and shown that the application of $B_\parallel = 10T$ hardly changes the magnitude of thermopower, as discussed in Sec.SI-8. Whereas in our experiment, we observe a reduction of 30%, consistent with the compressibility measurement [12] on MATBLG and thermopower measurements in other strongly correlated systems like oxides [14]. The atomic limit-based theory can qualitatively explain our in-plane magnetic field data, as shown in Fig. 1 of Ref. [11]. However, the reduction of thermopower with $B$ alone is not conclusive evidence. Instead, the combined effect of $B$ and the behavior of the crossing points in thermopower with temperature (as observed in our experiment, where three crossing points remain fixed from 5K to 60K for MATBLG) provide insight into the presence of local moments.

### SI-13: Estimation of entropy from the measured thermopower:

Entropy is calculated in units of $k_B$/moire by performing a definite integral of the measured $S_{th}$ over $-4 < \nu < 4$ as follows,

$$S_{en}^M = \int_{-4}^{+4} S_{th} d\nu \tag{21}$$

where $S_{en}^M$ is entropy per moire unit cell. Since we could estimate the temperature gradient at lower temperatures ($< 10K$) using noise thermometry, we convert the thermoelectric voltage into thermopower as shown in SI-Fig25a and the corresponding entropy shown in SI-Fig25b. Note that we could not convert the thermoelectric voltage into exact thermopower and entropy due to the lack of an exact measure of the temperature gradient at the elevated temperature $> 10K$. However, one can roughly estimate the value of entropy at the elevated temperature assuming a linear entropy increment with bath temperature as shown in SI-Fig. 25c for $20K$, giving a maximum value around $\sim 0.6 k_B$ per moire, which is a similar order of magnitude as measured in compressibility measurement in Ref [12, 13].

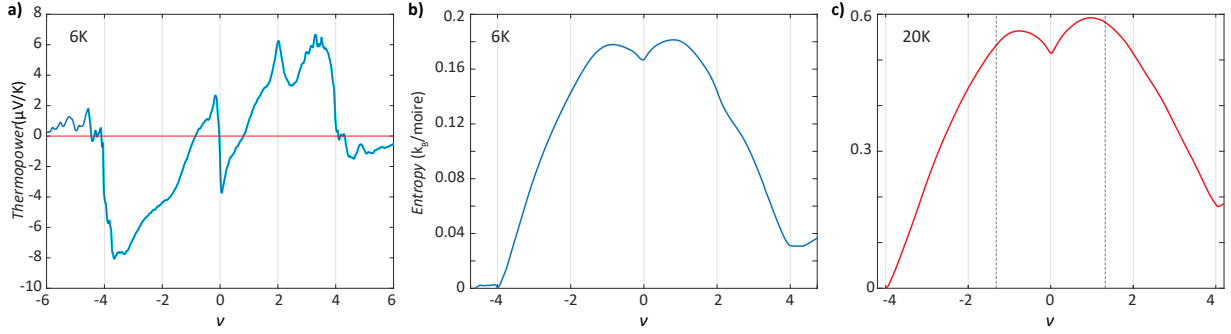

**SI-Fig. 25: Estimation of entropy from thermopower: (a)** Thermopower with $\nu$ at $6K$. **(b)** Entropy with $\nu$ extracted using Eq.21. **(c)** Approximate estimation of entropy at $20K$ assuming a linear entropy increment with temperature.

### SI-14: Entropy of non-interacting itinerant electrons:

For an ideal non-interacting electron gas entropy per unit area can be described by,

$$S_{en}^{A} = -k_{\mathrm{B}} \int_{\varepsilon} \{f(\varepsilon) \ln f(\varepsilon) + [1 - f(\varepsilon)] \ln [1 - f(\varepsilon)]\} g(\varepsilon) d\varepsilon \tag{22}$$

where, $S_{en}^{A}$ is entropy per unit area, $f(\epsilon)$ is the Fermi function and $g(\epsilon)$ is the DOS. Using the continuum DOS with bandwidth of 10meV (see SI-Fig 11), the calculated entropy (per moire unit cell) as a function of filling is shown in SI-Fig 26 at various temperatures. Note that the magnitude of the entropy is of the order of few $k_B$ per moire, which is in the similar range of atomic limit calculation.

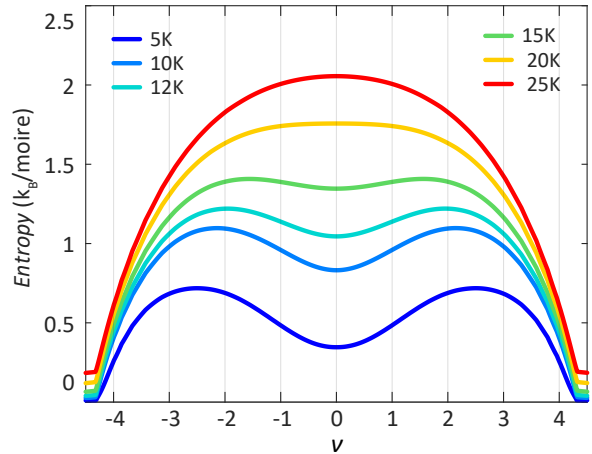

**SI-Fig. 26: Entropy of non-interacting electron gas:** Entropy versus filling at various temperature calculated using continuum DOS [15] for itinerant electron gas with $W \sim 10meV$.

39

**SI-15: Comparison of measured $V_{Th}$ with Mott formula:**

The semi-classical equation (described in Eq. 2) can be further simplified using the Somerfeld approximation to Mott's formula as;

$$S_{Mott} = -\frac{\pi^2 k_B^2 T}{3e} g(\epsilon)_{\epsilon=\varepsilon_f} \frac{dln(R)}{dn}, \tag{23}$$

According to Eq. 23, the term $dln(R)/dn$ is responsible for sign changes observed in Mott-thermopower ($S_{Mott}$). To know whether the Mott relation is valid, we will compare the experimentally measured ($V_{Th}$) and the Mott thermopower using $dln(R)/dn$. We present a detailed comparison of $V_{Th}$ and $dln(R)/dn$ as a function of filling $\nu$ at different temperature ranges for the four representative devices. In SI-Fig. 27,

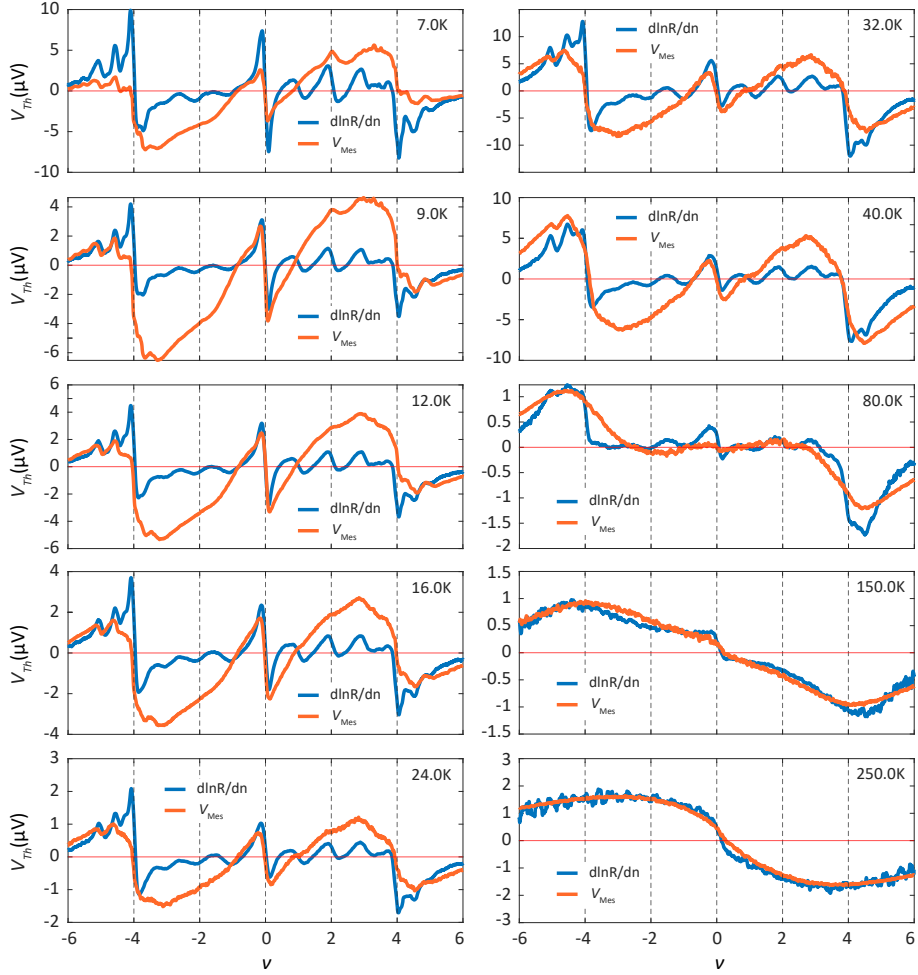

**SI-Fig. 27: Comparison between $dln(R)/dn$ and measured $V_{Th}$ for MATBLG ($1.05^0$) device:** Comparison between measured $V_{Th}$ signal and $dln(R)/dn$ (scaled to arbitrary units for visual clarity) at different temperatures. Blue and orange lines present the $dln(R)/dn$ and $V_{Th}$, respectively.

we demonstrate the evolution of thermoelectric voltage in comparison to the Mott formula ($dln(R)/dn$) at various temperature ranges for the MATBLG ($1.05^0$) device. As discussed in the main manuscript, below

40

$T < 120K$, the measured thermopower (in orange solid line) lacks many of the crossing points predicted by $dln(R)/dn$ (in blue solid line). Above $120K$ the $V_{Th}$ matches well with Mott relation with a graphene-like thermopower spectrum. Much like MATBLG, the aforementioned Mott violation can also be observed in near MATBLG ($0.95^0$) device. SI-Fig. 7a shows the resistance at different temperatures. As can be seen in figure SI-Fig. 28, below $50K$, $V_{Th}$ is generally more symmetric between the valence and conduction band, showing only one crossing point for each conduction and valence band, whereas the Mott relation, $dln(R)/dn$ shows many crossing points. Here also, the measured $V_{Th}$ matches quite well with the Mott relation, $dln(R)/dn$ at higher temperatures ($> 90K$). Similarly, the Mott violation is observed for $1.2^0$ graphite-gated TBLG device at lower temperatures. In TDBLG, there is a distinct departure of the $V_{Th}$ behaviour as compared to the MATBLG and near-MATBLG devices. Unlike the twisted bilayer (TBLG) devices, at low temperatures ($< 20K$), $V_{Th}$ shows qualitative agreement with the Mott relation (see SI-Fig. 29) in terms of general peak and the number of zero crossing points. At larger temperatures, the asymmetry in $R$ (see SI-Fig. 8a) also manifests in the $dln(R)/dn$ data, and $dln(R)/dn$ mostly maintains a positive value over the whole density range of the flat band, and Mott starts violating for the valence band in terms of sign changes. We believe that one needs to consider the higher energy bands to explain the data.

In summary, all these devices show Mott violation in terms of crossing points in some temperature ranges. In the next section, we discuss different possible origins for the Mott-law violation and discuss how the Mott relation can be violated even away from strongly correlated local moment regimes.

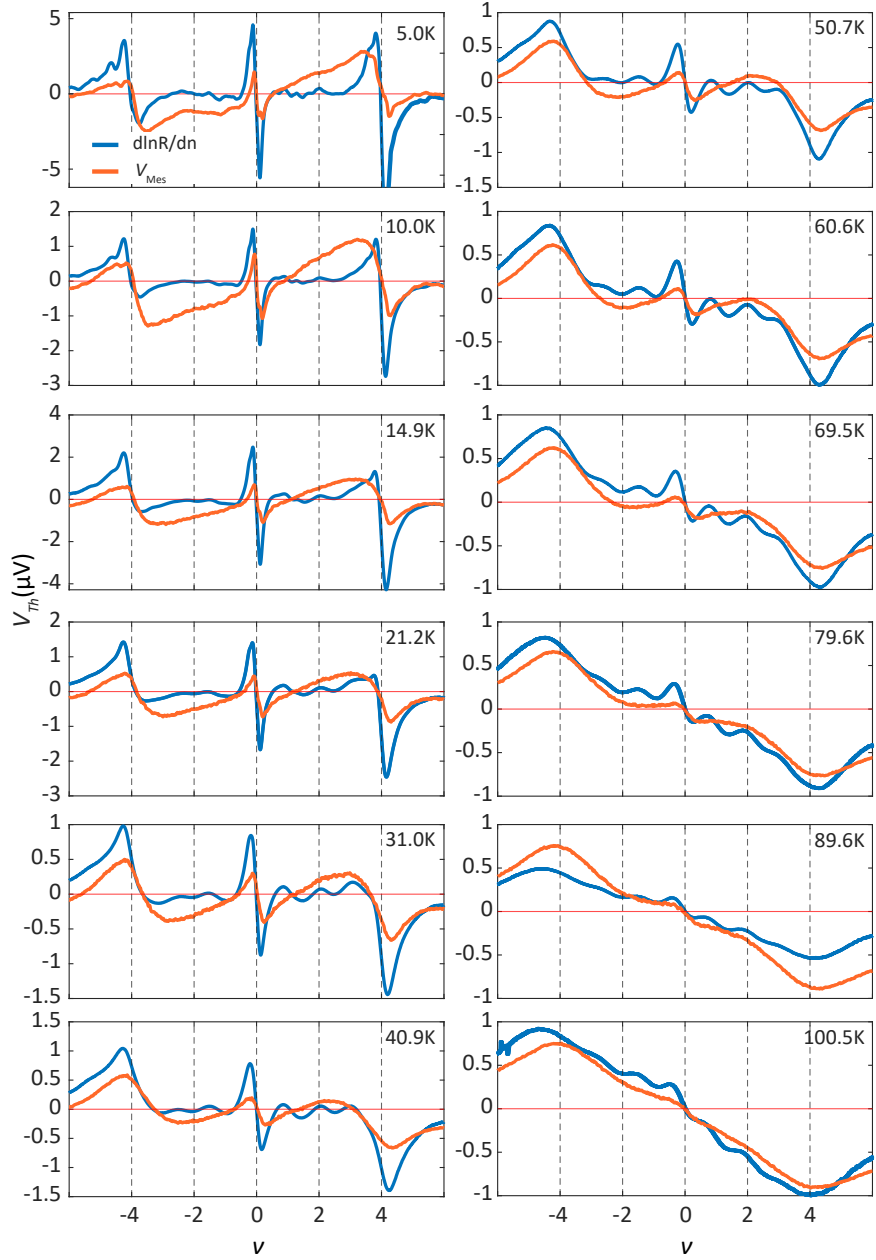

**SI-Fig. 28: Comparison between Mott relation ($dln(R)/dn$) and measured $V_{Th}$ for near MATBLG ($0.95^0$) device:** Comparison between measured $V_{Th}$ signal and $dln(R)/dn$ (scaled to arbitrary units for visual clarity) at different temperatures. Blue and orange lines present the $dln(R)/dn$ and $V_{Th}$, respectively.

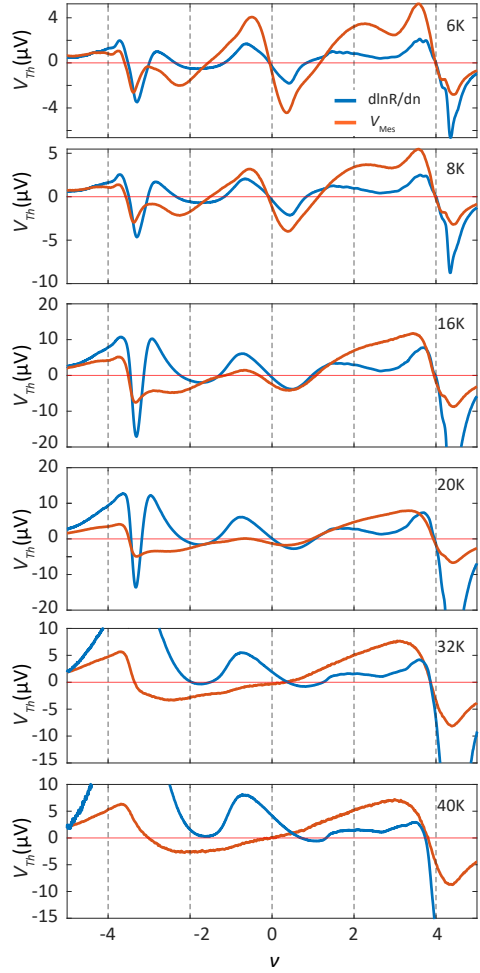

**SI-Fig. 29: Comparison between Mott relation ($dln(R)/dn$) and measured $V_{Th}$ for TDBLG ($1.1^0$) device:** Comparison between measured $V_{Th}$ signal and $dln(R)/dn$ (scaled to arbitrary units for visual clarity) at different temperatures. Blue and orange lines present the $dln(R)/dn$ and $V_{Th}$, respectively.

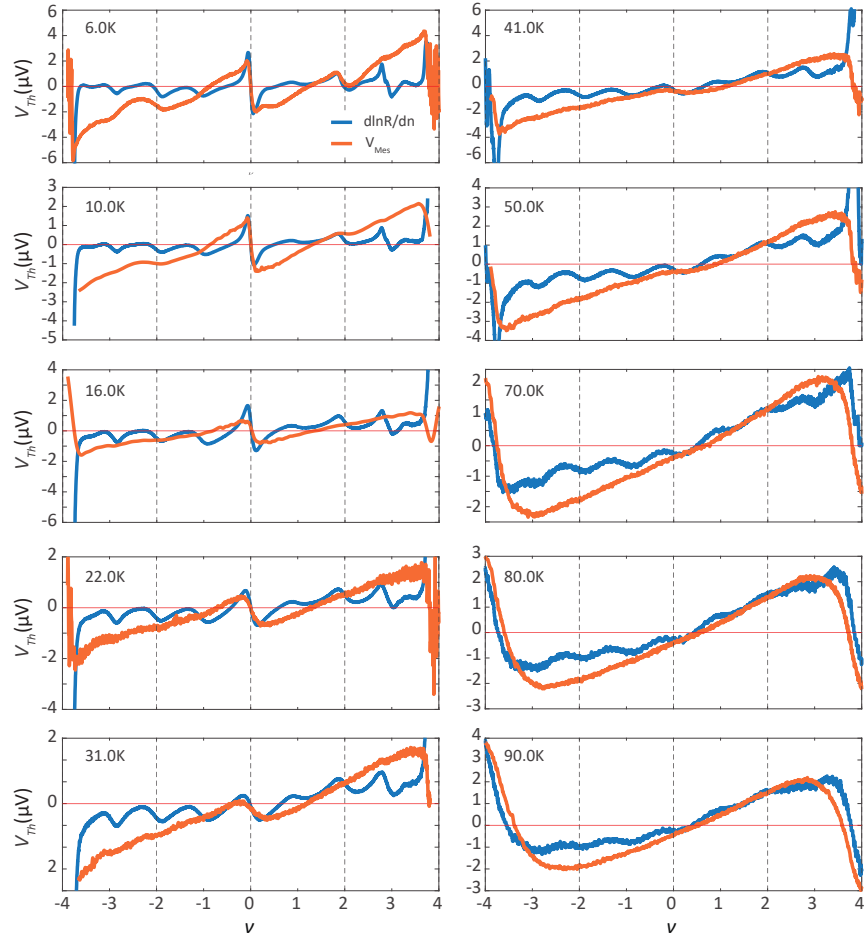

**SI-Fig. 30: Comparison between Mott relation ($dln(R)/dn$) and measured $V_{Th}$ for graphite-gated TBLG ($1.2^0$) device:** Comparison between measured $V_{Th}$ signal and $dln(R)/dn$ (scaled to arbitrary units for visual clarity) at different temperatures. Blue and orange lines present the $dln(R)/dn$ and $V_{Th}$, respectively.

## SI-16: Various origins for Mott violation:

In the linear-response regime, for weakly interacting electronic systems with well-defined quasiparticles, various transport coefficients, such as electrical conductivity, thermal conductivity, and thermopower coefficient, can be obtained using semi-classical equations and Boltzmann transport theory. The expression for thermopower can be written as follows:

$$S_{th} = -\frac{1}{|e|T} \frac{\int_{-\infty}^{\infty} (\epsilon - \mu)\sigma(\epsilon)\frac{\partial f(\epsilon)}{\partial \epsilon} d\epsilon}{\int_{-\infty}^{\infty} \sigma(\epsilon)\frac{\partial f(\epsilon)}{\partial \epsilon} d\epsilon} \tag{24}$$

$$\sigma(\epsilon) = \frac{2e^2}{3m_e} \rho(\epsilon)\epsilon\tau(\epsilon) \sim \rho(\epsilon)v^2(\epsilon)\tau(\epsilon) \tag{25}$$

Where, $\sigma(\epsilon)$ is the energy-dependent conductivity function, $\rho(\epsilon)$ is density of states, $v(\epsilon)$ is the band velocity and $\tau(\epsilon)$ is the scattering time. The equation can be further simplified using the Sommerfeld approximation. The latter assumes that the conductivity $\sigma(\epsilon)$ varies slowly with energy. This leads to the semi-classical Mott formula by retaining only the first-order correction and ignoring terms involving higher derivatives.

$$S_{th} = -\frac{\pi^2 k_B^2 T}{3e} \frac{d \ln \sigma(\varepsilon)}{d\varepsilon}\bigg|_{\varepsilon=\varepsilon_F} = \left(\pi^2 k_B T/3e\right)(d\ln(R)/dn)g(\epsilon) = \frac{\pi^2}{3} \frac{k_B^2 T}{|e|\mu} \tag{26}$$

Note that this condition is typically met when the temperature is much lower than the chemical potential ($K_B T < \mu$). In Fermi liquid metals, the measured thermopower generally aligns very well with the Mott formula.

When the measured thermopower for a system violates the Mott formula, there are the following ways to look at the violation;
(i) The deviation of the magnitude of the measured thermopower from the value predicted by the Mott formula.
(ii) The differences in sign change(s) of the thermopower or thermopower crossing point with carrier density from that predicted by the Mott formula ($dln(R)/dn = 0$).
(iii) The temperature dependence of thermopower, which is expected to be linear according to the Mott formula.
The verification of the first criteria is often challenging because it requires the exact expression of the density of states of the system. For the third criteria, one needs to know the temperature gradient to convert the thermoelectric voltage to thermopower. In practice, the second criterion is more convenient for detecting Mott violations, as it does not depend on the density of states or the temperature gradient. In our manuscript, we used the second criteria to verify whether the Mott formula is violated or not using the thermoelectric voltage only. This approach has been routinely used in the literature for graphene-based systems [2, 4, 43, 44]. It should be noted that the presence of one or the other form of Mott violation does not indicate the origin of the Mott violation are the same in all scenarios.

Thermopower crossing points occurs due to a change in the character of carriers from electron-like to hole-like or vice versa. More precisely the electron-hole or particle-hole symmetry determines the crossing point of thermopower, as seen from the semi-classical equation. For example, at the Dirac point of graphene, the thermopower is expected to be zero due to electron-hole symmetry, and the experimentally measured thermopower crossing point matches perfectly with the Mott formula based on the second criteria ($dR/dn = 0$). However, the magnitude of the measured thermopower around the Dirac point does not perfectly match the Mott formula (first criteria), as reported for extremely clean graphene devices in Ref [45]. This is not surprising since around the Dirac point within the energy range of $k_BT$, the Mott formula is not expected to be valid ($\mu < K_BT$).

For twisted bilayer graphene at an angle of $0.26^0$ and $1.66^0$, which is far from the magic angle of $1.1^0$, the crossing point predicted by the Mott formula matches well with the measured thermopower, as shown in SI-Fig. 21 of Ref [2]. However, for twisted bilayer graphene around the magic angle, many devices [2, 4] have reported violations of the Mott formula in terms of the crossing point (second criteria). The violation of the Mott formula (in the context of the crossing point) has been attributed to interaction effects, which can be summarized as follows:

1. Interaction-Induced Symmetry Breaking in Magic-Angle Twisted Bilayer Graphene:
Modification of Band Structure: Interactions alter the rigid, non-interacting band structure (BM model) and induce phase transitions around the integer fillings of the Moire unit cell due to symmetry breaking. This results in a highly particle-hole (PH) asymmetric band structure around the integer fillings, leading to a significantly large thermopower at the symmetry breaking points, in complete violation of the Mott formula. This is seen in our earlier work [2] at lower temperature $T < 5K$. This phenomenon was theoretically captured using weak interactions for band electrons using Hartree-Fock (HF), a mean-field theory that renormalizes the band structure of itinerant electrons and explains the experimental results.


2. Interaction-Induced Dynamical Effect in Twisted Bilayer Graphene:
Mott Violation in Non-Magic Angle devices: It has also been reported in the literature that twisted bilayer graphene with twist angles between $1.3^0$ and $1.6^0$, away from the magic angle, exhibits Mott violations in terms of the crossing point, as shown in Ref [4]. For these devices, there are no signatures of symmetry breaking (no resistance peaks at integer fillings). The Mott violation (crossing point) is understood in terms of interaction-induced dynamical effects that significantly modify the transport density of states of the itinerant electrons. This phenomenon was captured using Dynamical Mean Field Theory (DMFT) calculations.

It is important to note that the effects mentioned above are highly sensitive to temperature. For example, the symmetry-breaking effect in magic-angle twisted bilayer graphene disappears above 20K. Similarly, the crossing point resulting from the dynamical effect in non-magic angle twisted devices significantly changes with temperature [SI-Fig. 27 of [4]].

3. Strong Interaction Regime (Strongly Correlated Regime):
In this regime, the configurational and spin entropy from the localized carriers predominate in determining the thermopower. The atomic limit calculation demonstrates a thermopower crossing that violates the Mott formula[11, 34, 36, 39]. However, the crossing point of thermopower remains temperature-independent over

an order of magnitude variation. Further, in this regime, the thermopower is expected to exhibit reduction with the application of an in-plane magnetic field due to the reduction of the spin entropy of the local moments. In the strong interaction regime, the topological heavy fermion (THF) model also gives rise to Mott violation [43, 44, 46, 47].



\end{document}